\def\llncs{0}
\def\fullpage{1}
\def\anonymous{0}
\def\draft{0}
\def\submission{0}
\def\final{0}

\ifnum\submission=1
\def\llncs{1}
\def\draft{0}
\def\anonymous{0}
\def\fullpage{0}
\fi

\ifnum\llncs=1
    \documentclass{llncs}
    \ifnum\fullpage=1
    \usepackage{fullpage}
    \fi
\else
    \documentclass[letterpaper,hmargin=1.05in,vmargin=1.05in]{article}
    \ifnum\fullpage=1
    \usepackage{fullpage}
    \fi
    \usepackage{microtype}
\microtypesetup{
  final,        %
  activate={true,nocompatibility},  %
  kerning=true,
  spacing=true,
  protrusion=true
}
\microtypecontext{spacing=nonfrench}
\fi

\usepackage{verbatim}
\usepackage{authblk}
\usepackage{graphicx} %
\usepackage{physics} %
\usepackage{xcolor} %
\usepackage{amsmath}
\allowdisplaybreaks[4] %
\usepackage{amssymb}
\usepackage{mathtools}

 \usepackage{amsthm}
 \usepackage{thmtools}
\usepackage{enumitem} %
\usepackage{dsfont}
\usepackage[colorlinks=true,linkcolor=magenta,citecolor=blue,pagebackref=true,hypertexnames=false,pdftex,pdfpagelabels,bookmarks,hyperindex,hyperfigures]{hyperref}
\usepackage{dashbox}
\usepackage{fancybox,framed}
\usepackage[skins]{tcolorbox}
\usepackage{subcaption}
\usepackage{breakcites}
\usepackage{relsize}
\usepackage[
n,
advantage,
operators,
sets,
adversary,
landau,
probability,
notions,
logic,
ff,
mm,
primitives,
events,
complexity,
asymptotics,
keys,
]{cryptocode}
\usepackage{comment} %
\usepackage{tikz}
\usetikzlibrary{shadings}
\usetikzlibrary{arrows}
\usetikzlibrary{hobby,backgrounds,calc,trees,decorations.pathmorphing,arrows.meta,quotes}
\let\subparagraph\paragraph
\usepackage{titlesec}
\titleformat*{\paragraph}{\normalsize\bfseries}

\usepackage[capitalise,nameinlink]{cleveref}

\ifnum\draft=1
	\newcommand{\gm}[1]{\textcolor{orange}{\(\langle\langle\)Garazi: #1\(\rangle\rangle\)}}
	\newcommand{\minki}[1]{\textcolor{blue}{\(\langle\langle\)Minki: #1\(\rangle\rangle\)}}
	\newcommand{\huy}[1]{\textcolor{purple}{\(\langle\langle\)Huy: #1\(\rangle\rangle\)}}
	\renewcommand{\sam}[1]{\textcolor{olive}{\(\langle\langle\)Samuel: #1\(\rangle\rangle\)}} %
\else
	\newcommand{\gm}[1]{}
	\newcommand{\minki}[1]{}
	\newcommand{\huy}[1]{}
	\renewcommand{\sam}[1]{}
\fi

\ifnum\llncs=0
	\newtheorem{theorem}{Theorem}[section]
	\newtheorem{lemma}[theorem]{Lemma}
	\newtheorem{corollary}[theorem]{Corollary}
	
	\newtheorem{definition}[theorem]{Definition}

	\newtheorem{claim}[theorem]{Claim}

	\theoremstyle{remark}
	\newtheorem{remark}[theorem]{Remark}

	\newtheorem{algorithm}{Algorithm}
\else
	\spnewtheorem{algorithm}{Algorithm}{\bfseries}{\rmfamily}

	\spnewtheorem{claim}{Claim}{\bfseries}{\itshape}
\fi
\crefname{appendix}{Appendix}{Appendices}
\Crefname{appendix}{Appendix}{Appendices}

\newcommand{\Haar}{\mathcal{H}}
\newcommand{\cA}{\mathcal{A}}
\newcommand{\cB}{\mathcal{B}}
\newcommand{\N}{\mathbb{N}}

\newcommand{\cM}{\mathcal{M}}

\newcommand{\cS}{\mathcal{S}}

\newcommand{\eps}{\varepsilon}

\newcommand{\eqdef}{\vcentcolon=}

\renewcommand{\epsilon}{\varepsilon}

\renewcommand\prob[2][]{\Pr_{#1}\left[#2\right]}

\newcommand{\Exp}{\mathbb{E}}

\renewcommand{\secpar}{\lambda}

\newcommand{\gen}{\mathsf{Gen}}

\newcommand{\oracle}{\mathcal{O}}
\newcommand{\pspace}{\mathbf{PSPACE}}

\newcommand{\qpspace}{\mathbf{QPSPACE}}
\newcommand{\qcma}{\mathbf{QCMA}}
\newcommand{\coqcma}{\mathbf{coQCMA}}
\newcommand{\bqp}{\mathbf{BQP}}
\newcommand{\p}{\mathbf{P}}
\newcommand{\np}{\mathbf{NP}}
\newcommand{\sampqcma}{\mathbf{sampQCMA}}
\newcommand{\sampcoqcma}{\mathbf{sampcoQCMA}}

\createpseudocodeblock{pcb}{boxed,center}{}{}{} %

\newcommand{\bit}{\{0,1\}}

\newcommand{\ptest}{\mathtt{PTEST}}
\newcommand{\Tom}{\mathtt{Tom}}
\newcommand{\Ric}{\mathbf{Ric}}

\newcommand{\rA}{{\mathbf{A}}}
\newcommand{\rB}{{\mathbf{B}}}
\newcommand{\rC}{{\mathbf{C}}}

\ifnum\draft=1
	\newcommand{\modified}[1]{{\color{red}{#1}}}
\else
	\newcommand{\modified}[1]{#1}
\fi

\newcommand{\algoname}{ancilla-resetting}

\newcommand{\LPRSGs}{long PRSGs}

\newcommand{\SPRSGs}{short PRSGs}
\newcommand{\QPRG}{QPRG}
\newcommand{\QPRGs}{QPRGs}

\newcommand{\neglQPRGs}{\QPRGs{}}

\newcommand{\TV}{\mathrm{TV}}
\newcommand{\advb}{\mathcal B} %

\title{On Limits on the Provable Consequences of Quantum Pseudorandomness}
\date{}

\ifnum\llncs=1
    \ifnum\anonymous=1
        \author{}
        \institute{}
    \else
        \author{
        Samuel Bouaziz{-}{-}Ermann\inst{1} \and
        Minki Hhan\inst{2} \and
        Garazi Muguruza\inst{3} \and
        Quoc-Huy Vu\inst{4}
        }
        \institute{
        Okinawa Institute of Science and Technology Graduate University, Japan \\
        \email{samuel.bouazizermann@oist.jp}\and
        KAIST, Daejeon, Korea
        \\\email{minkihhan@kaist.ac.kr}\and
        Surf, The Netherlands \\\email{garazimugu@gmail.com} \and
        De Vinci Higher Education, De Vinci Research Center, Paris, France \\\email{quoc.huy.vu@ens.fr}
        }
    \fi
\else
    \author[1]{Samuel Bouaziz{-}{-}Ermann\footnote{This work was done while the author was affiliated at Sorbonne Université, CNRS, LIP6, France.}}
    \author[2]{\hskip 1em Minki Hhan\footnote{This work was done in part while the author was affiliated with KIAS, Korea, and UT Austin, USA.}}
    \author[3]{\hskip 1em Garazi Muguruza}
    \author[4]{\hskip 1em Quoc-Huy Vu}
    \affil[1]{{\small Okinawa Institute of Science and Technology Graduate University, Japan}
    \authorcr{\small samuel.bouazizermann@oist.jp}}
    \affil[2]{{\small KAIST, Daejeon, Korea}
    \authorcr{\small minkihhan@kaist.ac.kr}}
    \affil[3]{{\small Surf, The Netherlands}
    \authorcr{\small garazimugu@gmail.com}}
    \affil[4]{{\small De Vinci Higher Education, De Vinci Research Center, Paris, France}
    \authorcr{\small quoc.huy.vu@ens.fr}}
    
\fi

\begin{document}

\maketitle

\begin{abstract}
	There are various notions of quantum pseudorandomness, such as pseudorandom unitaries (PRUs), pseudorandom state generators (PRSGs), pseudorandom function-like state generators (PRFSGs), and quantum-computable pseudorandom generators (QPRGs).
	Unlike the different notions of classical pseudorandomness, which are known to be existentially equivalent to each other, the relations among quantum pseudorandomness have yet to be fully established.

	We present evidence suggesting that some forms of quantum pseudorandomness are unlikely to be constructed from the others.
	This indicates that quantum pseudorandomness behaves quite differently from classical pseudorandomness.

	Our main result is a unitary oracle separation where log-length output PRFSGs exist
	but QPRGs with negligible
	correctness error do not. This result suggests that the
	inverse-polynomial error in the state-of-the-art construction of QPRGs
	from log-length PRSGs is inherent.
	To achieve this, we prove a novel differential geometric
	\emph{barrier theorem} for the product Haar measure on quantum states
	which replaces the usual concentration
	inequalities by certifying a non-negligible ``gap'' between two large
	separated sets.
    
    Our separation is based on an oracle that outputs Haar random quantum states for each bit string, which can be viewed as a quantum state version of the random oracle model, where output strings are replaced by quantum states.

    Variations of this oracle can be used to study other relationships between quantum cryptographic primitives, and we use it to achieve partial separations, which highlight technical difficulties when dealing with ancillary
	registers, measurements, and adaptivity in the quantum setting.
	More precisely, we obtain separations showing limitations of: (i) deriving ancilla/measurement-free pseudorandom unitaries (PRUs) from PRFSGs, and (ii) a natural way of constructing super-log-length output PRSGs from log-length output PRFSGs.
	The latter partly
	complements the known hardness of shrinking the PRSG output lengths.
\end{abstract}
\ifnum\llncs=1
	\keywords{Quantum Pseudorandomness \and Quantum Cryptography \and Pseudorandom Unitaries \and Pseudorandom State Generators}
\fi

\ifnum\submission=1
\else
	\clearpage
	\newpage
	\setcounter{tocdepth}{2}
	\tableofcontents
	\newpage
\fi

\section{Introduction}
In classical cryptography, computational pseudorandomness generated by pseudorandom generators (PRGs) and functions (PRFs) serves as a central resource. It can be used for many applications, such as commitments~\cite{JC:Naor91}, digital signatures~\cite{STOC:Rompel90}, and symmetric key encryptions.
Furthermore, the existence of PRGs and PRFs is necessary for the existence of almost all cryptographic primitives with computational security, including one-way functions (OWFs)~\cite{HILL99}.

However, when we treat the world as operating under the laws of quantum mechanics, the notion of pseudorandomness must be revisited. Ji, Liu, and Song~\cite{C:JiLiuSon18} proposed the first two inherently
quantum pseudorandom primitives, pseudorandom state generators (PRSGs) and unitaries (PRUs).
Quantum pseudorandomness has been shown to be useful for constructing many (quantum) cryptographic primitives, for example, PRSGs imply quantum commitments and oblivious transfers~\cite{C:MorYam22,C:AnaQiaYue22}.
On the other hand, Kretschmer showed that PRSGs and PRUs (with super-logarithmic output lengths) are potentially weaker primitives than classical pseudorandomness by presenting an oracle separation~\cite{Kre21}.
The dramatic interest in fundamentally quantum cryptographic primitives emerged as a new cryptographic direction possible even in the world without one-way functions.

Quantum pseudorandomness turns out to be quite different from its classical counterparts. To begin with, the length becomes an important parameter in quantum cryptography.
Given an \(\secpar\)-bit input seed,
\cite{C:BraShm20} shows that \(c\log \secpar\)-length output PRSGs exist unconditionally for \(c\ll1\) whereas achieving \(c\ge1\) requires computational assumptions. We name \(c\log \secpar\)-length and superlogarithmic-length PRSGs by short PRSGs and long PRSGs, respectively.
\cite{BM24,C:ChuGolGra24} shows that \LPRSGs{} cannot be used to construct \SPRSGs{}, highlighting that the output lengths of PRSGs cannot be shrunk.
Note that in the classical setting, the output length of PRGs can easily be shrunk and lengthened arbitrarily.

Contrary to the \LPRSGs{} that are separated from classical cryptography, \cite{ITCS:AnaLinYue24} shows that \SPRSGs{} can be used to construct a variant of PRGs called pseudodeterministic quantum-computable PRGs (\QPRGs{}) with an inverse-polynomial pseudodeterminism error.
A \QPRG{} \(G\) with an \(\epsilon\) error is a QPT algorithm that, on \((1-\epsilon)\)-fraction of the input seed, outputs the same value with probability \(1-\epsilon\).\footnote{The second error (for the same seed) can be arbitrarily reduced by the repetition. We use this notion following the original definition \cite{ITCS:AnaLinYue24}.} When \(\epsilon\) is negligible, we simply call \(G\) a QPRG.
With an inverse-polynomial \(\epsilon\), one can construct digital signatures and IND-CPA encryptions~\cite{AC:BBOSS25}, although the proof technique is more complicated than the classical one.

\cite{ITCS:AnaLinYue24} questioned if the error \(\epsilon\) of \QPRG{} can be reduced to be negligible.
If it is possible, virtually
classical Minicrypt
can be recovered using the same technique.
Thus, our main question is:
\begin{center}
	\emph{Do \SPRSGs{} imply \neglQPRGs{} (with negligible errors)?}
\end{center}

In particular, if the answer is affirmative,
\LPRSGs{} can be constructed from \SPRSGs{} through \neglQPRGs{} by using known constructions \cite{JACM:GolGolMic86,C:JiLiuSon18}.
Yet, if the answer is no, there is still a possibility of lengthening PRSG outputs without relying on classical cryptographic primitives,
thus we ask:
\begin{center}
	\emph{Can we extend the length of PRSGs?}
\end{center}

We also consider another difference between classical and quantum in the landscape of pseudorandomness.
For example, we do not know how to construct PRUs from PRSGs or even pseudorandom function-like state generators (PRFSGs) \cite{C:AnaQiaYue22,TCC:AGQY22}, whereas classically PRGs can be used to construct PRFs and vice versa.
We are left with an unsatisfactory state of affairs; unlike in classical pseudorandomness, there is no single assumption unifying quantum pseudorandomness.
This makes us ask:
\begin{center}
	\emph{Are PRSGs, PRFSGs, and PRUs existentially equivalent?}
\end{center}

\subsection{Our results}
We provide negative evidence for the above questions by presenting new oracle worlds where one primitive exists, but the other does not exist, or at least is hard to construct, by showing that some natural constructions are insecure.
Our results suggest that quantum pseudorandomness could behave
fairly
differently from classical pseudorandomness.

\paragraph{Common Haar function-like state model.}
Our separations are based on variants of the common Haar function-like state (CHFS) oracles where for each input \(x\in \bit^*\) the oracle outputs a Haar\footnote{We sometimes choose the oracles from its sub-distribution, but we stick to use the name CHFS oracles for simplicity.} random state \(\ket{\phi_x}\) of length \(\ell(|x|)\) where \(|x|\) is the bit-length of \(x\). Note that this is an isometry. We also consider the unitary variants that instantiate this oracle.

Since the construction of PR(F)SGs with output length \(\ell(|x|)\) is straightforward with these oracles, our main contribution is to show that many other primitives are hard to construct even with the CHFS oracle (or its variants).

\paragraph{Separating \neglQPRGs{} from short PRFSGs.}
Using the CHFS oracles, our main contribution is to give a negative answer to
the open question asked in \cite{ITCS:AnaLinYue24}.\footnote{\cite{Bar25} concurrently resolves this question by showing a black-box impossibility.
See the concurrent work for more detailed discussion.} 
More formally, we prove the following theorem.

\begin{theorem}\label{thm_intro_conjsep}
	For \(c \log \lambda \le L \le C \lambda\) for some \(c,C>0\),
	there exists a unitary oracle relative to which PRFSGs with \(L\)-qubit outputs exist but \neglQPRGs{} do not.
\end{theorem}
Note that with the negligible errors, the classical constructions work well even for the quantum-computable counterparts. Therefore, this theorem states that QPRGs, as well as quantum-computable one-way functions (QOWFs) and quantum-computable pseudorandom functions (QPRFs), constructed from quantum primitives must suffer from inverse polynomial errors, as in the typical constructions that uses tomography.

This theorem has several interesting implications.
First, it complements the classical–quantum cryptography separation of \cite{Kre21} along two incomparable dimensions.
On the classical side, \cite{Kre21} considers primitives with perfect correctness (in particular, OWFs)\footnote{Technically, they showed the stronger statement \({\bf BQP=QMA}\).}, whereas our separation continues to hold even when the primitive is allowed to have a negligible correctness error.
On the quantum side, \cite{Kre21}'s separation requires long output length (in particular, \(\omega(\log \lambda)\)-length PRUs), while our oracle separation works even for short PRSGs. Thus, our result shows that the gap between classical and quantum cryptography persists even when correctness is relaxed and the quantum pseudorandomness assumption has a significantly shorter output length.

Secondly, it can be interpreted as finding a quantum-computable version of Pessiland, where there are hard-on-average languages in \(\qcma \cap \coqcma\) yet there are no \neglQPRGs{}. Previously, \cite{AC:BarNisYam25} gave a black-box separation of PRGs from EV-OWPuzz, thereby separating PRGs and hard-on-average languages in \(\qcma\) (from EV-OWPuzz).\footnote{Similarly, the worst-case hardness \(\bqp \neq \qcma\) was proven in \cite{BM24,C:ChuGolGra24}.} This result strengthens the result to include \(\coqcma\) and to unitary oracle separation.
To show this, we choose \(\ell = O(\log \lambda)\) and consider a state version of the decisional permutation inversion problem \cite{BBBV97} that decides if the given input \(({\sf str},z)\) satisfies \(x\le z\) where \({\sf str}\) specifies a state \(\ket{\psi_{\sf str}}\) that is sufficiently close to the CHFS oracle's output state \(\ket{\phi_x}\) for some \(x\).
By appropriately choosing the output length \(\ell\), the uniqueness of \(x\) is ensured with overwhelmingly high probability.
We outline this language and interpretation in 
\ifnum\final=1
the full version~\cite{EPRINT:BHMV25}.
\else
\cref{app:average-hard}.
\fi
In the classical setting, \cite{TCC:Wee06} found a Pessiland oracle.\footnote{The authors noted that an unpublished work by Impagliazzo and Rudich gave the first Pessiland oracle.}
\begin{corollary}
	There exists a unitary oracle relative to which
	a quantumly hard-on-average promise problem in \(\qcma\cap \coqcma\) exists,
	but no quantum-computable one-way functions or pseudorandom generators exist.
\end{corollary}

On the other hand, however, we found that this oracle world is not too pessimistic.
Recall that PRSGs with output length \(c\log \secpar\) for large enough \(c\) can be used to construct digital signatures and IND-CPA encryptions, with classical keys and outputs, using the notion of recognizable aborts \cite{AC:BBOSS25}.
This gives the following corollary in the world of quantum-computable classical-communication (QCCC) cryptography \cite{C:ChuGolGra24}, which was separated from (classical-computable) classical cryptography in \cite{STOC:KreQiaTal25}.\footnote{More precisely, \cite{STOC:KreQiaTal25} constructed the oracle world relative to which \(\p= \np\) yet quantum-computable trapdoor one-way functions exist.}
This strengthens a similar black-box separation proven in \cite{AC:BarNisYam25}.
\begin{corollary}
	Relative to the same oracle,
	quantum-computable digital signatures and IND-CPA symmetric-key encryptions exist, but no quantum-computable OWFs or PRGs exist.
\end{corollary}
We found this interesting, as it shows that OWFs and PRGs are \emph{stronger} than the main applications, encryptions and signatures.
This is possible because internal randomness in quantum machines cannot be extracted as random coins, unlike randomness in classical machines as observed in several previous works \cite{STOC:AarArk11,CCC:AarIngKre22,ITCS:BraCanQia23,AC:BarNisYam25}.
Consequently, our separation also separates EV-OWPuzz from QOWFs, confirming the belief of \cite{C:ChuGolGra24} that EV-OWPuzz could be the central primitive for QCCC cryptography.\footnote{We note that \cite{C:ChuGolGra24} does not exclude the possibility that QOWFs are equivalent to EV-OWPuzz. We refute this possibility at least in a black-box way.}
We refer the reader to~\cref{fig:minimalassumption} for further implications that can be derived from our main theorem.

\paragraph{Further results: Partial separations.}
Variations of the oracle used to prove \Cref{thm_intro_conjsep} can be used to study the relationship of other quantum pseudorandom primitives, for which a complete separation or construction remains out of reach, and appear to require additional ideas beyond the current techniques.
We include these partial separations to indicate where the obstructions come from, as this show what is left to prove for a full separation, or on the other hand, what a construction needs to overcome.

\noindent\emph{\underline{On constructing PRUs from PRFSGs, without ancilla registers.}}
We study the hardness of constructing PRUs from PRFSGs.
We prove that some restricted candidate PRU constructions fail to be secure.
The formal statement is as follows.
\begin{theorem}[Restricted PRU separation]
\label{thm_intro_PRFSG/PRU}
	There exists a {unitary} oracle\footnote{In this paper, we assume that the algorithms can access unitary oracles and its inverses. We do not consider the controls, conjugates or transposes of the oracles, but we believe our results can be extended to them using a similar idea from \cite{C:Zhandry25}.} relative to which adaptively-secure quantum-accessible PRFSGs exist, but non-adaptively secure (and inverseless) PRUs {whose implementation uses neither
ancillary registers nor intermediate measurements} do not.
\end{theorem}
Our oracle consists of the CHFS oracle for \(\ell(|x|)=|x|\) and the \(\qpspace\) oracle that computes the unitary polynomial-space circuit given as input.
Our impossibility shows that even the strongest form of PRFSGs cannot be used to construct the weakest form of PRUs in a black-box way without using ancillary registers and measurements.\footnote{The impossibility is shown by an explicit adversary that only uses PRU generation algorithms non-adaptively, so the other forms of PRUs are automatically impossible to construct without ancilla registers and measurements.}
We believe the black-box separation between PRFSGs and PRUs \emph{without the no-ancilla and measurement conditions} also holds in the same oracle world, and leave this open question for future work.

\noindent\emph{\underline{On the length extension of PRSGs.}}
We now turn to the problem of extending the output length of PRSGs.
There are already many approaches with some partial positive answers to this problem that we will summarize below.
However, we believe that there is no general black-box construction of long-PRSGs from short-PRSGs.
We note that proving such a black-box impossibility result would imply~\Cref{thm_intro_conjsep}, as long-PRSGs can be built from QPRGs, and thus studying this separation is a natural next step.
We show some partial impossibility result regarding this question, that highlights its difficulty.
We consider another natural class of length extensions, including the extension algorithm that takes small PRSs non-adaptively as input and applies a unitary (e.g., \(G_k \to U_k(\ket{\phi_1}\otimes \ket{\phi_2})\) for smaller PRSs \(\ket{\phi_1},\ket{\phi_2}\)).
\begin{theorem}[Conditional PRSGs length-extension separation]\label{thm_intro_short_to_long}
	There exists an isometry oracle relative to which short PRFSGs exist but long
  PRSGs whose generation algorithm makes non-adaptive oracle queries and uses no ancilla, do not.
\end{theorem}
In fact, our result is slightly stronger: any PRSG length extension of this form is impossible.\footnote{For example, for \(s>t\) and \(s=\omega(\log \lambda)\), PRSGs with output length \(s\) cannot be constructed from PRSGs with output length \(t\) if the longer PRSGs follow the described algorithms.}
Moreover, the impossibility of long PRSGs also holds with constant-size ancilla for a certain type of algorithm that revert the ancilla to \(0\).
We note that there are still multiple ways to extend the length of PRSGs that our result does not cover.
{Besides using superconstant-size ancilla, } two notable approaches\footnote{We remark that there is another recent work~\cite{LV24} that discusses the possibility of the length extension of the PRSGs, but only for very specific forms. Furthermore, their work only shows how to do length extension from PRSGs with \emph{super-log} output size.}  showing that some PRSG length extensions are \emph{possible} as follows:
\begin{itemize}
	\item Construct QPRGs first using tomography \cite{ITCS:AnaLinYue24} (with inverse polynomial errors) and then use them to construct new PRSGs. In \cite{AC:BarNisYam25}, the authors show that the PRSG length extension is possible in the log-length regime, albeit with quantum key sampling (i.e., the keys are not uniformly distributed but quantumly sampled). This strategy is excluded from our restricted model. Partial progress to construct long PRSGs from short PRSGs was also discussed in the same paper.
	\item Use quantum queries to the oracle. This is excluded because of our
  classical-accessible oracle model. In fact, for a length-\(\ell\) PRSG \(\{\phi_k\}\), the
  state \(\ket{0}\ket{\phi_a} + \ket{1}\ket{\phi_b}\) for two random keys \(a,b\) forms a length-\((\ell+1)\)
  PRSGs for \(\ell = \omega(\log \lambda)\).
\end{itemize}

We believe the above strategies, allowing length extension up to log-length, are optimal.\footnote{More explicitly, the first approach gives PRSGs with any log-length output using classical queries to the other log-length output PRSGs. The second approach gives a length-\(s+O(\log \lambda)\) output PRSG given quantum access to the other length-\(s\) output PRSG.}
The full impossibility of PRSG length extension
would complement the impossibility of shrinking the output length of PRSGs~\cite{BM24,C:ChuGolGra24}, and suggests that both primitives are in fact incomparable.
Moreover, given the construction of one-way state generators (OWSGs) from short PRSGs~\cite{MY22b,CGG+23}, it provides evidence for the hardness of constructing PRSGs from OWSGs,
while the other direction is possible~\cite{CGG+23}.
We believe that the PRSG length extension is the most challenging problem discussed in this paper, and any progress on this problem seems to give new interesting techniques.

A summary of our results is given in~\cref{fig:minimalassumption}.
\definecolor{darkblue}{RGB}{0, 0, 150}
\definecolor{darkred}{RGB}{150, 0, 0}
\tikzstyle{implication}=[-{Stealth[length=3mm]},thick,black]
\tikzstyle{equivalence}=[{Stealth[length=3mm]}-{Stealth[length=3mm]},thick,black]
\tikzstyle{separation}=[dotted,-{Stealth[length=3mm]},thick,black]
\tikzstyle{ourseparation}=[dotted,-{Stealth[length=3mm]},thick,red]
\tikzstyle{ourweakseparation}=[dotted,-{Stealth[length=3mm]},thick,orange]
\tikzstyle{citation}=[fill=white,midway]

\begin{figure}[!htbp]
  \centering
  \begin{tikzpicture}[scale=2.1]
    \node(pru) at (-.5,3)[inner sep=1.5pt] {PRU};
    \node(prfs) at (-.5,2)[inner sep=1.5pt] {PRFSG};
    \node(prs) at (-.5,1)[inner sep=1.5pt] {PRSG};
    \node(shortprs) at (-2,2)[inner sep=1.5pt] {short-PRSG};
     \node(nQPRG) at (-3.5,3)[inner sep=1.5pt] {negl-QPRG};
    \node(pQPRG) at (-3.5,1)[inner sep=1.5pt] {1/poly-QPRG};
    \node(evowp) at (-5.3,3)[inner sep=1.5pt] {ev-OWPuzz};
    \node(sig) at (-4.2,2)[inner sep=1.5pt] {SIG};
    \node(pke) at (-4.8,2.3)[inner sep=1.5pt] {SKE};
    \draw[implication][bend left=40] (prfs) edge (prs);
    \draw[implication][bend left=40] (pru) edge (prfs);
    \draw[implication][bend left=10] (nQPRG) edge (pQPRG);
    \draw[implication][bend left=10] (nQPRG) edge node[citation]{\tiny\cite{STOC:MaHua25}} (pru);
    \draw[implication][bend left=10] (nQPRG) edge %
    (evowp);
    \draw[implication][bend left=20] (pQPRG) edge node[citation]{\tiny\cite{AC:BBOSS25}} (sig);
    \draw[implication][bend left=35] (pQPRG) edge node[citation]{\tiny\cite{AC:BBOSS25}} (pke);
    \draw[implication][bend left=80] (pQPRG) edge node[citation]{\tiny\cite{C:ChuGolGra24}} (evowp);
    \draw[separation][bend left=10] (prs) edge node[citation]{\tiny\cite{BM24}} (shortprs);
    \draw[ourweakseparation][bend left=10] (shortprs) edge (prs);
    \draw[ourweakseparation][bend left=40] (prfs) edge
    (pru);
    \draw[ourseparation][bend left=10] (shortprs) edge (nQPRG);
    \draw[ourseparation][bend left=10] (pQPRG) edge (nQPRG);
    \draw[ourseparation][bend left=10] (evowp) edge (nQPRG);
    \draw[implication][bend left=20] (shortprs) edge node[citation]{\tiny\cite{ITCS:AnaLinYue24}} (pQPRG);
    \draw[implication][bend left=10] (nQPRG) edge node[citation]{\tiny{\cite{C:BraShm20}}} (shortprs);
    \draw[implication][bend left=14] (nQPRG) edge  (sig);
    \draw[ourseparation][bend left=14] (sig) edge  (nQPRG);

    \draw[implication][bend left=6] (nQPRG) edge  (pke);
    \draw[ourseparation][bend left=6] (pke) edge  (nQPRG);

  \end{tikzpicture}
  \caption{
    Implications between primitives are represented with an arrow, and separations with a dotted arrow.
    Our results are the arrows in red and orange, where red arrows indicate an oracle separation and orange arrows indicate conditional/restrictive black-box separations.}
  \label{fig:minimalassumption}
\end{figure}

\subsection{Related works}
\paragraph{Quantum oracle models.}
The isometry CHFS oracle was first studied in~\cite{TCC:AnaGulLin24} to show the (isometry oracle) separation between QCCC primitives and PRSGs.
Recent works suggest different quantum oracle models. The common Haar state model~\cite{EC:CheColSat25,TCC:AnaGulLin24} represents a world where copies of a random single Haar random quantum state are easily generated. A similar model without restricting Haar randomness was also studied in~\cite{DLS24,C:MorYam24,C:Qian24}.
The quantum Haar random unitary oracle model (QHROM) was suggested in~\cite{quantum:CM24,ITCS:BouFefVaz20}, and the applications are studied in~\cite{EC:ABGL25,TCC:HhaYam25}.

\paragraph{Quantum black-box impossibility.}
Recently, various black-box impossibilities have been shown based on new oracles and new techniques.
We briefly summarize this line of research.\footnote{For the full relations, we refer \href{https://sattath.github.io/microcrypt-zoo/}{Microcrypt-zoo}.}
The separation between OWFs and quantum primitives relative to a quantum oracle~\cite{Kre21} initiated this direction, and the same oracle later was shown to imply the hardness of shrinking PRSG output lengths~\cite{BM24,C:ChuGolGra24}.
This result was later strengthened relative to a classical oracle~\cite{STOC:KQST23} albeit for weaker quantum primitives.
A separation between classical and quantum-computable OWFs is shown in~\cite{STOC:KreQiaTal25}.

Relative to the common Haar state oracles, various separations are implied, e.g., commitments (and EFI pairs~\cite{ITCS:BraCanQia23}) and single-copy PRSGs exist but no OWSGs and (multi-copy) PRSGs~\cite{EC:CheColSat25,EC:BMMMY25,EC:BosCheNeh25}.
In~\cite{EC:BMMMY25}, they also show a black-box separation between quantum money and EFI pairs.

The isometry version of CHFS oracle provides a world with PRFSGs but without QCCC primitives~\cite{TCC:AnaGulLin24}.
On the other hand, an oracle world with QCCC key exchange where \({\bf BQP}={\bf QCMA}\) holds was introduced in~\cite{GMMY24}, with some more separations.

Finally, a very recent work~\cite{AC:BarNisYam25} shows the black-box impossibility of constructing OWSGs from \(\bot\)-(Q)PRGs (that can be seen as a weaker version of the QPRGs with negligible correctness errors, see~\cite{AC:BBOSS25}). A difference between quantum sampling of the keys and uniformly random keys is also explored in the same paper; in this work, we only assume the uniform key setting.

\paragraph{Concurrent work.} A concurrent and independent work~\cite{cryptoeprint:2025/1864} shows the oracle separation between PRFSGs and PRUs using similar oracles but with different techniques. The full separations remain open as both papers consider the bounded-length ancilla.
We, however, note that \cite{cryptoeprint:2025/1864} achieves a stronger result, as it separates PRFSG from PRU with a logarithmic number of ancilla qubits, whereas our paper requires no ancilla implementation.
They also consider the separations regarding the pseudorandom isometries~\cite{EC:AGKL24}.
The results about the log-length CHFS oracles are unique to this paper.

\cite{Bar25} concurrently addresses the main question of this paper.
They consider a black-box separation between QPRGs and PRSGs. Under their definition, the tomography-based construction of \cite{ITCS:AnaLinYue24} is not treated as black-box, whereas our unitary oracle separation rules out such a tomography approach as well. Ours also exclude the use of PRFSGs. On the other hand, \cite{Bar25} proves the separation regarding \(\bot\)-PRG and QPRG which we did not.

\ifnum\anonymous=1
\else
	\paragraph{Acknowledgments.}
	We thank Takashi Yamakawa for the helpful discussion on the definition of \(\qpspace\) oracle, and Shoga Yamada and Yao-Ting Lin for noting a minor error in the previous version.
	MH is supported by Schmidt Sciences Polymath award to David Soloveichik.
	SBE is supported by PEPR integrated project EPiQ ANR-22-PETQ-0007 part of Plan France 2030 and by ANR JCJC TCS-NISQ ANR-22-CE47-0004. Q-H.V was supported in part by the French ANR project CTQC (ANR-25-CE48-7926-01 CTQC).

	In the previous version of this work, \cref{thm:state_conjecture} was left as a geometric conjecture. We managed to prove this conjecture using tools from differential geometry and isoperimetric inequalities, with the assistance of {\tt ChatGPT 5.1 Thinking}.
	The proof in this paper is fully written by the authors, yet the main idea is from {\tt ChatGPT 5.1 Thinking}. Finally, ChatGPT 6 Astra pointed out a subtle bug in the previous PRFSG existence, and provided the new proof based on the nonuniform QROM bound (without changing the construction).
\fi

\section{Technical overview}\label{sec:overview}
\paragraph{(Unitarized) Common Haar function-like state oracles and PRFSGs.}
Our results are in a relativized world with (variants of) the common Haar function-like state (CHFS) oracles.
The CHFS oracles with length \(\ell\) are defined as follows: it is a family of unitaries \(\{S_x\}_{x\in\{0,1\}^*}\) defined as follows:
\[
	S_x:
	\begin{cases}
		\ket{0}\to \ket{\phi_x}  &                                                               \\
		\ket{\phi_x}\to \ket{0}  &                                                               \\
		\ket{\psi}\to \ket{\psi} & \text{ if }\ket{\psi} \notin {\sf span}(\ket0 ,\ket{\phi_x}),
	\end{cases}
\]
where \(\ket{\phi_x}\) is a predetermined Haar random state of length \(\ell(|x|)\), with \(|x|\) denoting the bit-length of \(x\). This oracle is inspired by the reflection/swap oracles in \cite{C:GolZha25,EC:BMMMY25}.

In this overview,
we assume that the algorithm accesses the unitaries \(S_x\) one by one, and also assume that \(\braket{0}{\phi_x}=0\) for simplicity, so that \(S_x\) can be understood as a reflection
\[
	S_x= I - 2\ketbra{\phi_x-},
\]
where \(\ket{\phi-}=\frac{\ket0 -\ket{\phi_x}}{\sqrt2}.\)

The construction of PRFSGs with the CHFS oracles is rather straightforward:
the generation algorithm, on input \((k,x)\) for key \(k\) and input \(x\) of length \(\lambda\), outputs \(\ket{\phi_{k||x}}\) by querying \(S_{k||x}\), where \(k||x\) is the concatenation of \(k\) and \(x\).
Note that the output length of the PRFSGs is \(\ell(k||x)\).
The security can be shown by the standard reduction to the unstructured search problem.

\subsection{Separating QPRGs from short PRFSGs}
We show that relative to \emph{some} CHFS oracles with output length \(\ell(n)\), together with the PSPACE oracle,
PRFSGs with output length \(\ell\) exist but QPRGs do not.
But first, we discuss a technical hurdle we met.

\paragraph{Concentration inequality fails.}
The concentration inequality for Haar measure (see \cref{thm: unitaryHaarconcentration})
is the most common tool currently used for oracle separations.
However, when the oracle outputs only \(\ell(n)=\Theta(\log n)\) qubits,
the ambient dimension is too small (because \(2^\ell = \poly[n]\)) for these bounds to force the type of
``almost-everywhere'' behavior needed to rule out QPRGs.

We instead start from an \emph{extreme} concentration phenomenon that any QPRG must satisfy.
Consider a single-bit-output QPRG \(G^O\) relative to CHFS oracles \(O\) with negligible errors.
For any fixed seed \(x\), the output distribution of \(G^O(x)\) must be almost deterministic
for all but a negligible fraction of oracles \(O\):
equivalently, defining \(f(O)=\Pr[G^O(x)\to 1]\), we must have
\(f(O)\approx 0\) or \(f(O)\approx 1\) for almost all \(O\).
These are the two extreme points in the concentration inequality.
A natural question is thus whether these two extreme points can be simultaneously concentrated.

\paragraph{The new Barrier theorem.}
We ask the following question: if \(f(O)\in[0,1]\) has substantial mass near both \(0\) and \(1\),
what can we say about the geometry of the preimages
\(f^{-1}([0,\epsilon])\) and \(f^{-1}([1-\epsilon,1])\) in oracle space?
If \(G^O(x)\) for a fixed \(x\) can output both \(0\) and \(1\) with non-zero probability, both pre-image regions are large.
We also expect the distance between the two pre-image regions to be large, since close oracles would likely induce close outputs.
Our Barrier theorem (\cref{thm:state_conjecture}) asserts that under such conditions, the intermediate region \(f^{-1}((\epsilon,1-\epsilon))\) must itself be large:

\begin{theorem}[Barrier theorem (Informal)]
	\label{thm:barrier_informal}
	Let \(X\) be the product space of pure quantum states with the corresponding product Haar measure \(\sigma\).
	If \(S_0,S_1\) are two measurable subsets of \(X\) such that \(\sigma(S_0),\sigma(S_1)\ge A\), and if \(d(S_0,S_1)\ge B\) for some distance \(d\) on \(X\), then \(\sigma(X\setminus(S_0\cup S_1))\ge cAB\) for a universal constant \(c>0\) not depending on \(X\).
\end{theorem}

Now we turn back to the QPRGs \(G^O\) with negligible pseudodeterminism error.
Leveraging the Barrier theorem, our key observation is that with such strong pseudodeterminism,
for a fixed seed \(x\) the output cannot meaningfully depend on the CHFS oracles,
otherwise we would see a noticeable ``intermediate'' mass of oracles where the output is not deterministic.
In more detail, consider the first output bit of \(G^O\) denoted by \(G_1^O\) and let \(f(O)=\Pr_G[G_1^O(x)\to 1]\).
The Barrier theorem rules out the case where \(S_0=f^{-1}([0,\epsilon])\) and \(S_1=f^{-1}([1-\epsilon,1])\) are both large, because it leads that \(G^O\) is not pseudodeterministic on the too large barrier \(X\setminus(S_0\cup S_1)\).
This leads to a dichotomous intuition: either
\begin{enumerate}
	\item \(G_1^O\) is not pseudodeterministic, or
	\item \(G_1^O\) is essentially constant independent of \(O\).
\end{enumerate}
That is, \(G_1^O\) must fail to achieve pseudodeterminism or security.

It turns out that this intuition is not quite true. One of the reasons is that the barrier theorem, even if the second statement miserably failed, shows that the first statement holds only for a mildly large fraction of \(O\).
A more correct dichotomy is as follows, considering all output bits. For a fixed security parameter,
one of the following holds:
\begin{enumerate}
	\item \(G^O\) is not pseudodeterministic on a (slightly) large fraction of \(O\), or
	\item \(G^O\) outputs a fixed value for a (very) large fraction of \(O\).
\end{enumerate}
We make an observation for each case.
For the first case, we observe that the PRFSG construction \(\gen:(k,x)\to\ket{\phi_{k||x}}\) is still secure even if we choose \(O\) from a slightly smaller sub-distribution of the product Haar random distributions (at parameter \(\lambda\)). This is because choosing \(O\) from a slightly sub-distribution corresponds to the non-uniform security and we can use the PRG security in the nonuniform QROM in~\cite{AC:HhaXagYam19,FOCS:CGLQ20,EC:Liu23}. 
It means we may hope to choose \(O\) such that \(\gen\) is a secure PRFSG yet \(G\) is not pseudodeterministic.

For the second case, we observe that \(G^O\) can be estimated by \(G^{O'}\) for randomly chosen \(O'\), in which case we know how to simulate without querying \(O\) using known techniques.\footnote{This is the final goal of the concentration inequality used in \cite{Kre21}!}
Let \(F\) be the simulated function, then given input \(y\), determining if there is \(x\) such that \(F(x)\) can be done in polynomial space (i.e., with the PSPACE oracle), breaking the pseudorandomness of \(G^O\).

Based on this observation, we sample oracle \(O\) (at parameter \(\lambda\)) depending on which case occurs: If \(G_i^O\) is pseudodeterministic with high probability over \(O\) for all \(i\), we just sample \(O\) from the product Haar random distribution, breaking the pseudorandomness.
For the other case, the dichotomy says that a (slightly) large fraction of \(O\) makes \(G^O\) not pseudodeterministic, thus
we sample \(O\) among one that makes \(G^O\) not pseudodeterministic yet \(\gen\) is secure, given that the fractions for \(G\) not being pseudodeterministic are larger than the fractions for insecure \(\gen\). This is indeed possible by adjusting parameters.

This, however, only ensures that a single candidate \(G\) is insecure or not pseudodeterministic at a single parameter.
Our final proof of~\cref{thm_intro_conjsep} proceeds by diagonalization, considering all uniform oracle-aided QPRG candidates
\(\{G_j\}_{j\in\mathbb{N}}\)
as follows.
\begin{enumerate}
	\item Define \(\{G'_j\}_{j\in\mathbb{N}}\) so that for each \(i\) there are infinitely many \(j\)'s such that \(G_i=G'_j\) holds. For example, one can consider \(\{G_i\}_i\) as \(G_1,G_1,G_2,G_1,G_2,G_3,\dots\).
	\item Choose a super-fast-growing sequence of security parameters
	      \(\{\lambda_i\}_{i\in\mathbb{N}}\) (e.g. \(\lambda_{i+1}=2^{\lambda_i^{200}}\)) and blocks of oracle input
	      lengths \(I_i=[\log(\lambda_i),\lambda_i^{100}]\) that are disjoint.

	      Intuitively, the properties of \(G'_i\) at parameter \(\lambda_i\) \emph{essentially} depend only on the oracle with input lengths in \(I_i\). This is because \(G'_i(1^{\lambda_i},\cdot)\) may never query too large input length \(|x|>\lambda_i^{100}\), and for the query \(|x|<\log (\lambda_i)\), it can be efficiently estimated by the adversary using tomography.
	\item Fix an oracle \(O_{<i}\) up to input length \(<a_i\). Over the random choice of \(O\) with respect to the input length in \(I_i\), %
	      the dichotomy holds: either \((G'_i)^O\) is not pseudodeterministic with high probability over \(O\), or \((G'_i)^O\) can be simulated with high probability over \(O\) so that it is insecure. We sample \(O\) over the input length in \(I_i\) as above, according to which case occurs.
	\item Proceed to the next block after sampling \(O\), by fixing \(O<{i+1}\).
\end{enumerate}

The above procedure results in an oracle \((O,\pspace)\) relative to which \(\gen\) is a secure PRFSG, while any QPRG candidate \(G\) must be infinitely often insecure or infinitely often non-pseudodeterministic.
Note that in the actual proof, we need to consider each oracle QPT adversary \(\cA\) instead of the one-shot security argument of \(\gen\), which further complicates the argument.

\subsection{On separating PRUs from PRFSGs}

We now give a brief overview on how we achieve~\Cref{thm_intro_PRFSG/PRU}.
We consider the unitary CHFS oracles with output length \(\ell(n)=n\). As discussed above, we can easily construct PRFSGs relative to this oracle, but breaking the PRU constructions is quite involved. We sketch the outline of the proof here.

\paragraph{Breaking PRUs without ancilla.}
To establish \cref{thm_intro_PRFSG/PRU}, we present an explicit attack for any PRU candidate without ancilla with respect to the CHFS oracle of length \(\ell(n)=n\).

We consider the following simplified form of the PRU algorithm \(\{G_k\}_{k\in \bit^*}\) on key \(k\in \bit^\secpar\) and input state \(\ket{\psi}\):
\[
	G_k:\ket{\psi} \mapsto U^{(k)}_T\cdot S_{x_T^{(k)}} \cdot U^{(k)}_{T-1} \cdot \ldots \cdot U^{(k)}_1 \cdot S_{x_1^{(k)}} \cdot U^{(k)}_0 \ket{\psi},
\]
where \(U^{(k)}_T,\dots,U^{(k)}_0\) are some unitaries and \(S_{x_T^{(k)}},\dots,S_{x_1^{(k)}} \) are the CHFS oracle queries.
In the main body of the paper, we consider a more general form of \(G_k\) where queries may be in superposition or adaptive.

Our main observation is as follows:
for a Haar random state \(\ket{\rho}\) independently chosen from the oracle, the application of the reflection oracle does not change the state much, i.e.,
\begin{align}\label{eqn_intro:approx_ref}
	S_x \ket{\rho} \approx \ket{\rho}.
\end{align}
This is because the reflection \(S_x\) only makes a change on the tiny space spanned by \(\{\ket{\phi_x},\ket{0}\}\). Therefore, one may argue that
\[
	G_k\ket{\rho} \approx U^{(k)}_T\cdot U^{(k)}_{T-1} \cdot \ldots \cdot U^{(k)}_1 \cdot U^{(k)}_0 \ket{\rho}
\]
because \( \ket{\rho_t} := U^{(k)}_t \cdot ...\cdot U^{(k)}_0 \ket{\rho}\) is a Haar random state independent of the oracle due to the invariant property of Haar measure.
Unfortunately, this is not the case in general, as the loss in \cref{eqn_intro:approx_ref} is proportional to \(1/2^{|x|}\), so we cannot ignore \(S_x\) for small \(|x|\).

We instead learn all \(S_x\) to obtain \(S'_x\) for small \(|x|\) using process tomography~\cite{FOCS:HKOT23}. We define \(\tilde S_x\) by \(S'_x\) for small \(|x|\) and \(I\) for large \(|x|\), and define
\[
	F_k:\ket{\psi} \mapsto U^{(k)}_T\cdot \tilde S_{x_T^{(k)}} \cdot U^{(k)}_{T-1} \cdot \ldots \cdot U^{(k)}_1 \cdot \tilde S_{x_1^{(k)}} \cdot U^{(k)}_0 \ket{\psi}
\]
which now satisfies \(F_k\ket{\rho} \approx G_k\ket{\rho}\).

Now we describe the adversary that given oracle \(V\), distinguishes whether it is one of \(\{G_k\}\) or a true Haar random unitary.
The adversary first prepares
\(\Phi=(\ket{\rho}\otimes V\ket{\rho})^{\otimes M}\) for some large \(M\) and Haar random state \(\ket{\rho}\) (or a \(t\)-design for sufficiently large \(t\)) and defines:
\begin{description}
	\item[\(P_k\):] on input \(\Phi=(\ket{\rho}\otimes V\ket{\rho})^{\otimes M}\), it applies \((F_k\otimes I)^{\otimes M}\), applies \(M\) swap tests on each copy; if sufficiently many copies pass the swap test, it returns 1. Otherwise, it returns 0.
\end{description}
We can show that \(P_k\) returns \(1\) if \(V=G_k\) with high probability, but \(P_k\) almost always returns \(0\) if \(V\) is a Haar random unitary.
This satisfies the setting where the quantum OR tester~\cite{HLM17} can be run with the \(\qpspace\) oracle\footnote{This oracle, roughly, takes a quantum state \(\ket{\phi}\) and a succinct description of a quantum circuit \(C\) computable in polynomial space, and returns \(C\ket{\phi}\).
	See \cref{def:qpspace} for the formal definition.} as observed in~\cite{EC:CheColSat25}.
By augmenting our world with the \(\qpspace\) oracle, we obtain a relativized world where PRFSGs exist
but PRUs without ancilla and measurements do not, proving \cref{thm_intro_PRFSG/PRU}.

The attack even breaks the non-adaptive PRU security as \(V\) is only used to prepare \(\Phi\). Extending this to the quantum-accessible PRFSGs security requires considering the coherent version of CHFS oracles, which can be done similarly with some more computation.

\ifnum\submission=1
	Due to space limitations, the formal presentation of this result is deferred to
    \ifnum\final=1
    the full version~\cite{EPRINT:BHMV25}.\else
    ~\cref{sec:separation_pru}.
    \fi
\fi

\subsection{On the length extension of PRSGs}

We now give an overview on how to prove~\cref{thm_intro_short_to_long}. 
We again consider the CHFS oracle with \(\ell(|x|)=\lfloor\log |x|\rfloor\) together with the \(\qpspace\) oracle.
Here, we consider the classical-accessible isometry version: a family of isometries \(\{O_x\}_{x\in \bit^*}\) where \(O_x\) takes input \(\ket{0}\) and outputs an \(\ell(|x|)\)-qubit Haar random state \(\ket{\phi_x}\). We do not allow querying the other input states.

Consider the following PRSG candidate that makes nonadaptive queries to the oracle and outputs on key \(k\)
\begin{equation}\label{eqn: UsmallPRS}
	\rho_k = U_k
	\left(\ket{\phi_{x_1^{(k)}}}\otimes\dots\otimes  \ket{\phi_{x_t^{(k)}}}\right),
\end{equation}
where we assume that the parameter \(t\) and the lengths of \(x_i^{(k)}\)'s are all the same for different keys for simplicity in this overview.
Here \(\ket{\phi_{x_1^{(k)}}}\),\dots,\(\ket{\phi_{x_t^{(k)}}}\) are shorter PRSG outputs.
We have that the state
\[
	U_k^\dagger \rho_k U_k = \ket{\phi_{x_1^{(k)}}}\otimes \dots\otimes \ket{\phi_{x_t^{(k)}}}
\]
is a product of many pure states.
On the other hand, for a Haar random state \(\ket\psi\), \(\tilde U_k^\dagger \ket\psi\) definitely does not have such a product structure, as it is also Haar random by definition.
Given the efficient product test algorithm~\cite{FOCS:HarMon10}, we can run the quantum OR tester with the \(\qpspace\) oracle as in the separation between PRUs and PRFSGs.

The formal proof considers more general algorithms allowing partial traces with constant-size ancilla that can be reset to \(0\).
\ifnum\submission=1
	Due to space limitations, the formal presentation of this result is deferred to
    \ifnum\final=1
    the full version~\cite{EPRINT:BHMV25}.\else
    ~\cref{sec:separation_prs}.
    \fi
\fi

\section{Preliminaries}
\paragraph{Notations.}
We use \(\secpar \in \mathbb{N}\) to denote the security parameter.
For any \(m \in \mathbb{N}\), we use the notation \([m]\) to refer to the set \(\{1, \ldots, m\}\).
For any finite set \(U\), we write \(x \gets U\) to denote that \(x\) is sampled
uniformly at random from \(U\).
For a distribution \(\mathcal D\),
\(x\gets \mathcal D\) denotes that \(x\) is sampled from
\(\mathcal D\).
For a bit string \(x \in \left\{0,1\right\}^{*}\), we denote its bit-length by \(|x|\).
We assume that all functions used to represent the lengths of the cryptographic primitives are QPT-computable.
We write \(\TV\) the total variation distance between two classical probability measures.
We assume the reader is familiar with the basics of quantum computation, and refer to~\cite{nielsen2010quantum} otherwise.

\ifnum\submission=0

  We will also use standard notations from quantum information and cryptography.
\subsection{Quantum states, operations, and trace}

A \(d\)-dimensional quantum state is a positive semi-definite Hermitian density matrix \(\rho = \sum_{x\in[d]} p_x\ketbra{\phi_x}\), where the pure states \(\ketbra{\phi_x}\) have trace one, and \(p_1,\dots,p_d\) is a probability distribution, i.e., \(p_1,\dots,p_d \ge 0\) and \(p_1+\dots+p_d=1\).
Pure states are the rank-1 quantum state that can be written as \(\ketbra{\phi}\). We sometimes write \(\ket{\phi}\) or just \(\phi\) to denote the pure state \(\ketbra{\phi}\) for simplicity.
We can consider any positive semi-definite Hermitian matrix (with arbitrary unit trace) as an unnormalized quantum state, e.g., \(\Pi \rho \Pi\) for some projection \(\Pi\) and quantum state \(\rho\), and call them unnormalized states.

A quantum operation \(\Phi\) is a completely positive and trace-nonincreasing operator, that can be represented by matrices \(B_1,\dots,B_k\) satisfying
\[
	I-\sum_{i=1}^k B_i^\dagger B_i \ge0.
\]

The matrices \(B_1,\dots,B_k\) are the Kraus operators of the operations, and with this notation, \(\Phi\) maps a quantum state \(\rho\) to
\(
\Phi(\rho) = \sum_{i=1}^k B_i \rho B_i^\dagger
\).
Quantum operations can represent unitary operations, projective measurements, or applying a projection \(\Pi\). We write the composition of two quantum operations \(\Phi,\Psi\) by \(\Phi\circ\Psi.\)
For a unitary \(U\), the corresponding operation is represented by \(U(\rho)=U \rho U^{\dagger}\) or sometimes the calligraphic font \(\mathcal U(\rho)\).

The trace norm of a Hermitian matrix \(A\) is defined by
\(\|A\|_1 := \sum_{i=1}^d |\lambda_i|\),
where \(\lambda_1,\dots,\lambda_d\) are the eigenvalues of \(A\). If \(A\) is positive semi-definite, we can write \(\|A\|_1 = {\Tr(A)}.\)
This induces the \emph{trace distance} \(\|\rho-\sigma\|_{tr} = \frac 12\|\rho - \sigma\|_1\) between two (possibly unnormalized) mixed states, which forms a distance over (unnormalized) mixed states.
A quantum operation \(\Phi\) does not increase the trace norm. That is, for any Hermitian matrix \(A\), it holds that \(\|\Phi(A)\|_1 \le \|A\|_1\).
In particular, we have \(\Tr(\Phi(A)) \le \Tr(A)\) for any positive semi-definite matrix \(A\).
For any two (possibly unnormalized) states \(\rho,\sigma\),
\begin{align}\label{eqn:channel_does_not_decrease_trace_distance}
	\|\Phi(\rho)-\Phi(\sigma)\|_{tr}=\frac{1}{2}\|\Phi(\rho-\sigma)\|_1 \le
	\frac12\|\rho-\sigma\|_1 = \|\rho-\sigma\|_{tr}.
\end{align}

For a positive semi-definite matrix \(A\), it holds that
\begin{align}\label{eqn: TrA2}
	\Tr(A^2) \le \Tr(A)^2.
\end{align}

We stress that most of the facts on the trace norm and distance also hold for unnormalized states, i.e., positive semi-definite Hermitian matrices.

\else

We will also use standard notations from quantum information and cryptography\ifnum\final=1
.
\else
and defer the relevant definitions and basic facts on quantum states, operators, and
trace distance to~\cref{app:pre}.
\fi
\fi
\subsection{Haar random states and unitaries}
We write \(\mathbb S(N)\) and \(\mathbb U(N)\) to denote the set of \(N\)-dimensional pure quantum states and the group of \(N \times N\) unitary matrices.
We denote by \(\sigma_{n}\) (or sometimes \(\Haar_{n(\secpar)}\)) and \(\mu_n\) the Haar distribution over \(n\)-qubit states and \(n\)-qubit unitaries, i.e., over \(\mathbb S(2^n)\) and \(\mathbb U(2^n)\), respectively. When the dimension is clear from the context, we drop the parameter and use \(\sigma\) or \(\mu\).
The Frobenius norm \(\|A\|_F\) of a matrix \(A\) is defined by \(\sqrt{\Tr(A^\dagger A)}.\)

\begin{theorem}[{\cite[Theorem 5.17]{Mec19}}]\label{thm: unitaryHaarconcentration}
	Let \(n_1,\dots,n_k \in \mathbb N\) and \(\mu = \mu_{n_1}\times \dots \times \mu_{n_k}\) be the product of Haar unitary measures over \(X=\mathbb U(2^{n_1}) \times \dots \times \mathbb U(2^{n_k})\). Suppose that \(f:X \to \mathbb R\) is \(L\)-Lipschitz in the Frobenius norm.
	Let \(N=\min(2^{n_1},\dots,2^{n_k})\).
	For every \(t>0\), it holds that
	\[
		\Pr_{U\gets \mu}\left[
			f(U) \ge \Exp_{V\gets \mu}[f(V)] + t
			\right]\le
		\exp\left(
		-\frac{(N-2)t^2}{24L^2}
		\right).
	\]
\end{theorem}

\begin{corollary}\label{cor: stateHaarconcentration}
	Let \(C^U\) be an \(m\)-query quantum oracle algorithm for the product of Haar random unitaries \(U\) chosen from \(X\) according to \(\mu\) defined above.
	Let \(g(U):=\Pr[1 \gets C^U]\).
	Then it holds that
	\[
		\Pr_{U\gets \mu}\left[
			g(U) \ge \Exp_{V\gets \mu}[g(V)] + t
			\right]\le
		\exp\left(
		-\frac{t^2(N-2)}{24m^2}
		\right).
	\]
\end{corollary}
\begin{proof}
	In \cite{Kre21}, the following statement is shown.
	\begin{lemma}[{\cite{Kre21}}]\label{lem: Lipschitz}
		Let \(A^U\) be a quantum algorithm that makes \(T\) queries to the unitary
		oracle \(U\).
		Define \(f(U):=\Pr[1 \gets A^U]\).
		Then \(f\) is \(T\)-Lipschitz in the Frobenius norm, i.e., \(|f(U)-f(V)| \le T\cdot \|U-V\|_F.\)
	\end{lemma}
	This lemma ensures that \(C\) is \(m\)-Lipschitz, thus
	applying \cref{thm: unitaryHaarconcentration}, we obtain the desired result.
\end{proof}

\begin{lemma}\label{lem: Haarproject}
	For any rank-\(D\) projection \(\Pi\) on \(m\) qubits for \(m\ge n\),
	\[\Exp_{\ket{\phi}\gets \sigma_n} \bra{\phi,0^{m-n}}\Pi\ket{\phi,0^{m-n}}\le\frac{D}{2^n}.\]
	If \(m=n\), the equality holds.
	In particular, for any \(n\)-qubit mixed state \(\rho\),
	\(\Exp_{\ket{\phi}\gets \sigma_n} \bra{\phi}\rho\ket{\phi}=\frac{1}{2^n}\).
\end{lemma}
\begin{proof}
	We simply write \(0\) to denote \(0^{m-n}\).
	We can write \(\Exp_{\ket{\phi} \gets \sigma_n} \bra{\phi,0}\Pi\ket{\phi,0} \) by
	\[
		\Exp_{\ket{\phi} \gets \sigma_n}\Tr(\Pi\cdot\ketbra{\phi,0} )
		=\Tr(\Pi\cdot \frac{I\otimes \ketbra{0}}{2^{n}} )
		\le \frac{1}{2^n}\Tr(\Pi) = \frac{D}{2^n},
	\]
	where
	the last equality follows from the fact that
	\(\Trace(\Pi) = \rank(\Pi)\).
	If \(m=n\), the inequality is saturated.
	The last statement can be shown by writing \(\rho = \sum_i p_i \ketbra{\psi_i}\) for \(\sum_i p_i=1\).
\end{proof}

\subsection{Cryptographic primitives}
\ifnum\submission=1
  In this work, we study quantum pseudorandom primitives in the oracle model:
  pseudorandom state generators (PRSGs), pseudorandom function-like state
  generators (PRFSGs), pseudorandom unitaries (PRUs), and quantum-computable
  pseudorandom generators (QPRGs). 
  Our definitions follow prior work~\cite{C:JiLiuSon18,C:AnaQiaYue22,Kre21,TCC:AGQY22,ITCS:AnaLinYue24}.
  We recall below the definitions of PRFSGs and QPRGs\ifnum\final=1.\else, and refer~\cref{app:pre} for the other primitives.\fi

We by default consider the adaptively-secure PRFSGs defined as follows.
\begin{definition}[PRFSGs]\label{def:PRFSG}
	We say that a QPT algorithm \(\gen^O\) is a secure pseudorandom
	function-like state generator (PRFSG) in the CHFS model if the following
	holds for some functions \(\kappa,m,n:\mathbb N\to \mathbb N\) such that \(\kappa,m=\omega(\log \secpar)\):
	\begin{itemize}
		\item \textbf{State Generation:} For any \(\secpar \in \NN\) and
		      \(k \in \bin^{\kappa(\secpar)}\), the algorithm \(\gen^O_k\) takes as input \(x \in \bin^{m(\secpar)}\) and outputs
		      \(n(\secpar)\)-qubit (possibly mixed) state \(\gen^O_k(x)\) stored in a new register.
		\item \textbf{Pseudorandomness:} For any
		      oracle QPT adversary
		      \(\adv^O = \{\adv^O_{\secpar}\}_{\secpar \in \NN}\),
		      there exists a negligible function \(\varepsilon(\cdot)\) such that for
		      all \(\secpar \in \NN\):
		      \begin{equation*}
			      \abs{\prob[k \gets \bin^{\kappa(\secpar)}]{1 \gets \adv_{\secpar}^{O,\gen^O(k, \cdot)}} -
			      \prob[G_{\sf Haar}]{1 \gets \adv_{\secpar}^{O,G_{\sf Haar}(\cdot)}}} \leq \varepsilon(\secpar),
		      \end{equation*}
		      where \(G_{\sf Haar}(\cdot)\) on input \(x \in \bin^{m(\secpar)}\), output
		      \(\ket{\psi_{x}}\) stored in a new register, where, for every \(x \in \bin^{m(\secpar)}\),
		      \(\ket{\psi_{x}} \gets \Haar_{n(\secpar)}\).
	\end{itemize}
	When the adversary always measures the input register before making queries to \(\gen^O_k\) or \(G_{\sf Haar}\), we say that \(\gen^O_k\) is classical-accessible.
	Otherwise, we say that it is quantum-accessible.

	We say that \(\gen\) is a \((\kappa(\secpar), m(\secpar), n(\secpar))\)-PRFSG to indicate that its key length is \(\kappa(\secpar)\), its input length is \(m(\secpar)\), and its output
	length is \(n(\secpar)\).
	We say that a PRFSG is a \emph{short PRFSG} when \(n=\bigTheta{\log \secpar}\), and a \emph{(long) PRFSG} when \(n=\omega(\log \secpar)\).
\end{definition}

  We define QPRGs as quantum-computable algorithms whose outputs are
  computationally indistinguishable from uniform random strings and, on almost
  every seed, are fixed with all but negligible probability.
\begin{definition}[QPRGs]
	\label{def:QPRG}
	We say that an oracle QPT algorithm \(F^O\) that outputs an \(m(\secpar)\)-bit classical string on \(n(\secpar)\)-bit input
	is a \emph{quantum pseudorandom generator} (QPRG) with \((1-\epsilon)\)-pseudodeterminism if the following conditions hold for a function \(\epsilon\). If \(\epsilon\) is negligible, we just call \(F\) a QPRG.
	\begin{itemize}
		\item \textbf{\((1-\epsilon)\)-Pseudodeterminism}.
		      For every \(\secpar \in \NN\),
		      the following holds:
		      \begin{enumerate}
			      \item There exists a set \(K_{\secpar} \subseteq \bin^{n(\secpar)}\) of ``good seeds'' such that
			            \[\Pr[x \in K_{\secpar}:x \gets \bin^{n(\secpar)}]\ge 1-\epsilon.\]
			      \item There exists a deterministic function \(f_{\secpar}:\{0,1\}^{n(\secpar)}\to\{0,1\}^{m(\secpar)}\)
			            such that for every \(x \in K_{\secpar}\), it holds that
			            \begin{align*}
				            \prob{F^O(x)=f_{\secpar}(x)} \ge 1-\varepsilon,
			            \end{align*}
			            where the probability is over the randomness of \(F\).
		      \end{enumerate}
		\item \textbf{Security}. For any oracle QPT algorithm \(\adv^O = \{\adv^O_{\secpar}\}_{\secpar \in \NN}\), there exists a negligible
		      function \(\varepsilon\) such that
		      \begin{equation*}
			      \abs{\prob[y \gets \bin^{m(\secpar)}]{1 \gets \adv_{\secpar}^O(y)} -
				      \prob[x \gets \bin^{n(\secpar)}]{1 \gets \adv_{\secpar}^O(F^O(x))}} \leq \varepsilon(\secpar),
		      \end{equation*}
		      where the probability is over the randomness of \(F\) and \(\adv_{\secpar}\).
		\item \textbf{Length extension.} \(n(\secpar)<m(\secpar)\) holds for all \(\secpar\in\NN\).
	\end{itemize}
\end{definition}

\subsection{QPSPACE oracle}
We recall the definition of the QPSPACE oracle that implements the arbitrary unitary operation described by polynomial size input \cite{EC:CheColSat25,EC:BMMMY25}.
\begin{definition}[\(\qpspace\) Oracle]\label{def:qpspace}
	The \emph{unitary} QPSPACE machine oracle, denoted by \(\qpspace\), is defined as follows: it takes a pair \((\rho,M,t,b)\) of an \(\ell\)-qubit quantum state \(\rho\), a classical Turing machine \(M\), and an integer \(t\in\mathbb N\), and \(b\in\bit\) denotes the abort bit.
	The oracle runs \(M\) for \(t\) steps to obtain the description of a unitary quantum circuit \(C\) that operates on \(\ell\) qubits; if \(M\) does not terminate after \(t\) steps or the output is not described as above, the oracle flips the abort register and halts. Otherwise, the oracle applies \(C\) on \(\rho\) and returns the output quantum state without measurement.
\end{definition}

The quantum access to the QPSPACE oracle is done by allowing coherent \((M,t)\).
For any unitary quantum circuit \(C\) that is output by a machine \(M\) after \(t\) steps, there is a QPT algorithm with \(\qpspace\) oracle that implements \(C^{-1}(\rho)\) on input \(\rho\) \cite[Proposition 3.5]{EC:BMMMY25}.

\else

We define cryptographic primitives relative to an oracle \(O\).
\begin{definition}[PRSGs]
	We say that an oracle QPT algorithm \(\gen^O\) is a secure pseudorandom
	state generator (PRSG) in the CHFS model if the following holds for some functions \(\kappa,n:\mathbb N\to \mathbb N\) such that \(\kappa=\omega(\log \secpar)\):
	\begin{itemize}
		\item \textbf{State Generation:} For any \(\secpar \in \NN\) and
		      \(k \in \bin^{\kappa(\secpar)}\), the algorithm \(\gen^O(k)\) outputs
		      an \(n(\secpar)\)-qubit state.
		\item \textbf{Pseudorandomness:} For any polynomial \(t(\cdot)\) and any
		      oracle QPT adversary \(\adv^O = \{\adv^O_{\secpar}\}_{\secpar \in \NN}\), there
		      exists a negligible function \(\varepsilon(\cdot)\) such that for all
		      \(\secpar \in \NN\):
		      \begin{equation*}
			      \abs{\prob[k \gets \bin^{\kappa(\secpar)}]{1 \gets \adv^O_{\secpar}(\gen^O(k)^{\otimes t(\secpar)})} -
			      \prob[\ket{\psi} \gets \sigma_{n(\secpar)}]{1 \gets \adv^O_{\secpar}(\ket{\psi}^{\otimes t(\secpar)})}} \leq \varepsilon(\secpar).
		      \end{equation*}
	\end{itemize}
	We say that \(\gen^O\) is a \(n(\secpar)\)-PRSG to indicate that
	its output length is \(n(\secpar)\).
	We further say that a PRSG is a \emph{short PRSG} when its output length is \(\bigTheta{\log \secpar}\), and a \emph{(long) PRSG} when its output length is \(\omega(\log \secpar)\).
\end{definition}
From now on we will use PRSGs to refer to long PRSGs and short PRSGs for logarithmic output.

We by default consider the adaptively-secure PRFSGs defined as follows.
\begin{definition}[PRFSGs]\label{def:PRFSG}
	We say that a QPT algorithm \(\gen^O\) is a secure pseudorandom
	function-like state generator (PRFSG) in the CHFS model if the following
	holds for some functions \(\kappa,m,n:\mathbb N\to \mathbb N\) such that \(\kappa,m=\omega(\log \secpar)\):
	\begin{itemize}
		\item \textbf{State Generation:} For any \(\secpar \in \NN\) and
		      \(k \in \bin^{\kappa(\secpar)}\), the algorithm \(\gen^O_k\) takes as input \(x \in \bin^{m(\secpar)}\) and outputs
		      \(n(\secpar)\)-qubit (possibly mixed) state \(\gen^O_k(x)\) stored in a new register.
		\item \textbf{Pseudorandomness:} For any
		      oracle QPT adversary
		      \(\adv^O = \{\adv^O_{\secpar}\}_{\secpar \in \NN}\),
		      there exists a negligible function \(\varepsilon(\cdot)\) such that for
		      all \(\secpar \in \NN\):
		      \begin{equation*}
			      \abs{\prob[k \gets \bin^{\kappa(\secpar)}]{1 \gets \adv_{\secpar}^{O,\gen^O(k, \cdot)}} -
			      \prob[G_{\sf Haar}]{1 \gets \adv_{\secpar}^{O,G_{\sf Haar}(\cdot)}}} \leq \varepsilon(\secpar),
		      \end{equation*}
		      where \(G_{\sf Haar}(\cdot)\) on input \(x \in \bin^{m(\secpar)}\), output
		      \(\ket{\psi_{x}}\) stored in a new register, where, for every \(x \in \bin^{m(\secpar)}\),
		      \(\ket{\psi_{x}} \gets \Haar_{n(\secpar)}\).
	\end{itemize}
	When the adversary always measures the input register before making queries to \(\gen^O_k\) or \(G_{\sf Haar}\), we say that \(\gen^O_k\) is classical-accessible.
	Otherwise, we say that it is quantum-accessible.

	We say that \(\gen\) is a \((\kappa(\secpar), m(\secpar), n(\secpar))\)-PRFSG to indicate that its key length is \(\kappa(\secpar)\), its input length is \(m(\secpar)\), and its output
	length is \(n(\secpar)\).
	We say that a PRFSG is a \emph{short PRFSG} when \(n=\bigTheta{\log \secpar}\), and a \emph{(long) PRFSG} when \(n=\omega(\log \secpar)\).
\end{definition}

For pseudorandom unitaries, we only consider the super-logarithmic output length and without inverse oracle access. Unlike~\cite{C:JiLiuSon18}, we allow PRUs to not be unitary.

\begin{definition}[PRUs]\label{def: PRU}
	We say that
	an oracle QPT algorithm \(G^O\) is a pseudorandom unitary in the CHFS model if
	the following holds for some \(n:\mathbb N \to \mathbb N\) such that \(n=\omega(\log \secpar)\):
	\begin{itemize}
		\item \textbf{Quantum operation:}
		      For any \(\secpar\in\NN\) and \(k\in \bit^\secpar\),
		      \(G^O_k\) takes as input an \(n(\secpar)\)-qubit (mixed) state \(\rho\) and outputs
		      an \(n(\secpar)\)-qubit state \(G^O_k(\rho)\).
		\item \textbf{Pseudorandomness}: For any oracle QPT adversary \(\adv^O = \{\adv^O_{\secpar}\}_{\secpar \in \NN}\), there exists a negligible
		      function \(\varepsilon\) such that for all \(\secpar \in \NN\),
		      \begin{equation*}
			      \abs{\prob[k \gets \bin^{\secpar}]{1 \gets \adv_{\secpar}^{O,G^O_k}} -
			      \prob[\mathcal{U} \gets \mu_{n(\secpar)}]{1 \gets \adv_{\secpar}^{O,\mathcal{U}}}} \leq \varepsilon(\secpar).
		      \end{equation*}
	\end{itemize}
	When the adversary makes non-adaptive queries to \(G^O\) and \(\mathcal U\), we say that \(G\) is non-adaptively secure.
\end{definition}

We also define quantum pseudorandom generators (QPRGs), which are algorithms whose output is indistinguishable from random, and is always the same with probability negligibly close to one.

\begin{definition}[QPRGs]
	\label{def:QPRG}
	We say that an oracle QPT algorithm \(F^O\) that outputs an \(m(\secpar)\)-bit classical string on \(n(\secpar)\)-bit input
	is a \emph{quantum pseudorandom generator} (QPRG) with \((1-\epsilon)\)-pseudodeterminism if the following conditions hold for a function \(\epsilon\). If \(\epsilon\) is negligible, we just call \(F\) a QPRG.
	\begin{itemize}
		\item \textbf{\((1-\epsilon)\)-Pseudodeterminism}.
		      For every \(\secpar \in \NN\),
		      the following holds:
		      \begin{enumerate}
			      \item There exists a set \(K_{\secpar} \subseteq \bin^{n(\secpar)}\) of ``good seeds'' such that
			            \[\Pr[x \in K_{\secpar}:x \gets \bin^{n(\secpar)}]\ge 1-\epsilon.\]
			      \item There exists a deterministic function \(f_{\secpar}:\{0,1\}^{n(\secpar)}\to\{0,1\}^{m(\secpar)}\)
			            such that for every \(x \in K_{\secpar}\), it holds that
			            \begin{align*}
				            \prob{F^O(x)=f_{\secpar}(x)} \ge 1-\varepsilon,
			            \end{align*}
			            where the probability is over the randomness of \(F\).
		      \end{enumerate}
		\item \textbf{Security}. For any oracle QPT algorithm \(\adv^O = \{\adv^O_{\secpar}\}_{\secpar \in \NN}\), there exists a negligible
		      function \(\varepsilon\) such that
		      \begin{equation*}
			      \abs{\prob[y \gets \bin^{m(\secpar)}]{1 \gets \adv_{\secpar}^O(y)} -
				      \prob[x \gets \bin^{n(\secpar)}]{1 \gets \adv_{\secpar}^O(F^O(x))}} \leq \varepsilon(\secpar),
		      \end{equation*}
		      where the probability is over the randomness of \(F\) and \(\adv_{\secpar}\).
		\item \textbf{Length extension.} \(n(\secpar)<m(\secpar)\) holds for all \(\secpar\in\NN\).
	\end{itemize}
\end{definition}
\fi

\ifnum\submission=0

\subsection{QPSPACE oracle}
We recall the definition of the QPSPACE oracle that implements the arbitrary unitary operation described by polynomial size input \cite{EC:CheColSat25,EC:BMMMY25}.
\begin{definition}[\(\qpspace\) Oracle]\label{def:qpspace}
	The \emph{unitary} QPSPACE machine oracle, denoted by \(\qpspace\), is defined as follows: it takes a pair \((\rho,M,t,b)\) of an \(\ell\)-qubit quantum state \(\rho\), a classical Turing machine \(M\), and an integer \(t\in\mathbb N\), and \(b\in\bit\) denotes the abort bit.
	The oracle runs \(M\) for \(t\) steps to obtain the description of a unitary quantum circuit \(C\) that operates on \(\ell\) qubits; if \(M\) does not terminate after \(t\) steps or the output is not described as above, the oracle flips the abort register and halts. Otherwise, the oracle applies \(C\) on \(\rho\) and returns the output quantum state without measurement.
\end{definition}

The quantum access to the QPSPACE oracle is done by allowing coherent \((M,t)\).
For any unitary quantum circuit \(C\) that is output by a machine \(M\) after \(t\) steps, there is a QPT algorithm with \(\qpspace\) oracle that implements \(C^{-1}(\rho)\) on input \(\rho\) \cite[Proposition 3.5]{EC:BMMMY25}.

\subsection{State property tests}
\subsubsection{Swap test}
We review the basic results of the swap test, which can be used to test the purity of a state.
We provide some lemmas about the swap test on a state that is close to pure states, which are essential to obtain our results.

For two quantum states \(\sigma,\rho\) stored in two different registers \(\rA,\rB\), the swap test is executed on the registers \(\rA,\rB\) and a control register \(\rC\) initialized to \(\ketbra{1}\).
It applies Hadamard on \(\rC\), swaps \(\rA\) and \(\rB\) conditioned on \(\rC\), and measures \(\rC\) on the Hadamard basis.
\begin{lemma}[Swap test]
	\label{lem: swap}
	The swap test on input \((\sigma,\rho)\) outputs 1 with probability
	\[
		\frac{1+\Tr(\rho\sigma)}{2},
	\]
	in which case we say that it passes the swap test. For pure states \(\ket\sigma,\ket\rho\), it equals \(\frac{1+|\braket{\rho}{\sigma}|^{2}}{2}\).

\end{lemma}

When \(\sigma=\rho\), we sometimes call it a \text{purity test} on \(\rho\), which outputs \(1\) with certainty if and only if \(\rho\) is a pure state.
\begin{lemma}\label{lem: purity_test}
	Suppose that \(\Tr(\rho^{2}) \le 1-1/T\) for some state \(\rho\) and \(T\in \mathbb N\).
	Let \(\lambda\in \mathbb N\).
	If we run the purity test \(16T\lambda\) times on \(\rho\), then the probability that at least \(4\lambda\) tests fail among \(16T\lambda\) is at least \(1-2^{-\lambda}\).
\end{lemma}
\begin{proof}
	Note that each test succeeds with probability \((1+\Tr(\rho^{2}))/2\le 1-1/2T\), and is independent of each other.
	Applying Chernoff's inequality (\cref{lem:Cher}) for \(\delta=1/2\), we obtain the desired result.
\end{proof}

\subsubsection{Product test}
We first recall the product test to determine whether an \(n\)-partite state
\(\ket{\phi}\) is a product state or far from any product state from
\cite{FOCS:HarMon10}, then give a bound on the success of the product test on Haar-random
states.
\begin{lemma}[{\cite[Lemma 3]{FOCS:HarMon10}}, Product test for mixed states]\label{lemma:product_test_mixed}
	Let \(m\in\mathbb{N}\) and \(d_{1},\ldots,d_{m}\) be the local dimensions of a \(n\)-qubit system,
	i.e.\ \(\prod_{i\in[m]}d_{i}=2^{n}\).
	Let \(\rho\) be a mixed state of \(n\)-qubits and for every \(S\subseteq[m]\), denote
	by \(\rho_{S}\) the state after tracing out the subsystem \(\overline{S}:=[m]\setminus S\).
	Let \(\mathcal{A}_{\ptest}\) denote the algorithm that, given two copies of \(\rho\), performs the swap test on each
	of the \(m\) pairs of corresponding subsystems of the two copies of \(\rho\), and
	that outputs \(1\) if all the tests succeed, and \(0\) otherwise.
	Then, the probability that the algorithm \(\mathcal{A}_{\ptest}\) outputs \(1\) when
	applied to two copies of \(\rho\) is equal to
	\begin{equation*}
		\Pr(1\gets\mathcal{A}_{\ptest}(\ket{\phi}^{\otimes 2}))=\frac{1}{2^{m}}\sum_{S\subseteq[m]}\Tr[\rho_{S}^{2}].
	\end{equation*}
\end{lemma}

For Haar-random states, the above formula is explicitly calculated
for any partition \(S\cup\overline{S}\) of \([m]\) by~\cite{Lubkin}:
\begin{equation*}
	\underset{\ket{\psi}\gets\sigma}{\mathbb{E}}\Tr[\rho_{S}^{2}]=\frac{d_{S}+d_{\overline{S}}}{d_{S}\cdot d_{\overline{S}}+1}.
\end{equation*}
	As a consequence, we have the following bound for the success of the product
	test on Haar-random states.
\begin{lemma}[Product test for Haar-random states]\label{lemma:product_test_haar}
	Let \(m\in\mathbb{N}\) and \(\{d_{i}\}_{i\in[m]}\) be the local dimensions of a \(n\)-qubit system,
	i.e.\ \(\prod_{i\in[m]}d_{i}=2^{n}\).
	Then, the probability that the algorithm \(\mathcal{A}_{\ptest}\) outputs \(1\) when applied
	to two copies of a \(n\)-qubit Haar-random state \(\ket{\psi}\) satisfies:
	\begin{equation*}
		\underset{\ket{\psi}\gets\sigma}{\mathbb{E}}\Pr(1\gets\mathcal{A}_{\ptest}(\ket{\psi}^{\otimes 2}))\leq 2\left(\frac{3}{4}\right)^{m}.
	\end{equation*}
\end{lemma}
\ifnum\llncs=1
\begin{proof}[{Proof of~\cref{lemma:product_test_haar}}]%
\else
\begin{proof}
\fi
      For every partition \(S\cup\overline{S}\) of \([m]\), the local dimension of each partition is given by \(d_S=\prod_{i\in S}d_i\).
      \begin{align*}
        \underset{\ket{\psi}\gets\sigma}{\mathbb{E}}\Pr(1\gets\mathcal{A}_\ptest(\ket{\psi}^{\otimes 2}))&
        =\underset{\ket{\psi}\gets\sigma}{\mathbb{E}}\left[\frac{1}{2^m}\sum_{S\subseteq[m]}\Tr[\rho_S^2]\right]
        \\
        &=\frac{1}{2^m}\sum_{S\subseteq[m]}\frac{d_S+d_{\overline{S}}}{d_S\cdot d_{\overline{S}}+1}\le \frac{1}{2^m}\sum_{S\subseteq[m]}\frac{d_S+d_{\overline{S}}}{d_S\cdot d_{\overline{S}}}
        \\&=\frac{1}{2^m}\left(\sum_{S\subseteq[m]}\frac 1{d_S} + \frac1{d_{\overline{S}}}\right)
        =\frac{2}{2^m}\left(\sum_{S\subseteq[m]}\frac 1{d_S} \right)
        \\&=\frac{2}{2^m}\prod_{i\in[m]} \left(1+\frac1{d_i}\right)
        \leq \frac2{2^m}\prod_{i=1}^m\left(\frac{3}{2}\right)=2\left(\frac{3}{4}\right)^m,
  \end{align*}
  where we use the fact that each \(d_i\geq 2\) to obtain the last inequality.
\end{proof} %

We also need the following simple extension of the Haar product-test bound, in
which the Haar-random input is restricted to a subspace obtained by appending
zero ancillas.
\begin{lemma}
  \label{lemma:product-test-subspace}
  Let \(\mathcal H_0\) be a Hilbert space of dimension \(d=2^{n-t}\), and let
  \(\mathcal H=\mathcal H_0\otimes \mathbb C^{2^t}\), so that
  \(\dim(\mathcal H)=D=2^n\) for some \(0 \leq t < n\). 
  Let \(P\) be any quantum protocol that takes two copies of a pure state in
  \(\mathcal H\) and outputs a bit.
  Suppose that
  \begin{equation*}
    \prob[\ket{\phi} \leftarrow \sigma_{\mathcal H}]{P\left(\ket{\phi}^{\otimes 2}\right)=1} 
    \ge 1-\varepsilon.
  \end{equation*}
  Then
  \begin{equation*}
    \prob[\ket{\psi} \leftarrow \sigma_{\mathcal H_{0}}]{P\left(\left(\ket{\psi} \otimes \ket{0^{t}}\right)^{\otimes 2}\right)=1}
    \ge 1- 4^{t}\varepsilon .
  \end{equation*}
\end{lemma}

\begin{proof}
  Since \(P\) is a two-copy quantum protocol with binary output, there exists a
  POVM element \(0\le M\le I\) on \(\mathcal H^{\otimes 2}\) corresponding to
  the event that \(P\) outputs \(0\) such that, for every pure state
  \(\ket{\eta}\in\mathcal H\),
  \begin{equation*}
    \prob{P\left(\ket{\eta}^{\otimes 2}\right)=0}
    =
    \operatorname{Tr}\left(
      M \ketbra{\eta}^{\otimes 2}
    \right).
\end{equation*}
Let
\(\Omega_{\mathcal H} \coloneqq \mathbb E_{\ket{\phi}\leftarrow \sigma_{\mathcal H}} \left[ \ketbra{\phi}^{\otimes 2} \right]\).
By the Haar second-moment formula \cite{mele2024introduction},
\begin{equation*}
  \Omega_{\mathcal H}
  =
  \frac{\Pi_{\mathrm{sym}}^{\mathcal H}}{\binom{D+1}{2}}
  =
  \frac{2}{D(D+1)}\Pi_{\mathrm{sym}}^{\mathcal H},
\end{equation*}
where \(\Pi_{\mathrm{sym}}^{\mathcal H}\) is the projector onto the symmetric
subspace of \(\mathcal H^{\otimes 2}\).

Similarly, define the embedded subspace
\(W \coloneqq \mathcal H_0\otimes \ket{0^t}\subseteq \mathcal H\) and let
\(\Omega_{W} \coloneqq \mathbb E_{\ket{\psi}\leftarrow \sigma_{W}} \left[ \ketbra{\psi}^{\otimes 2} \right]\).
Again by the Haar second-moment formula, now applied inside the subspace \(W\),
\begin{equation*}
  \Omega_{W}
  =
  \frac{\Pi_{\mathrm{sym}}^{W}}{\binom{d+1}{2}}
  =
  \frac{2}{d(d+1)}\Pi_{\mathrm{sym}}^{W},
\end{equation*}
where \(\Pi_{\mathrm{sym}}^{W}\) is the projector onto the symmetric subspace of
\(W^{\otimes 2}\).

Since \(W\subseteq \mathcal H\), the symmetric subspace of \(W^{\otimes 2}\) is
a subspace of the symmetric subspace of \(\mathcal H^{\otimes 2}\). 
Hence
\(\Pi_{\mathrm{sym}}^{W}\le \Pi_{\mathrm{sym}}^{\mathcal H}\).
Therefore,
\begin{equation*}
  \Omega_{W}
  =
  \frac{2}{d(d+1)}\Pi_{\mathrm{sym}}^{W}
  \le
  \frac{2}{d(d+1)}\Pi_{\mathrm{sym}}^{\mathcal H}
  =
  \frac{D(D+1)}{d(d+1)}\Omega_{\mathcal H}.
\end{equation*}
By assumption,
\begin{equation*}
  \operatorname{Tr}(M\Omega_{\mathcal H})
  =
  \prob[\ket{\phi}\leftarrow \sigma_{\mathcal H}]
  {P\left(\ket{\phi}^{\otimes 2}\right)=0}
  \le \varepsilon .
\end{equation*}
Thus,
\begin{align*}
  \prob[\ket{\psi}\leftarrow \sigma_{\mathcal H_{0}}]
  {P\left(\left(\ket{\psi} \otimes \ket{0^{t}}\right)^{\otimes 2}\right) = 0}
  &=
  \operatorname{Tr}(M\Omega_{W}) \\
  &\le
  \frac{D(D+1)}{d(d+1)}
  \operatorname{Tr}(M\Omega_{\mathcal H}) \\
  &\le
    \frac{D(D+1)}{d(d+1)}\varepsilon \\
  &\le
    \left(\frac{D}{d}\right)^{2}\varepsilon = 4^{t}\varepsilon. \qedhere
\end{align*}
\end{proof}

\subsection{Quantum OR lemma}
\begin{lemma}[{\cite[Corollary 3.1]{HLM17}}, Quantum OR lemma]\label{lemma:quantum_or}
	Let \(\{\Pi_{i}\}_{i\in[N]}\) be binary-valued POVMs.
	Let \(0<\eps<1/2\) and \(\delta>0\).
	Let \(\Psi\) be a quantum state such that either
	\begin{enumerate}[label=\roman*)]
		\item there exists \(i\in[N]\) such that \(\Tr[\Pi_{i}\Psi]\geq1-\eps\), or
		\item for all \(i\in[N]\), \(\Tr[\Pi_{i}\Psi]\leq\delta\).
	\end{enumerate}
	Then, there is a quantum circuit \(C\), called ``OR tester'', such that
	measuring the first qubit in case \(i)\) yields
	\begin{equation*}
		\Pr(1\gets C(\Psi))\geq\frac{(1-\eps)^{2}}{7},
	\end{equation*}
	and in case \(ii)\),
	\begin{equation*}
		\Pr(1\gets C(\Psi))\leq 4N\delta.
	\end{equation*}

	Moreover, the circuit \(C\) can be implemented by a unitary quantum poly-space machine as long as each POVM \(\Pi_{i}\) can be implemented by a quantum poly-space machine and the set of measurements has a concise polynomial description.
	In other words, the quantum OR tester can be executed by a \(\qpspace\)-aided
	BQP algorithm, where the oracle \(\qpspace\) is defined in~\cref{def:qpspace}.
\end{lemma}
\begin{remark}\label{rem: POVMqpspace}
	``Moreover'' part of the above theorem for the projective measurements is shown in \cite[Appendix A]{EC:CheColSat25}, and the extension to the POVMs is observed in \cite[Lemma 5.2]{EC:BMMMY25}.
\end{remark}

\subsection{Useful lemmas}
\begin{lemma}[Almost as good as new lemma {\cite{aaronson2004limitations,aaronson2016complexity}}]\label{lem:almost-as-good-as-new}
	Let \(\cM=(\Pi_{0},\Pi_{1})\) be a binary measurement that acts as \(\cM(\rho)=\Pi_{0} \rho \Pi_{0} + \Pi_{1} \rho \Pi_{1}\). If \(\Tr[\Pi_{0} \rho]\ge 1-\epsilon\) for \(\epsilon>0\), then it holds that
	\(\|\rho - \cM(\rho)\|_{tr} \le \sqrt{\epsilon}.\)
\end{lemma}
\begin{corollary}\label{cor:gentle proj}
	In the same setting, \(\|\rho-\Pi_{0}\rho\Pi_{0}\|_{tr}\le \epsilon+\sqrt{\epsilon}\le 2\sqrt{\epsilon}.\)
\end{corollary}
\begin{proof}
	We have \(\|\cM(\rho)-\Pi_{0}\rho\Pi_{0}\|_{tr}=\|\Pi_{1}\rho \Pi_{1}\|_{tr}\le \epsilon,\) which gives the result.
\end{proof}

\subsubsection{Norms and Process tomography.}
For a matrix \(M\), the operator norm is defined by
\[
	\|M\|_{op}:= \sup_{\|\ket{\phi}\|_{2}=1} \|M\ket{\phi}\|_{2},
\]
which satisfies \(\|M+N\|_{op}= \max(\|M\|_{op},\|N\|_{op})\) if \(M\) and \(N\) act on the orthogonal space. In particular, for \(M=\sum_{x} \ketbra{x} \otimes M_{x}\), it holds that
\begin{align}\label{eqn:operator_norm_maximum}
	\|M\|_{op} = \max_{x} \|M_{x}\|_{op}.
\end{align}

The diamond norm of an operator \(A\), denoted by \(\|A\|_{\diamond}\), is defined by:
\[
	\|A(\cdot)\|_{\diamond}:= \sup_{\Tr(\rho)=1,\rho \ge 0} \|A\otimes I(\rho)\|_{1},
\]
where \(I\) denotes the identity acting with the same dimension as \(A\). We sometimes omit \((\cdot)\) if it is clear from the context.
We use the following fact about the diamond norm: for quantum channels
\(A,B\) and a density matrix \(\rho\), it holds that
\[
	\|A\otimes I(\rho) - B\otimes I(\rho)\|_{tr} \le\frac12 \|A(\cdot) -B(\cdot) \|_{\diamond}.
\]
We will also use the fact that for unitaries \(U,V\) and the corresponding channels \(\mathcal U,\mathcal V\), it holds that
\begin{equation}\label{eqn: diamond_bound_by_operator_norm}
	\|\mathcal U(\cdot) - \mathcal V(\cdot) \|_{\diamond} \le 2 \|U-V\|_{op}
\end{equation}
for the operator norm \(\|\cdot\|_{op}\)
because
\begin{align*}
	\|\mathcal U(\cdot) - \mathcal V(\cdot)\|_{\diamond} & =\sup_{\Tr(\rho)=1,\rho \ge 0} \|(U\otimes I)(\rho)(U^{\dagger} \otimes I)-(V\otimes I)(\rho)(V^{\dagger} \otimes I)\|_{1}                                                         \\
	                                                   & \le \sup_{\Tr(\rho)=1,\rho \ge 0} \|(U-V)\otimes I\|_{op} \|\rho\|_{1} \|U^{\dagger} \otimes I\|_{op} + \|V\otimes I\|_{op} \|\rho\|_{1} \|(U^{\dagger}- V^{\dagger}) \otimes I\|_{op}
	\\&\le 2\|U-V\|_{op}
\end{align*}
where we use \(\|A\rho B\|_{1} \le \|A\|_{op} \|\rho\|_{1} \|B\|_{op}\).

\begin{theorem}[{\cite{FOCS:HKOT23}}]\label{thm: process_tomography}
	There exists a quantum algorithm \(\Tom\) that, given black-box access to a unitary \(Z\) acting on the \(d\)-dimensional space, satisfies the following for any input \(\epsilon,\delta \in (0,1)\):
	\begin{description}
		\item[Accuracy:] It outputs a classical description of a unitary \(Z\) such that
		      \[
			      \Pr_{Z'\gets \Tom}\left[
			      \|\mathcal Z(\cdot ) - \mathcal Z' (\cdot) \|_{\diamond} \le \epsilon
			      \right] \ge 1-\delta.
		      \]
		\item[Efficiency:] It makes \(O\left(\frac{d^{2}}{\epsilon} \log \frac{1}{\delta}\right)\) queries to \(Z\), and takes \({\sf poly}(d,\frac{1}{\epsilon},\log\frac{1}{\delta})\) time.
	\end{description}
\end{theorem}

\subsubsection{Chernoff bounds.}

We use the following concentration inequalities.
\begin{lemma}[Multiplicative Chernoff bound]\label{lem:Cher}
	Let \(X_{1},\dots,X_{n}\) be some independent random variables over \(\bit\). Let \(X=\sum_{i=1}^{n} X_{i}\) and \(\mu=\Exp[X]\).
	It holds that
	\begin{itemize}
		\item \(\Pr[X \ge (1+\delta)\mu] \le \exp\left(-\frac{\mu\delta^{2} }{2+\delta}\right)\) for \(\delta\ge 0\), and
		\item \(\Pr[X \le (1-\delta) \mu] \le \exp\left(-\frac{\mu\delta^{2} }{2}\right)\) for \(0<\delta<1\).
	\end{itemize}
\end{lemma}
 \fi

\section{Common Haar Function-like State Oracles}\label{sec:oracle_model}

\subsection{CHFS oracles and unitarization}
We first recall the definition of \emph{swap (or reflection) oracles}.

\begin{definition}
	\label{def:swap_oracle}
	For a \(n\)-qubit pure quantum state \(\ket{\phi}\), the swap (or reflection) unitary is defined by
	\begin{equation*}
		S_{\ket{\phi}} := \ketbra{0^n}{\phi} + \ketbra{\phi}{0^n} + I_{\bot} = I - 2 \ketbra{ \phi-},
	\end{equation*}
	where we assume w.l.o.g. that \(\ket{\phi}\) is orthogonal to \(\ket{0^{n}}\),
	since if not, we can always append a single \(\ket{1}\) to it in order to make
	it orthogonal.
	Here, \(I_\bot\) is the identity on the subspace orthogonal to
	\({\sf span}\{\ket{0^n},\ket{\phi}\}\) and \(\ket{\phi-} = \frac{\ket{0^n} -\ket{\phi}}{\sqrt{2}}\).

	The last equality implies that \(S_{\ket{\phi}}\) is actually the reflection
	unitary with respect to \(\ket{\phi-}.\)
\end{definition}

We proceed to define the length-\(\ell\) common Haar-random function-like state (CHFS) oracle
and its ``unitarized'' oracle.
We fix a (QPT-computable) function \(\ell: \NN \to \NN\) representing the output length for each oracle, where we
typically consider \(\ell(\lambda)=\Theta(\log \lambda)\) or \(\ell(\lambda)= \lambda\).
We define two versions of the CHFS oracles as follows.
\begin{definition}[The isometry CHFS oracle]
	\label{def:state_chfs}
	We denote by \(\mathcal O_\ell\) the distribution
	over the family of isometry oracles where
	\begin{itemize}
		\item {\bf Randomness:} Choose a \(\ell(|x|)\)-qubit Haar random quantum state \(\ket{\phi_x}\) for each
		      \(x\in\bit^*\) and define \(\Phi=\{\ket{\phi_x}\}_{x\in \bit^*}\).
		\item {\bf Setup:} A family of oracles
		      \(O^\Phi = (O^\Phi_x)_{x\in \bit^*} \gets \mathcal O_\ell\)
		      is chosen by randomly sampling \(\Phi\), where \(O_x^\Phi := \ketbra{\phi_x}{0}\) denotes
		      the isometry operator.
		      Here \(\ket{0}\) denotes the trivial quantum state of dimension \(1\).
		\item {\bf Query:} It takes a quantum state \(\rho_{\bf XZ}\) as input and applies the isometry
		      \[
			      O^\Phi:= \sum_{x\in \bit^{|{\bf X}|}}\ketbra{x}_{\bf X} \otimes O_x^\Phi= \sum_{x\in \bit^{|{\bf X}|}}\ketbra{x}_{\bf X} \otimes \ket{\phi_x}_{\bf Y}\bra{0},
		      \]
		      on \(\rho_{\bf XZ}\), where \({\bf Y}\) denotes a new \(\ell(|{\bf X}|)\)-qubit
		      register, i.e., appending a new register \({\bf Y}\).
	\end{itemize}
	We say the CHFS oracle is \emph{classical-accessible} if the register \({\bf X}\)
	must always be measured in the computational basis before applying the query.
	Otherwise, we call the oracle \emph{quantum-accessible}.
\end{definition}

\begin{definition}[The unitarized CHFS oracle]
	\label{def:unitary_chfs}
	We denote by \(\mathcal S_\ell\) the distribution
	over the family of unitary oracles where
	\begin{itemize}
		\item {\bf Randomness:} Choose a \(\ell(|x|)\)-qubit Haar random quantum state \(\ket{\phi_x}\) for each
		      \(x\in\bit^*\) and define
		      \(\Phi=\{\ket{\phi_x}\ket{1}\}_{x\in \bit^*}\).\footnote{Here we explicitly append
			      \(\ket{1}\) to make the unitary CHFS oracle well-defined
			      w.r.t.~\cref{def:swap_oracle}. We occasionally omit \(\ket{1}\) if there is no confusion.}
		\item {\bf Setup:} A family of oracles
		      \(S^\Phi= (S^\Phi_x)_{x\in \bit^*} \gets \mathcal S_\ell\)
		      is chosen by randomly sampling \(\Phi\), where \(S^\Phi_x:= S_{\ket{\phi_x}}\)
		      denotes the reflection operator as defined in~\cref{def:swap_oracle}.
		\item {\bf Query:} It takes a quantum state \(\rho_{\bf XYZ}\) as input such that
		      \(|{\bf Y}|=\ell(|{\bf X}|)+1\) and applies the unitary
		      \[
			      S^\Phi:= \sum_{x\in \bit^{|{\bf X}|}}\ketbra{x}_{\bf X} \otimes S_x^{\Phi} = \sum_{x\in \bit^{|{\bf X}|}}\ketbra{x}_{\bf X} \otimes S_{\ket{\phi_x}},
		      \]
		      on \(\rho_{\bf XYZ}\), where \(S_x\) is applied on the register \({\bf Y}\).
	\end{itemize}
	The classical-accessible and quantum-accessible unitarized CHFS oracles are defined analogously.
\end{definition}

The \emph{(length-\(\ell\)) CHFS model} is defined as follows.
The randomness \(\Phi\) is chosen as an initialization.
We note that the sets of randomness \(\Phi\) used to define the isometry and
unitarized CHFS oracles are the same.
We omit the superscript \(\Phi\) if the context makes it clear.
Then, all parties have oracle access to the CHFS oracle \(O=O^\Phi\) or \(S=S^\Phi\).
We call this the log-length CHFS model for \(\ell(\lambda) = \bigO{\log \lambda}\), and the standard
CHFS model for \(\ell(\lambda) = \omega{\log \lambda}\).\footnote{We usually consider the
	standard CHFS model with \(\ell(\secpar) = \secpar\) for simplicity.}

We call it the \emph{state} (or \emph{isometry}) CHFS model when the oracle is
\(O^\Phi\), and the \emph{unitary} (or \emph{swap/reflection}) CHFS model when the oracle
is \(S^{\Phi}\). We, however, occasionally use \(O\) to denote the (any) CHFS oracle if the context is clear.

\subsection{Construction of PRFSGs in the CHFS model}\label{sec:existence_prfsg}
We show that PRFSGs with output length \(\ell\) exist in the
length-\(\ell\) CHFS model. Again, we stress that the PRFSGs are adaptively-secure by default. 

\begin{theorem}
	\label{thm:PRFS_in_CHFS}
	Quantum-accessible (resp.
	classical-accessible) \((\kappa, m, L)\)-PRFSGs exist in the length-\(\ell\)
	quantum-accessible (resp.
	classical-accessible) CHFS model with probability 1 for any key size \(\kappa=\omega(\log \lambda)\) and
	input size \(m =\poly\) and \(L=\ell(\kappa+m)\), regardless of the choice of unitary or isometry models.
	The same statement even holds relative to the \(\pspace\) and \(\qpspace\) oracles.
\end{theorem}
\begin{proof}
	We define the following \((\kappa,m,L)\)-PRFSGs. We explain the construction in the isometry CHFS model, but modifying it to the unitary CHFS model is obvious.
	\begin{description}
		\item[\(\gen^O(k,\cdot)\):] On the \(m\)-qubit input register \({\bf X}\),
		      it applies the map
		      \[
			      \ket{x}_{\bf X} \to \ket{x}_{\bf X}\otimes\ket{\phi_{k,x}}.
		      \]
		      This is done by, on input \(\rho_{\bf XZ}\), appending the
		      \(\kappa\)-qubit register \(\ket{k}_{\bf K}\) and making a query
		      to the oracle \(O^\Phi\) on the register \({\bf KX}\) and discarding the
		      registers \({\bf K}\).
	\end{description}

	We have that \(|k|+|x|=m+\kappa=\poly\) thus \(\gen\) can be implemented by a BQP algorithm with a single query to the CHFS oracle with \(m+\kappa\) length input.

    We claim that this construction is a secure PRFSG.
More precisely, we prove the following statement: For any algorithm \(A\) that
makes \(q\) queries,
define \({\sf Adv}(A,O)\) by
\begin{align}\label{eqn: goal_BBBVreduction}
\Pr\left[ A^{\gen(k,\cdot),O}(1^\secpar)\to 1 \right]
-
\Pr\left[ A^{G_{\sf Haar}(\cdot),O}(1^\secpar)\to 1 \right],
\end{align}
where \(G_{\sf Haar}(\cdot)\) is sampled independently of \(O\), and
\(G_{\sf Haar}(x)\) outputs an \(L=\ell(\kappa+m)\)-qubit Haar random state
\(\ket{\psi_x}\).
Then, it holds that
\begin{align*}
	\Exp_O\left[{\sf Adv}(A,O)^2\right]
	=
	\bigO{\frac{q^2}{2^\kappa}}.
\end{align*}
    We prove this using the Yao's indistinguishability to unpredictability \cite{FOCS:Yao82} following \cite[Corollary~32]{STOC:KQST23}.
    Formally, we consider a quantum oracle algorithm \(B^s\) for
    \(s\in\bit^{2^\kappa}\), whose goal is to determine whether \(s=0^{2^\kappa}\) or \(s=e_k\) for a uniformly random \(k\in\bit^\kappa\).

The algorithm \(B\) samples its own CHFS oracle \(O'\), together with two
independent instances \(G_{\sf Haar}(\cdot)\) and
\(G'_{\sf Haar}(\cdot)\) as defined in \cref{def:PRFSG}.
For \(z=(k',x)\in\bit^{\kappa+m}\), queries to the simulated CHFS oracle
are answered using \(G_{\sf Haar}(x)\) if \(s_{k'}=1\), and using the state
originally sampled for \(O'(z)\) if \(s_{k'}=0\).
Queries on all other input lengths are answered using \(O'\) without
modification.
We denote the resulting CHFS oracle by \(O'_s\).
Each query to \(O'_s\) can be processed coherently using a constant number
of queries to \(s\).

We identify an output bit \(b\in\bit\) with
\(2b-1\in\{-1,1\}\).
The algorithm \(B\) proceeds as follows using \(A\) twice.
\begin{enumerate}

	\item It runs
	\(A^{G_{\sf Haar}(\cdot),O'_s}(1^\secpar)\), and denotes its output by
	\(a\in\{-1,1\}\).

	\item It samples \(c\leftarrow\{-1,1\}\) uniformly.

	\item If \(c=1\), it samples \(k'\leftarrow\bit^\kappa\), runs 
	\(A^{\gen(k',\cdot),O'_s}(1^\secpar)\), where \(\gen\) is evaluated
	relative to \(O'_s\), and denotes its output by
	\(d\in\{-1,1\}\).
	If \(c=-1\), it runs
	\(A^{G'_{\sf Haar}(\cdot),O'_s}(1^\secpar)\), and again denotes its
	output by \(d\in\{-1,1\}\).

	\item It outputs \(1\) if \(a=cd\), and outputs \(0\) otherwise.
\end{enumerate}
The two executions of \(A\) use independent randomness.

For a fixed CHFS oracle \(O\), let \(p(O)\) denote the second probability
in \cref{eqn: goal_BBBVreduction}, and let
\(\Delta(O):={\sf Adv}(A,O)\).
Conditioned on the simulated CHFS oracle being \(O\), the second execution
of \(A\) satisfies
\[
	\Exp[cd\mid O]
	=
	\frac{1}{2}\left(2\big(p(O)+\Delta(O)\big)-1\right)
	-
	\frac{1}{2}\left(2p(O)-1\right) =
	\Delta(O).
\]

If \(s=0^{2^\kappa}\), then \(G_{\sf Haar}(\cdot)\) is independent of
\(O'_s\), and the first execution of \(A\) corresponds to the ideal
experiment.
Moreover, \(O'_s\) has the same distribution as a CHFS oracle.
It follows that
\begin{align*}
	\Pr\left[B^{0^{2^\kappa}}\to 1\right]
	=
	\frac{1}{2}
	+
	\frac{1}{2}
	\Exp_O\left[
		\left(2p(O)-1\right)\Delta(O)
	\right].
\end{align*}

On the other hand, if \(s=e_k\), then
\(G_{\sf Haar}(x)\) is the state corresponding to \((k,x)\) in \(O'_s\)
for every \(x\).
Thus, for a uniformly random \(k\), the first execution of \(A\)
corresponds to the real experiment.
The oracle \(O'_s\) still has the same distribution as a CHFS oracle, and
\(k\) is independent of \(O'_s\).
Therefore,
\begin{align*}
	\Pr_k\left[B^{e_k}\to 1\right]
	=
	\frac{1}{2}
	+
	\frac{1}{2}
	\Exp_O\left[
		\left(2\big(p(O)+\Delta(O)\big)-1\right)\Delta(O)
	\right].
\end{align*}
Consequently,
\[
\Pr_k\left[B^{e_k}\to 1\right]
	-
	\Pr\left[B^{0^{2^\kappa}}\to 1\right] =
	\frac{1}{2}
	\Exp_O\left[
		2\Delta(O)^2
	\right]=
	\Exp_O\left[{\sf Adv}(A,O)^2\right].
\]

Since \(B\) runs \(A\) twice,
by the BBBV theorem~\cite{BBBV97},
\[
	\Exp_O\left[{\sf Adv}(A,O)^2\right]=
	\Pr_k\left[B^{e_k}\to 1\right]
	-
	\Pr\left[B^{0^{2^\kappa}}\to 1\right] =
	\bigO{\frac{q^2}{2^\kappa}}.
\]

By Markov inequality, we have
\[
\Pr_O\left[
		\left|{\sf Adv}(A,O)\right|
		\ge 2^{-\kappa/4}
	\right]=\Pr_O\left[
		{\sf Adv}(A,O)^2
		\ge 2^{-\kappa/2}
	\right] \le
	\bigO{\frac{q^2}{2^{\kappa/2}}}.
\]
Because \(q\) is polynomially bounded and
\(\kappa=\omega(\log\secpar)\), the last quantity is at most
\(2^{-\kappa/4}\) for all sufficiently large \(\secpar\).
Moreover,
\(\sum_\secpar 2^{-\kappa(\secpar)/4}\) converges.
The Borel-Cantelli lemma ensures that, with probability \(1\) over \(O\),
\begin{align*}
	\left|{\sf Adv}(A,O)\right|
	\le
	2^{-\kappa/4}
\end{align*}
holds for all but finitely many \(\secpar\in\N\).
As there are only countably many uniform adversaries \(A\) with
polynomially many queries, this concludes the existence of the PRFSGs in
the isometry CHFS model.
The security proof for the unitary CHFS model works by replacing \(O\)
with \(S\).
\end{proof}

\section{Oracle Separation of QPRGs from PRFSGs}
\label{sec:conj_sep}
This section presents a separation between QPRGs and
PRFSGs with various output lengths.

\begin{theorem}
	\label{lemma:impossibility_qprg}
	Fix \(L\le \lambda\) such that \(L = \Omega(\log \lambda)\).\footnote{A standard padding argument extends this for all arbitrary polynomially bounded \(L\) with \(L=\Omega(\log \lambda).\)}
	There exists a unitary oracle \(O\) such that relative to \((O,\pspace)\), there exist (adaptively-secure
	quantum-accessible) PRFSGs with output length \(L\) but QPRGs do not exist.
\end{theorem}
In the proof of this theorem, we use
a novel geometric theorem about the
product Haar measure on states, which we call the barrier theorem.
We first present the oracle space which we work on, then present the barrier theorem and its proof in~\cref{sec:conjecture}.
The proof of the main separation is in~\cref{sec:qprg-short-prfs}.

\paragraph{Oracles and PRFSGs.}
Fix \(\ell\le \lambda\) such that \(\ell=\Omega(\log \lambda)\).
We set \(L=\ell(2\lambda)\) and consider the length-\(\ell\) unitary CHFS oracle \(O=O^{\Phi}\) for \(\Phi=\{\ket{\phi_x}\}_{x\in \{0,1\}^*}\)\footnote{For convenience, we omit the ancilla \(\ket{1}\).} sampled from a specific distribution to be specified and the \(\pspace\) oracle for a fixed PSPACE-complete problem.

We consider the following PRFSGs suggested in \cref{sec:existence_prfsg} with \(\kappa=m=\lambda\):
\begin{description}
	\item[\(\gen^O(k,x)\):] It outputs \(\ket{\phi_{k,x}}\).
\end{description}

We excerpt the following lemma from the proof of~\cref{thm:PRFS_in_CHFS} for later use. The final statement is proven using the series \(\sum_{i=n}^{\infty} a^i = a^n/(1-a)\) for \(a=2^{-1/4}\).
\begin{lemma}\label{lem: interval_PRFS}
	Fix \(\kappa=m=\lambda\).
	For any oracle QPT adversary \(\adv\) against \(\gen^O\), let \(\mathrm E_{\adv}[O,\lambda]\) be the event that
	\[
		\left|
		\Pr\left[ \adv^{\gen(k,\cdot),O}\to 1 \right] -\Pr_{G_{\sf Haar}}\left[ \adv^{G_{\sf Haar}(\cdot),O}\to 1 \right]
		\right|
		\ge 2^{-\lambda/4}
	\]
	happens. 
    Then, \(\Pr_{O_{2\lambda}}[\mathrm{E}_\adv[O,\lambda]|O_{\neq 2\lambda}] \le 2\cdot 2^{-\lambda/4}\) for sufficiently large \(\lambda\) for any fixed \(O_{\neq 2\lambda}= (O_n)_{n\neq 2\lambda}\). 
    Furthermore, for any \(n<m\) and 
    for any fixed \(O_{<n} \cup O_{>m}=(O_i)_{i<n \text{ or }i>m}\), 
	\[
		\Pr_O\left[
		\mathrm E_\adv[O,i] \text{ occurs for some }n/2 \le i< m/2
		| O_{<n}\cup O_{>m}\right] \le 11\cdot 2^{-n/8}.
	\]
\end{lemma}

\subsection{The barrier theorem}
\label{sec:conjecture}
Let \(X=\mathbb S(2^{n_1})\times \cdots \times \mathbb S(2^{n_k})\) be the product space of quantum states (plus global phases) equipped with the product Haar measure \(\sigma:=\sigma_{n_1}\times \cdots \times \sigma_{n_k}\).
For two elements \(\Phi=(\ket{\phi_1},\ldots,\ket{\phi_k}),\Psi=(\ket{\psi_1},\ldots,\ket{\psi_k})\) in \(X\), we define the max-\(\ell_2\) distance \(d_{\infty}(\Phi,\Psi)\eqdef\max_{i\in [k]} \|\phi_i - \psi_i\|_{2}\).
For two subsets \(S,T\) of \(X\), we define their distance as \(d_{\infty}(S,T):=\inf_{\Phi\in S,\Psi \in T} d_{\infty}(\Phi,\Psi)\).

We will show the following theorem.
\begin{theorem}[Barrier theorem]
	\label{thm:state_conjecture}
	Let
	\(X=\mathbb S(2^{n_1})\times \cdots \times \mathbb S(2^{n_k})\)
	with the corresponding product Haar measure \(\sigma=\sigma_{n_1}\times \cdots \times \sigma_{n_k}\), and let
	\(S_0,S_1\) be two measurable subsets of \(X\).
	If \(d_{\infty}(S_0,S_1)\ge \Delta\) and \(\min(\sigma(S_0),\sigma(S_1))\ge\Gamma\), then
	\(\sigma(X\setminus(S_0\cup S_1)) =\Omega(\Delta \Gamma)\).
\end{theorem}
Intuitively, \Cref{thm:state_conjecture} states that
if two sets
have a distance gap between them, then there must be a non-negligible \emph{barrier} of the
whole space that they do not cover regardless of their shape.

\paragraph{Proof overview.} A worst-case candidate is to take a very small \(S_0\) and the maximal \(S_1\) subject to the conditions. Given this, a natural approach is
\begin{enumerate}[label=(\roman*)]
	\item to show that the surface area of \(S_0\) or larger regions is bounded below, say by \(\epsilon\);\label{item1_con_overview}
	\item to ``integrate'' the surface area along a path from \(S_0\) to \(S_1\) of length \(\Delta\), obtaining a \(\Omega(\epsilon\Delta)\) lower bound.
\end{enumerate}
This approach indeed works with some results from differential geometry. The most technical part is \cref{item1_con_overview}, which is formalized by the notion of Cheeger isoperimetric constant. Roughly, the Cheeger isoperimetric constant \(h_X\) ensures that:
\begin{center}
	\text{any ``surface'' area between \(S_0\) and \(S_1\) is bounded below by }\((\epsilon=)h_X \cdot \Gamma \)
\end{center}
given \(\sigma(S_0),\sigma(S_1)\ge \Gamma\). This yields a lower bound of \(h_X \Gamma \Delta\).

The proof is completed by showing the dimension-free lower bound of \(h_X\). To do so, we use the results of Lichnerowicz and Buser, which assert that the Ricci curvature lower bound \(\Ric_X \ge c\cdot g_X\) for \(c>0\) implies the lower bound \(h_X =\Omega(\sqrt{c})\), where \(g_X\) is the Riemannian metric of the ambient space. The condition can be easily shown using some basic properties of the Ricci curvature of the product space\ifnum\final=0 (\cref{eqn:Ricci_product})\fi. This differential geometric analysis shows the result for the \(\ell_2\) of the arc-cosine distance\footnote{which is the geodesic distance corresponding to the natural round Riemannian metric on \(X\).}, and the desired result with the max-\(\ell_2\) distance follows by simple calculation.

\ifnum\llncs=1
	The formal proof is given in 
    \ifnum\final=1
    the full version~\cite{EPRINT:BHMV25}
    \else\cref{sec:formal_proof_conjecture}.\fi
\else
	The formal proof is given below.
\begin{proof}

	Recall that \(\mathbb S(2^n):=\{\ket{\psi}\in \mathbb C^{2^n}:\|\psi\|_2=1\}\) denotes the space of \(n\)-qubit quantum states together with the global phase, i.e., the complex hypersphere in \(\mathbb C^{2^n}\). Our theorem is stated as follows.
	For two elements \(\Phi=(\ket{\phi_1},\ldots,\ket{\phi_k}),\Psi=(\ket{\psi_1},\ldots,\ket{\psi_k})\) in \(X=\mathbb S(2^{n_1})\times \cdots \times \mathbb S(2^{n_k})\), the max-\(\ell_2\) distance is defined by \(d_{\infty}(\Phi,\Psi)\eqdef\max_{i\in [k]} \|\phi_i - \psi_i\|_{2}\).
	\paragraph{Distances.}
	We consider various distances of quantum states.
	For two states in \(\mathbb S(2^{n_i})\), the distance is
	\[
        \|\phi -\psi\|_
		{2} = \sqrt{2-2\Re\braket{\phi}{\psi}}.
	\]
	On the complex unit sphere \(\mathbb S(2^{n_i})\), the round geodesic distance is defined by
	\[
		d_{geo}(\phi,\psi) = \arccos(\Re\braket{\phi}{\psi}).
	\]
	We have \(\|\phi -\psi\|_{2}= 2\sin (\frac{d_{geo}(\phi,\psi)}2) \le d_{geo}(\phi,\psi)\) in \(\mathbb S(2^{n_i})\).

	In the product space \(X\), we consider two distances. The first distance is the max-\(\ell_2\) distance \(d_{\infty}\) defined above. The other distance is the \(\ell_2\)-geodesic distance
	\begin{align}\label{eqn:defell2distance}
		d^{geo}_{2}(\Phi,\Psi) = \sqrt{\sum_{i=1}^k d_{geo,i} (\phi_i,\psi_i)^2}
	\end{align}
	where \(d_{geo,i}\) is the angular geodesic distance in \(\mathbb S(2^{n_i})\).
	We stress that the distance \(d^{geo}_2\) is defined for the geodesic distance, while \(d_\infty\) is defined for the \(\ell_2\) distance. We will later use the following inequality:
	\[d_{\infty}(\Phi,\Psi)  = \max_i \|\phi_i-\psi_i\|_2 \le \max_i d_{geo,i}(\phi_i,\psi_i) \le d^{geo}_2(\Phi,\Psi).\]

	\paragraph{Riemannian manifolds and metrics, and Ricci curvature.}
	A Riemannian manifold is a pair of \((M,g)\), where \(M\) is a smooth manifold and \(g\) is a Riemannian metric\footnote{We stress that two notions, \emph{distance} and \emph{metric}, are used differently; the metric is used only for the Riemannian metric.}, which assigns a positive-definite symmetric bilinear form \(g_p: T_pM\times T_pM \to \mathbb R\) for each point \(p\in M\), where \(T_pM\) denotes the tangent space of \(M\) at \(p\). The Riemannian manifold gives rise to the geodesic distance \(d_g\). In the complex sphere \(\mathbb{S}(2^n) \subseteq \mathbb C^{2^n} \subseteq \mathbb R^{2^{n+1}}\), we use the round
    Riemannian metric \(g_{\mathbb S}\) induced by the real Euclidean inner product
    \[
        \langle u,v\rangle_{\mathbb R}:=\Re\braket{u}{v}.
    \]
    The corresponding geodesic distance is, as discussed above,
    \[
        d_{\mathbb S}(\phi,\psi)
        =
        \arccos(\Re\braket{\phi}{\psi}).
    \]
    There is a natural notion of the products of Riemannian manifolds; the geodesic distance of the products of the complex spheres with the round Riemannian metric gives the \(\ell_2\) distance defined in \cref{eqn:defell2distance}. Looking ahead, we will prove the statement for \(d^{geo}_2\) (instead of \(d_\infty\)) using the results from differential geometry below.

	On the Riemannian manifold, the Ricci curvature tensor \(\Ric\) is uniquely determined and gives a symmetric bilinear form \(\Ric_p: T_pM\times T_pM \to \mathbb R\) for each point \(p\in M\). In the product of Riemannian manifolds \(S=\prod_{1\le i \le r}(M_i,g_i)\) and the point \({\bf p} \in S\), 
    the Ricci curvature on a tangent vector \({\bf v}=(v_1,...,v_r) \in T_{\bf p}S\) satisfies
	\begin{align}\label{eqn:Ricci_product}
		\Ric_S({\bf v},{\bf v}) = \sum_{i=1}^r {\Ric}_{M_i} (v_i,v_i).
	\end{align}
	The intrinsic Riemannian metric \({\bf g}=(g_1,...,g_r)\)\footnote{Occasionally written \(\oplus_i g_i\).} satisfies a similar equality \(g({\bf v},{\bf v}) = \sum_{i=1}^{r} g_i(v_i,v_i)\).

	For a constant \(c\), we occasionally use the notation
	\begin{align}\label{eqn:Lic_condition}
		\Ric \ge c \cdot g \Longleftrightarrow \Ric_p - c\cdot  g_p \geqslant 0~~ \forall p \in M
	\end{align}
	where \(\geqslant\) denotes the positive semi-definite inequality.
	Equivalently, it means
	\[
		\Ric_p(v,v) \ge c\cdot g_p(v,v)
	\]
	holds (as an inequality over real numbers) for all \(p \in M\) and \(v \in T_pM\).
	We note that \(\Ric \ge c \cdot g\) is usually denoted by \(\Ric \ge c\) in the literature.

	We only use Riemannian manifolds and Ricci curvatures for our underlying space \(X=\mathbb S(2^{n_1})\times \cdots \times \mathbb S(2^{n_k})\) regarding the above relations, where we recall \(\mathbb S(2^{n})\) can be understood as the real unit sphere of dimension \(m_n:=2^{n+1}-1\). 
    The round sphere of real dimension \(m\) satisfies \(\Ric = (m-1)\cdot g\) (i.e., an Einstein manifold)~\cite{kobayashi1996foundations}.
    Therefore \(\Ric_{\mathbb S(2^{n_i})} = (2^{n_i+1}-2)\cdot g_{i}\).

	For our main interest \(X\) together with the product round Riemannian metric, for \({\bf v} = (v_1,...,v_k) \in T_{\bf p} X\) for some \({\bf p} \in X\),
	it holds that
    \begin{align*}
        \Ric_X({\bf v},{\bf v}) &= \sum_{i=1}^k \Ric_{\mathbb S(2^{n_i})} (v_i,v_i)\\
        &=\sum_{i=1}^k (2^{n_i+1}-2)\cdot g_{i}(v_i,v_i) \\
        &\ge 2 \sum_i g_{i} (v_i,v_i) = 2{\bf g}({\bf v},{\bf v})
    \end{align*}
    where we use \(2^{n_i+1}-2 \ge 2\) for \(n_i \ge 1\). This means \(X\) satisfies \cref{eqn:Lic_condition} for \(c=2.\)

	Lichnerowicz's theorem~\cite{Lic58} states that (See e.g., \cite[Theorem 4.19]{Aub98}) for an \(n\)-dimensional compact Riemannian manifold \((M_n,g)\), if \(\Ric \ge c\cdot g\) for some \(c>0\), then the first nonzero eigenvalue of the Laplacian\footnote{Here, the Laplacian refers the Laplace-Beltrami operator of Riemannian manifold is defined as the negative of the divergence of the gradient, i.e., \(-\Delta f =-{\sf div}({\sf grad} f) .\) Negative sign is the analysts' convention which ensures the non-negativeness of the eigenvalues. The first nonzero eigenvalue is the smallest positive eigenvalue of \(-\Delta\), i.e., the smallest \(c>0\) such that \(-\Delta f = c f\) for some function \(f\).} satisfies
	\begin{align}
		\lambda_1 \ge \frac{nc}{n-1} \ge c.
	\end{align}
	In particular, for \(X\) together with the product round Riemannian metric, it holds that
	\begin{align}\label{eqn:minimum_nzeign}
		\lambda_1 \ge 2.
	\end{align}

	\paragraph{Cheeger isoperimetric constant.}
	We use the modern exposition for isoperimetric inequalities from \cite{Rit23}.
	Consider a Riemannian manifold \((M,g)\) with the geodesic distance \(d\) and probability measure \(\sigma\) (i.e., \(\sigma(M)=1\)).
	We define the outer Minkowski boundary measure (or the \emph{area}) of a measurable set \(E\) by
	\[
		\sigma^+(E)=\liminf_{r \downarrow 0} \frac{\sigma(E[r]) - \sigma(E)}{r}
	\]
	where \(E[r]:= \{p \in M: d(p,E)\le r\}\) is a closed neighborhood of \(E\).
	The Cheeger isoperimetric constant is
	\begin{equation}\label{def:Cheeger}
		h_M := \inf_{E:\text{measurable}, 0<\sigma(E)<1} \frac{\sigma^+(E)}{\min(\sigma(E),1-\sigma(E))}.
	\end{equation}

	Let \(\lambda_1\) be the smallest nonzero eigenvalue of the Laplacian of \(M\).
	Buser's inequality~\cite{Bus82} says that if \(\Ric\ge -(n-1)a^2\) for some \(a\ge 0\), then it holds that \(\lambda_1\le 2a(n-1)h_M + 10h_M^2\).
	If \(\Ric \ge 0 \), we have \(\lambda_1 \le 10h_M^2\).
	This implies that, for our \(X\),
	\begin{align}\label{eqn: hXminimum}
		h_X \ge \sqrt{1/5}
	\end{align}
	holds because \(\Ric \ge 2 {\bf g} \ge 0\) in \(X\) and \cref{eqn:minimum_nzeign}.

	\paragraph{Proof of~\cref{thm:state_conjecture}.}
	We first work on the product space \(X\) together with the product round Riemannian metric and the corresponding \(\ell_2\) distance \(d_2\).
	Let \(v(r):= \sigma(S_0[r])\) for \(0\le r\le \Delta\)
    where \(S_0[r]=\{x: d(x,S_0)\le r\}.\)
	Here, we assume that \(v\) is differentiable so that \(v'(r)=\sigma^+(S_0[r])\). This significantly simplifies the proof of \cref{eqn: after_integral}, and we provide the formal proof without this assumption in \cref{subsec: differentiability}.

	Note that in the definition of \(S_0[r]\) in \(X\), we use the natural distance \(d_2\) induced by the Riemannian manifold \(X\).
	Given the condition \( d_\infty(S_0,S_1) \ge \Delta\), we also have \(d^{geo}_2(S_0,S_1) \ge d_\infty(S_0,S_1) \ge \Delta\). For any \(0\le r <\Delta\), 
    \(v(r)\ge v(0) \ge \Gamma\). Also, 
    \(S_0[r] \cap S_1 = \emptyset\) holds so that \(v(r)= \sigma(S_0[r]) \le 1-\sigma(S_1)\) and \(1-v(r) \ge \sigma(S_1) \ge \Gamma\). This implies \(\min(v(r),1-v(r)) \ge \Gamma.\)

	The Cheeger constant ensures
	\begin{equation}\label{eqn:v'lowerbound}
		v'(r) \ge h_X \min (v(r),1-v(r)) \ge h_X \Gamma.
	\end{equation}
	Integrating \(v'\) from \(0\) to \(r<\Delta\) gives:
	\begin{align}\label{eqn: after_integral}
		1-\sigma(S_1) -\sigma(S_0) \ge \sigma(S_0[r]) - \sigma(S_0) = v(r)-v(0) \ge \int_0^{r} v'(x) dx \ge h_X \Gamma r.
	\end{align}
    By taking \(r\to \Delta\) from below, 
	this gives:
	\[
		\sigma(X\setminus(S_0 \cup S_1))=1-\sigma(S_0) - \sigma(S_1) \ge h_X \Gamma \Delta
	\]
	where \(h_X \ge  \sqrt{1/5}\) by \cref{eqn: hXminimum}. This proves the desired result.
\end{proof} \fi

\subsection{Impossibility of QPRGs}
\label{sec:qprg-short-prfs}
We can now prove the main result of this section.

\begin{proof}[Proof of~\cref{lemma:impossibility_qprg}]
	We will construct (by diagonalization) a single unitary oracle \(O\) such that, relative to
	\((O,\pspace)\), (i) short-PRFSGs exist for all sufficiently large \(\secpar\),
	yet (ii) no pseudodeterministic QPRG exists (i.e., for every candidate pseudodeterministic QPRG \(G\),
	there are infinitely many security parameters \(\secpar\) for which \(G\) is either non-pseudodeterministic or insecure).
	In the following, the unitary CHFS oracle \(S\) is indexed by an \emph{oracle-input length} \(t\in\mathbb N\):
	on query inputs \(x\in\bin^t\) (possibly in superposition), the oracle applies a reflection
	\(S_{\ket{\psi_x}\ket{1}}\) on \(\ell(t)+1\) qubits (cf.~\cref{def:unitary_chfs}).

	Without loss of generality, we work in the uniform quantum circuit model
	with oracle gates for \(S\) and access to \(\pspace\).
	Thus, the set of uniform oracle-aided QPRG candidates is countable,
	and we fix an enumeration over these candidates as \(\{G_j\}_{j\in\NN }\).
	For each \(G_j\), let \(p_j(\secpar) =\max(T_j(\secpar),\secpar)\) where \(T_j(\secpar)=\poly\) is a polynomial upper bound
	on the running time of \(G_j(1^\secpar, \cdot)\) (including oracle queries).
	Fix a surjection \(u:\NN\to\NN\) such that every \(j\in\NN\) has infinitely many preimages:\footnote{E.g., any \(u\) such that \(u(n^2+j)=j\) for \(j\le n\) works.}
	\[
		\forall j\in\NN,\quad |\{i\in\NN : u(i)=j\}|=\infty.
	\]
	Intuitively, we diagonalize against the candidate \(G_{u(i)}\) at security parameter \(\lambda_i\).

	We will define \(O\) for each index \(i \in \NN\).
	For index \(i\), we choose a security parameter \(\lambda_i\) and then fix
	the oracle \(O\) on a block of oracle-input lengths \(t\in I_i \coloneqq [a_i,b_i]\),
	which are defined as follows.
	Choose \(\lambda_1\) sufficiently large, and set \(\lambda_{i+1} \coloneqq 2^{2^{\lambda_i}}\) for \(i\ge 1\). 

	Define the parameters for each \(i\) as follows:
	\begin{itemize}
		\item \(\epsilon_{u(i)} (\lambda)\coloneqq \frac{d_{u(i)}}{p_{u(i)}(\lambda)^8}\) for a constant \(d_{u(i)}>0\) to be determined later.
		\item \(c_{u(i)}\ge 1\) be a constant so that \(\lambda_i^2 \cdot 2^{-c_{u(i)}/8} < \epsilon_{u(i)}(\lambda_i)\) 
		\item \(a_i \coloneqq \left\lceil c_{u(i)} \log \lambda_i\right\rceil\)
		\item \(b_i \coloneqq \max(p_{u(i)}(\lambda_i),\,a_i)\)
	\end{itemize}
	Note that  \(b_i= \max(p_{u(i)}(\lambda_i),a_i)\) while \(a_{i+1}=\Theta(\log\lambda_{i+1})=\Theta(2^{\lambda_i})\),
	hence \(b_i<a_{i+1}\) for all \(i\), so the blocks \(I_i\) are disjoint and strictly increasing.

	In the following, we assume that an oracle \(O_i\) is sampled so that it has the desired properties (to be explained) up to the \(i\)-th block.
	Our goal is to sample a new oracle that satisfies the properties up to the \((i+1)\)-th block, without hurting the previous blocks.

	Let \(S^{(<t)}\) denote the restriction of \(O\) to oracle-input lengths \(<t\), and define the conditional distribution \(\mathcal D_{S^{(<t)}}\) in which we sample the CHFS oracles according to the product Haar random distribution conditioned on the fixed prefix \(S^{(<t)}\).

	The procedure is as follows.
	Given \(O_i\), we fix \(S^{(<a_i)}\). This will never change in the later updates. We will sample \(O=O_{i+1}\) from \(\mathcal D_{S^{(<a_i)}}\).
	All probabilities below are with respect to this conditional distribution.

	Now we analyze the candidate \(G_{u(i)}\) relative to \(O\gets \mathcal D_{S^{(<a_i)}}\).
	We consider its security at the security parameter \(\lambda_i\).
	Define the following event.
	\begin{description}
		\item[\(B_i{[O]}\):]
			      This is the event that \(G_{u(i)}\) is pseudodeterministic
			      with probability at least \(1- \epsilon_{u(i)}(\lambda_i)\) over the random choice of an input
			      (as in \cref{def:QPRG}) at parameter \(\lambda_i\).
			      In other words, this event occurs only when:
			      for at least a \((1-\epsilon_{u(i)}(\lambda_i))\)-fraction of \(x \in \{0,1\}^{n(\lambda_i)}\), there exists \(y_{x,O}\) such that
		      \(\Pr[G^O(x) \to y_{x,O}] \ge 1-\epsilon_{u(i)}(\lambda_i)\).
	\end{description}
	Note that \(\lnot B_i[O]\) means that \(G_{u(i)}\) fails to achieve the \((1-\epsilon_{u(i)})\)-pseudodeterminism at least for the parameter \(\lambda_i\).
	We will need the following lemma whose proof is deferred to \cref{sec:proof_QPRGnotsecure}.

	\begin{lemma}
		\label{lemma:QPRGnotsecure}
		Suppose that \(O\) is sampled from \(\mathcal D_{S^{(<a_i)}}.\)
		Let \(G=(G_\lambda)_\lambda\) be a candidate oracle QPRG that takes an \(n=n(\secpar)\)-bit seed and outputs an \(m=m(\secpar)\)-bit string relative to \((O,\pspace)\) that makes at most \(p = \poly\) queries.

		There exists an oracle QPT algorithm \(\cB_G\) having access to \(\pspace\) against the QPRG security such that:
		If \(G\) is \((1-\epsilon(\lambda))\)-pseudodeterministic with probability at least \((1-\epsilon(\lambda))\) over \(O\) for parameter \(\lambda=\lambda_i\),
		then, it holds that
		\[
			\Pr_O\left[\cB_G^{\pspace}\text{ against }G^O\text{ has advantage }0.1\text{ at parameter }\lambda_i\right]\ge 1-\frac{1}{\lambda_i^2}.
		\]
	\end{lemma}
	Roughly, it says that there is an adversary breaking the pseudorandomness of \(G_{u(i)}\) with very high probability over the oracle, as long as \(\Pr[B_i[O]]\) is big enough.

	Regarding this, we consider the following two cases.
	We choose the oracle in the next block depending on them.
	\begin{enumerate}[label=\textbf{Case \arabic*:}, leftmargin=*]
		\item \(\Pr[\neg B_i[O]] \ge\epsilon_{u(i)}(\lambda_i)\).
		      In this case, we sample the oracle \(O\) from \(\mathcal D_{S^{(<a_i)}}\) conditioned on \(\lnot B_i[O]\).
              This condition enforces that \(G_{u(i)}\) fails at being pseudodetermistic.
		\item \(\Pr[\neg B_i[O]] < \epsilon_{u(i)}(\lambda_i)\).
		      In this case, we sample the oracle \(O\) from \(\mathcal D_{S^{(<a_i)}}\) (without any condition).
		      In this case, \cref{lemma:QPRGnotsecure} says that the distinguisher \(\cB=\cB_G\) achieves an advantage 0.1 at parameter \(\lambda_i\) with probability at least \(1-\frac{1}{\lambda_i^2} \ge 1-\frac{1}{i^2}\) for sufficiently large \(i\).
	\end{enumerate}

	In any case, we define \(S^{(<a_{i+1})}\) by the chosen \(O\) up to input length less than \(a_{i+1}\).
	Our final oracle will be the oracle matching to \(S^{<a_{i}}\) up to length less than \(a_i\) for all \(i\in \mathbb N\), which is iteratively constructed by the above procedure.
    The following claim states that PRFSGs exist relative to this oracle, and we defer the proof to~\Cref{sec:prfsgexists}.
	\begin{claim}
    \label{claim:prfsgexists}
		Relative to \(O\) sampled above and \(\pspace\), \(\gen^O\) is a secure PRFSG with probability 1.
	\end{claim}

	Fix a QPRG candidate \(G=G_i\), and write \(u^{-1}(i)=\{i_1,i_2,...\}\). Because \(\lnot B_{i_j}[O]\) means \(G_i\) does not satisfy \((1-\epsilon_{u(i)})\)-pseudodeterminism, the following is clear.
	\begin{claim}
		Relative to \(O\) sampled above and \(\pspace\), if {\bf Case 1} occurs infinitely many times in \(u^{-1}(i)\), \(G\) does not satisfy the \((1-\epsilon_{u(i)})\)-pseudodeterminism. In particular, it cannot be a QPRG with negligible errors.
	\end{claim}

	On the other hand, \cref{lemma:QPRGnotsecure} shows the following.
	\begin{claim}
		Relative to \(O\) sampled above and \(\pspace\), if {\bf Case 2} occurs infinitely many times in \(u^{-1}(i)\), \(\cB_G\) breaks the pseudorandomness of \(G\).
	\end{claim}
	This completes the proof.
\end{proof}

\subsection{Proof of~\Cref{claim:prfsgexists}}
\label{sec:prfsgexists}
We will need the following lemma, who shows that our oracle construction defined in the previous section stay close to the true CHFS oracle, in the sense that an adversary cannot distinguish between the two, even when given an advice.

\begin{lemma}%
\label{lem:transfer}
Fix \(\lambda\in\mathbb N\). 
Let \(\mu_r\) be the product Haar law of the CHFS
components on input lengths at most \(r\), with any permanently fixed initial
prefix held fixed. 
Let \(\nu_r\ll\mu_r\) and suppose that
\[
  \left\|\frac{\dd\nu_r}{\dd\mu_r}\right\|_\infty\le 2^s
\]
for some \(s\in \mathbb N\).
Let \(\adv\) make at most \(q\) oracle queries, all of whose CHFS
query strings have length at most \(r\).
For the PRFSG construction with
\(\kappa=m=\lambda\), define
\[
  \Delta_{\adv}(O,\lambda)
  :=
  \Pr\!\left[\adv^{\operatorname{Gen}^{O}(k,\cdot),O}(1^\lambda)=1\right]
  -
  \Pr\!\left[\adv^{G_{\Haar}(\cdot),O}(1^\lambda)=1\right],
\]
where \(k\leftarrow\bit^\lambda\), and where \(G_{\Haar}\) is an independent
Haar function-like state oracle with domain \(\bit^\lambda\).  Then
\begin{equation}
  \Exp_{O\leftarrow\nu_r}\!\left[|\Delta_{\adv}(O,\lambda)|\right]
  \le
  C\left(
       \frac{q+1}{2^{\lambda/2}}
       +
       \left(\frac{(s+1)(q+1)}{2^\lambda}\right)^{1/3}
    \right)
  \label{eq:desired-bound}
\end{equation}
for a universal constant \(C\).
This holds even if \(\adv\) also has
access to a fixed PSPACE oracle.
\end{lemma}

We will use the following adaptation of~\cite[Theorem~7.2]{EC:Liu23}.

\begin{lemma}
\label{lemma:liu}
There is a universal constant \(C>0\) such that, for a
uniformly random function \(H:[N]\to[M]\), every quantum algorithm having
\(S\) bits of classical advice depending arbitrarily on \(H\) and making at most
\(T\) quantum queries to \(H\) has signed acceptance-probability gap at most
\begin{equation}
  C\left(
      \frac{T}{\sqrt N}
      +\left(\frac{ST}{N}\right)^{1/3}
    \right)
  \label{eq:liu}
\end{equation}
between the experiments in which its challenge is \(H(K)\) for uniform
\(K\leftarrow[N]\) and in which its challenge is an independent uniform
element of \([M]\).
\end{lemma}

To prove~\Cref{lem:transfer}, we first prove it when every CHFS component
takes values in a finite set and the product reference law is generated by
uniform finite tokens.
We then pass to Haar-distributed states by a finite approximation and a query hybrid.

\begin{proof}[Proof of~\Cref{lem:transfer}]
Let \(N=2^\lambda\).  
The idea of the proof is as follows. 
The inequality
\(\dd\nu_r/\dd\mu_r\le 2^s\) means that a sample from \(\nu_r\) can be obtained
by rejection sampling from the product law \(\mu_r\) with acceptance
probability at least \(2^{-s}\). 
We sample about \(2^s\) independent candidates
and use only their selected index, which has length about \(s\), as
oracle-dependent advice. 
Then,~\Cref{lemma:liu} bounds the distinguishing advantage associated with that selected candidate. 
The finite-token formulation in Step~1 makes this reduction exact, and in Step~2 we remove the finiteness assumption.

\medskip
\noindent\emph{Step 1: the finite-token case.}
Write \(\mu\) and \(\nu\) for the finite versions of \(\mu_r\) and \(\nu_r\), and let
\[
  R:=2^s,
  \qquad
  f(O):=\frac{\dd\nu}{\dd\mu}(O),
\]
and note that \(0\le f\le R\) and \(\Exp_\mu[f]=1\). 
The target part of a CHFS oracle is indexed by pairs \((k,x)\in\bit^\lambda\times\bit^\lambda\).
For each \(k\), let \(Z_k(O)\) denote the block consisting of all target
components indexed by \((k,x)\) as \(x\) ranges over \(\bit^\lambda\).
Under
\(\mu\), the variables \((Z_k)_{k\in\bit^\lambda}\) are independent and
identically distributed and are independent of the remaining components,
which we denote by \(W(O)\).

By the finite-token assumption, there is a finite set \(\mathcal T\), a
uniform random variable \(U_{\mathcal T}\) on \(\mathcal T\), and a deterministic
decoder \(D\) such that \(D(U_{\mathcal T})\) has the same law as \(Z_k\) under
\(\mu\). 
A token \(t\in\mathcal T\) is only a classical label: \(D(t)\) specifies
the complete quantum state-oracle block
\((|\phi_{k,x}\rangle)_{x\in\bit^\lambda}\). 
We package the entire block into
one token because the random-image challenge supplies one value \(H(K)\),
which must determine the whole challenge oracle \(x\mapsto|\phi_{K,x}\rangle\).
Sample independently \(R\) candidate oracles
\(O^{(1)},\ldots,O^{R}\leftarrow\mu\).
Equivalently, sample their non-target
parts \(W^{(1)},\ldots,W^{(R)}\) independently, sample a uniformly random
function
\[
  H:\bit^\lambda\longrightarrow\mathcal T^R,
  \qquad
  H(k)=(T_{k,1},\ldots,T_{k,R}),
\]
and set \(Z_k(O^{(j)})=D(T_{k,j})\).
Thus \(H\) is a uniformly random function with domain size \(N\) and finite range size \(M=|\mathcal T|^R\).

Independently sample thresholds \(U_1,\ldots,U_R\) uniformly from \([0,1]\) and
define
\[
  J:=\min\left\{j\in[R]:
       U_j\le \frac{f(O^{(j)})}{R}
     \right\},
\]
with \(J=0\) if the set is empty.  Each candidate is accepted with probability
\[
  \Exp_{O\leftarrow\mu}\left[\frac{f(O)}R\right]=\frac1R.
\]
It follows that
\begin{equation}
  p_{\mathrm{succ}}:=\Pr[J\ne0]
  =1-\left(1-\frac1R\right)^R
  \ge 1-e^{-1}.
  \label{eq:success-probability}
\end{equation}
This is the point of the rejection-sampling step: although directly changing
measure from \(\mu\) to \(\nu\) would lose the multiplicative factor \(R=2^s\),
selection among \(R\) candidates records the same change of measure in
an index of only \(\log_2 R=s+O(1)\) bits.

Moreover, conditioned on \(J\ne0\), the selected oracle \(O^{(J)}\) has law
\(\nu\).  Indeed, for every set \(B\) of finite oracles and every \(j\in[R]\),
\begin{align*}
  \Pr[J=j\ \text{and}\ O^{(j)}\in B]
  &=\left(1-\frac1R\right)^{j-1}
    \int_B\frac{f(O)}R\,\dd\mu(O)\\
  &=\left(1-\frac1R\right)^{j-1}\frac{\nu(B)}R.
\end{align*}
Summing this identity over \(j\) and dividing by
\(p_{\mathrm{succ}}\) proves the assertion.

Fix an adversary \(\adv\) for the PRFSG experiment under the modified distribution \(\nu\).
We construct an auxiliary nonuniform distinguisher \(\advb\).
The advice given to \(\advb\) identifies a candidate oracle distributed according to \(\nu\), while the challenge determines whether \(\adv\) receives the real keyed block of that oracle or an independent block.
We will show that the advantage of \(\advb\) is a constant multiple of \(\mathbb E_{O\leftarrow\nu}[|\Delta_{\adv}(O,\lambda)|]\), and then apply~\Cref{lemma:liu}.
Note that the range of \(H\) can be superpolynomial: a single value contains descriptions of
all \(x\)-indexed states for all \(R\) candidates.  
This is not an issue, as~\Cref{lemma:liu} is a query-complexity statement uniform in the finite range size
\(M\).
Moreover, these tokens are oracle answers, not
advice, and they are never revealed to \(\adv\).
The only \(H\)-dependent data that will be supplied as advice are the selected index and one sign bit.

Fix for the moment the values
\[
  w=(W^{(1)},\ldots,W^{(R)})
  \quad\text{and}\quad
  u=(U_1,\ldots,U_R).
\]
Once \((w,u)\) is fixed, the non-target oracle components and
the rejection thresholds become part of a fixed, possibly very large,
online circuit, whereas all remaining target-block randomness is exactly the
classical random function \(H\).

For these fixed values, both \(J\) and \(O^{(J)}\) are functions of \(H\). 
Give to \(\advb\) the following \(H\)-dependent classical advice:
the value \(J\in\{0,1,\ldots,R\}\) and, when \(J\ne0\), the bit
\[
  \sigma(H):=
  \begin{cases}
    0,&\Delta_{\adv}(O^{(J)},\lambda)\ge0,\\
    1,&\Delta_{\adv}(O^{(J)},\lambda)<0.
  \end{cases}
\]
There is no computational restriction on the map producing the advice, so
this sign bit is legitimate.
The number of advice bits is at most
\begin{equation}
  S_0:=\left\lceil\log_2(R+1)\right\rceil+1
  \le \lceil s\rceil+3.
  \label{eq:advice-length}
\end{equation}

We now define \(\advb\).
If \(J=0\), it outputs an independent fair bit.
Otherwise it simulates \(\adv\) with public oracle \(O^{(J)}\).
A query
to a non-target component is answered using the hardwired value \(W^{(J)}\).
To answer a query to a target component \((k,x)\), \(\advb\) queries \(H\) on
\(k\), coherently selects the \(J\)-th token and the \(x\)-th component of its decoded
block, applies the corresponding finite oracle operation, and uncomputes the
token by a second query to \(H\).
This procedure works coherently when \(k\) and
\(x\) are in superposition.
Hence all public target-oracle queries of \(\adv\) cost
at most \(2q\) queries to \(H\).

Let \(Y\) be the random-image challenge.
Since \(Y\in\mathcal T^R\), its \(J\)-th
coordinate decodes to a complete function-like state block indexed by \(x\).
\(\advb\) uses that block as the challenge oracle given to \(\adv\).
This
requires no additional query to \(H\): the challenge value is part of the input given to \(\advb\).
Finally, \(\advb\) outputs \(\adv\)'s output when \(\sigma=0\)
and its complement when \(\sigma=1\).

If \(Y=H(K)\) for uniform \(K\leftarrow\bit^\lambda\), then the selected block
is exactly the real PRFSG block \((\phi_{K,x}^{(J)})_x\) belonging to the public
oracle \(O^{(J)}\).
If \(Y\) is an independent uniform element of
\(\mathcal T^R\), then its selected block is independent of \(O^{(J)}\) and has
the ideal block distribution.
It follows, for the fixed pair \((w,u)\), that
the signed acceptance-probability gap of \(\advb\) is
\begin{align*}
  \Exp_H\!\left[
      \mathbf 1_{\{J\ne0\}},
      |\Delta_{\adv}(O^{(J)}\lambda)|
    \right] &= \Pr[J\ne0]\Exp_H\!\left[|\Delta_{\adv}(O^{(J)},\lambda)| | J\neq0\right]\\
    &= p_{\mathrm{succ}} \Exp_{O\leftarrow\nu_r}\!\left[|\Delta_{\adv}(O,\lambda)|\right].
\end{align*}
The number of \(H\)-queries is at most \(T_0:=2q\) (and may harmlessly be bounded
by \(2(q+1)\)).  Applying~\Cref{lemma:liu}, and then averaging its bound over the
previously fixed \((w,u)\), gives
\begin{align}
  \Exp\!\left[
      \mathbf 1_{\{J\ne0\}},
      |\Delta_{\adv}(O^{(J)},\lambda)|
    \right]
  &\le
  C\left(
      \frac{2(q+1)}{\sqrt N}
      +
      \left(\frac{2S_0(q+1)}N\right)^{1/3}
    \right).
  \label{eq:liu-applied}
\end{align}
By the conditional-law calculation above, the left-hand side
is equal to
\( p_{\mathrm{succ}}\,\Exp_{O\leftarrow\nu}[|\Delta_{\adv}(O,\lambda)|].\)

Equations \eqref{eq:success-probability} and \eqref{eq:advice-length}, with
\(N=2^\lambda\), therefore imply \eqref{eq:desired-bound} in the finite-token
case. 

\medskip
\noindent\emph{Step 2: reduction of Haar states to finite tokens.}
The preceding argument cannot be applied directly to exact Haar states, since a
generic Haar state has no finite classical description.
Thus we will approximate the continuous case by a finite-token experiments, apply the
bound to each approximation, and then take a limit.
The two properties that
must survive this passage are the density-ratio bound and the adversary's
acceptance probabilities.

Only finitely many CHFS components have input length at most \(r\).  Each
component takes values in a compact pure-state space.
For
\(\eta>0\), choose a finite Borel partition of each such space into sets of
diameter at most \(\eta\) in Euclidean norm and choose one representative from
every nonempty cell.  Let \(\pi_\eta\) replace every state by the representative
of its cell, and define the pushforward measures
\[
  \mu_\eta^0:=(\pi_\eta)_*\mu_r,
  \qquad
  \nu_\eta^0:=(\pi_\eta)_*\nu_r.
\]
For every subset \(B\) of the resulting finite oracle space,
\[
  \nu_\eta^0(B)
  =\nu_r(\pi_\eta^{-1}(B))
  \le R\mu_r(\pi_\eta^{-1}(B))
  =R\mu_\eta^0(B).
\]
Hence \(\nu_\eta^0\ll\mu_\eta^0\) and
\begin{equation}
  \left\|\frac{\dd\nu_\eta^0}{\dd\mu_\eta^0}\right\|_\infty\le R.
  \label{eq:pushforward-density}
\end{equation}
Thus coarse-graining cannot increase the density ratio: every event in the
finite experiment is the preimage of an event in the original experiment.

The finite component probabilities in \(\mu_\eta^0\) do not need to be rational.
Choose, on the same supports and consistently for identically distributed
components, rational probability vectors converging to those component
probabilities.  Their product defines a product measure
\(\widetilde\mu_{\eta,n}\) such that
\begin{equation}
  \TV(\widetilde\mu_{\eta,n},\mu_\eta^0)\longrightarrow0
  \quad(n\to\infty).
  \label{eq:mu-rational-convergence}
\end{equation}
The consistency condition preserves the fact that the target blocks are
i.i.d. over \(k\).  Put
\[
  f_\eta:=\frac{\dd\nu_\eta^0}{\dd\mu_\eta^0},
  \qquad
  Z_{\eta,n}:=\Exp_{\widetilde\mu_{\eta,n}}[f_\eta],
  \qquad
  \dd\widetilde\nu_{\eta,n}
  :=Z_{\eta,n}^{-1}f_\eta\,\dd\widetilde\mu_{\eta,n}.
\]
Because the space is finite, \eqref{eq:mu-rational-convergence} and
\(0\le f_\eta\le R\) imply
\[
  Z_{\eta,n}\longrightarrow\Exp_{\mu_\eta^0}[f_\eta]=1
  \quad\text{and}\quad
  \TV(\widetilde\nu_{\eta,n},\nu_\eta^0)\longrightarrow0.
\]
For all sufficiently large \(n\), \(Z_{\eta,n}\ge1/2\), and therefore
\begin{equation}
  \left\|\frac{\dd\widetilde\nu_{\eta,n}}
                   {\dd\widetilde\mu_{\eta,n}}
  \right\|_\infty
  \le 2R=2^{s+1}.
  \label{eq:approximated-density}
\end{equation}
Every rational finite probability vector is the image of a uniform variable
on a finite set: take a common denominator and assign to each outcome as many
tokens as its numerator.  Thus the target-block law under
\(\widetilde\mu_{\eta,n}\) has the exact finite-token representation required
in Step~1. 
Applying
Step~1 with density exponent \(s+1\) gives, uniformly
in all sufficiently large \(n\),
\begin{equation}
  \Exp_{O\leftarrow\widetilde\nu_{\eta,n}}
    [|\widetilde\Delta_{A,\eta,n}(O,\lambda)|]
  \le
  C_1\left(
       \frac{q+1}{2^{\lambda/2}}
       +
       \left(\frac{(s+1)(q+1)}{2^\lambda}\right)^{1/3}
     \right),
  \label{eq:discrete-bound}
\end{equation}
where changing \(s+1\) to \(s+O(1)\) has been absorbed into the universal
constant \(C_1\).

For completeness, the required continuity estimate follows from the standard
query hybrid.  Its role is to ensure that a \(q\)-query adversary cannot notice
the replacement of every Haar state by an \(\eta\)-close representative except
with total error \(O(q\eta)\).  If unit vectors \(|\psi\rangle\) and
\(|\widetilde\psi\rangle\)
satisfy \(\|\,|\psi\rangle-|\widetilde\psi\rangle\|_2\le\eta\), then their
preparation isometries differ by at most \(\eta\) on the prescribed blank
input, and their reflections satisfy
\[
  \bigl\|
    (I-2|\psi\rangle\!\langle\psi|)
    -(I-2|\widetilde\psi\rangle\!\langle\widetilde\psi|)
  \bigr\|_\infty
  \le 4\eta.
\]
The same estimate holds for a controlled direct sum, because its operator norm
is the maximum of the norms of its blocks.  Replacing oracle calls one at a
time and using the triangle inequality therefore changes any acceptance
probability of a \(q\)-query algorithm by at most \(4q\eta\) (or by another
universal constant times \((q+1)\eta\) if the challenge preparation is counted
separately).  Apply this hybrid once in the real experiment and once in the
ideal experiment, coupling every Haar state with the representative of its
partition cell.  If \(\Delta_{A,\eta}^0\) denotes the gap for
\((\mu_\eta^0,\nu_\eta^0)\), then, for a universal constant \(C_2\),
\begin{equation}
 \left|
  \Exp_{O\leftarrow\nu_r}|\Delta_A(O,\lambda)|
  -\Exp_{O\leftarrow\nu_\eta^0}
       |\Delta_{A,\eta}^0(O,\lambda)|
 \right|
 \le C_2(q+1)\eta.
 \label{eq:hybrid-error}
\end{equation}
Here we used \(||a|-|b||\le|a-b|\) after bounding the change in each of the two
acceptance probabilities.

It remains only to compare the two finite experiments.  Let
\(P_\eta^0\) and \(\widetilde P_{\eta,n}\) be their respective one-key target-block
laws.  Since a marginal cannot have larger total-variation distance than the
joint law,
\[
 \TV(\widetilde P_{\eta,n},P_\eta^0)
 \le \TV(\widetilde\mu_{\eta,n},\mu_\eta^0).
\]
For a fixed public oracle, replacing the independent ideal block changes its
acceptance probability by at most this distance.  Since the real experiment
is otherwise the same function of the public oracle, and since every absolute
gap lies in \([0,1]\), we obtain
\begin{align}
 &\left|
  \Exp_{O\leftarrow\widetilde\nu_{\eta,n}}
       |\widetilde\Delta_{A,\eta,n}(O,\lambda)|
  -\Exp_{O\leftarrow\nu_\eta^0}
       |\Delta_{A,\eta}^0(O,\lambda)|
 \right| \\
 &\qquad\le
 \TV(\widetilde\nu_{\eta,n},\nu_\eta^0)
 +\TV(\widetilde\mu_{\eta,n},\mu_\eta^0)
 \longrightarrow0.
 \label{eq:finite-continuity}
\end{align}
Taking \(n\to\infty\) in \eqref{eq:discrete-bound} and using
\eqref{eq:finite-continuity}, and then taking \(\eta\to0\) and using
\eqref{eq:hybrid-error}, proves \eqref{eq:desired-bound} for the original
Haar-state oracles.

\medskip
\noindent\emph{Step 3: a fixed PSPACE oracle.}
For fixed \(\lambda\), the computation of \(A\) can address only finitely many
strings in its PSPACE-oracle register.
The corresponding query unitary is
fixed and independent of \(H\).
Since the online computation in~\Cref{lemma:liu} is not
size-bounded, that finite unitary may be hardwired into every inter-query
operation of the simulator. 
It uses neither an \(H\)-query nor an
\(H\)-dependent advice bit.
The preceding proof is therefore unchanged.

\end{proof}

We can now prove~\Cref{claim:prfsgexists}.
    \begin{proof}[Proof of~\Cref{claim:prfsgexists}]   
View the inductive sampling procedure as a sequence of probability kernels, and let \(\nu\) be the resulting law of the final oracle.
Let \(\nu_r\) be its marginal on input lengths at most \(r\).
At a stage in Case~1, the construction conditions the fresh product-Haar extension on \(\neg B_i[O]\), and the definition of Case~1 gives
\[
  \Pr[\neg B_i[O]\mid S^{(<a_i)}]\ge\varepsilon_{u(i)}(\lambda_i).
\]
Consequently, that conditioning step has density ratio at most
\[
  \varepsilon_{u(i)}(\lambda_i)^{-1}
\]
with respect to the unconditioned extension.
A Case~2 step has density ratio one.
Later stages sample a normalized conditional distribution after the earlier prefix has been fixed, so they do not change the marginal law of that prefix.
Multiplying the losses from the relevant Case~1 stages gives
\begin{equation}\label{eq:density-product}
  \left\|\frac{d\nu_r}{d\mu_r}\right\|_\infty
  \le
  \prod_{i:a_i\le r}\varepsilon_{u(i)}(\lambda_i)^{-1}.
\end{equation}

The existing choice of parameters can be stated as
\[
  11\cdot 2^{-a_i/8}
  <\frac{\varepsilon_{u(i)}(\lambda_i)}{\lambda_i^2}.
\]
In particular,
\[
  \log_2\frac{1}{\varepsilon_{u(i)}(\lambda_i)}\le\frac{a_i}{8}.
\]
Since the \(a_i\) are distinct positive integers, Equation~\eqref{eq:density-product} implies
\begin{align*}
  s(r)
  &:=\log_2\left\|\frac{d\nu_r}{d\mu_r}\right\|_\infty \\
  &\le\frac18\sum_{i:a_i\le r}a_i
  \le\frac{r(r+1)}{16}.
\end{align*}
Thus every finite marginal of the final oracle differs from product Haar by a density ratio of the form \(2^{\poly(r)}\).

Fix a uniform PRFSG adversary \(\adv\). Let \(q_{\adv}(\lambda)\) be a polynomial query bound, and let \(r_{\adv}(\lambda)\ge 2\lambda\) be a polynomial upper bound on the input length of every CHFS query made by \(\adv(1^\lambda)\).
The value \(\Delta_{\adv}(O,\lambda)\) is determined by the marginal of \(O\) up to length \(r_{\adv}(\lambda)\).
We showed just above that
\[
  s(r_{\adv}(\lambda))=\poly(\lambda).
\]
Lemma~\ref{lem:transfer} therefore gives
\[
  \Exp_{O\leftarrow\nu}\!\left[\lvert\Delta_{\adv}(O,\lambda)\rvert\right]
  \le \poly[\lambda]\,2^{-\lambda/3}.
\]
For example, Markov's inequality yields
\[
  \Pr_{O\leftarrow\nu}\!\left[
    \lvert\Delta_{\adv}(O,\lambda)\rvert\ge 2^{-\lambda/12}
  \right]
  \le \poly[\lambda]\,2^{-\lambda/4}.
\]
The right-hand side is summable in \(\lambda\).
The Borel--Cantelli lemma shows that, for this fixed \(\adv\), the bound \(\lvert\Delta_{\adv}(O,\lambda)\rvert<2^{-\lambda/12}\) holds for all sufficiently large \(\lambda\) with probability one over the final oracle.
There are only countably many uniform oracle QPT adversaries, so the countable intersection of these probability-one events proves PRFSG security for the final oracle.
\end{proof}

\subsection{Proof of \cref{lemma:QPRGnotsecure}}
\label{sec:proof_QPRGnotsecure}
We will first state and prove the following lemma, which will be used in the proof of \cref{lemma:QPRGnotsecure}.
It essentially says that if a quantum algorithm outputs a bit \(b_\Phi\) with high probability
over the choice of the CHFS oracle \(\Phi\),
then this output bit is independent of the oracle with high probability over \(\Phi\).

As in the setting of \cref{lemma:QPRGnotsecure}, we fix the CHFS oracle \(O\) up to the length \(<a_i\). For convenience, we write \(\Phi\gets \sigma\) to denote \(S=S^{\Phi}\) is sampled from the conditional distributions \(\mathcal D_{S^{(<a_i)}}\).

\begin{lemma}
	\label{lemma:conj_learning_queries}
	Let \({\mathcal S}\) be the (unitarized)
	quantum-accessible CHFS oracle with \(\ell(\secpar) 
    \leq \secpar\) and let \(A^\cS\) be a
	polynomial-query oracle algorithm making at most \(T = T(\secpar)\) oracle queries.
	Let \(p=\poly\) be the maximal length of the CHFS oracles that \(A\) accesses.\footnote{This premise holds for any polynomial space algorithms.}
	Suppose that there exist a function \(\varepsilon = \varepsilon(\secpar)\) and a bit \(b_\Phi \in \bit\) such that
	\begin{align}\label{eqn: condition_removingoracle}
		\Pr_{\Phi\leftarrow\sigma}\left[\Pr[A^{S^\Phi}(1^\lambda) \to b_\Phi ] \ge \frac 23\right]=1-\varepsilon.
	\end{align}
	Then there exists \(b\in\bit\)\footnote{The bit \(b\) might depend on the parameter \(\secpar\), but we suppress this dependency for simplicity.} such that
	\(\Pr_{\Phi\leftarrow\sigma}\left[b=b_\Phi \right] = 1-\bigO{T\varepsilon}\).

	In particular, if \(\varepsilon(\secpar)=\negl\) then
	\(\Pr[b=b_\Phi]=1-\negl\), and if
	\(\varepsilon(\secpar)=1/\poly\) then
	\(\Pr[b=b_\Phi]=1-1/\poly\).
\end{lemma}
\begin{proof}
	Given the running time bound \(p\), the algorithm only accesses the oracle up to the maximum query length \(p\).
	Let
	\(X=\mathbb S(2^{n_1})\times\cdots \times \mathbb S(2^{n_k})\)
	be the states\footnote{This is implicitly parameterized by \(\secpar\).}
	to define the CHFS oracle up to the length \(p\), with the corresponding product
	Haar measure \(\sigma=\sigma_{n_1}\times\cdots\times \sigma_{n_k}\).
	Let \(S_0,S_1\subseteq X\) be defined as \ifnum\llncs=1
		\begin{align*}
			S_0\eqdef \left\{ \Phi\in X \:\colon\: \Pr(A^{S^\Phi}(1^\secpar)\rightarrow 0)\geq2/3\right\}, \\
			S_1\eqdef \left\{ \Phi\in X \:\colon\: \Pr(A^{S^\Phi}(1^\secpar)\rightarrow 1)\geq2/3\right\}.
		\end{align*}
	\else
		\begin{align*}
			S_0\eqdef \left\{ \Phi\in X \:\colon\: \Pr(A^{S^\Phi}(1^\secpar)\rightarrow 0)\geq2/3\right\},~~
			S_1\eqdef \left\{ \Phi\in X \:\colon\: \Pr(A^{S^\Phi}(1^\secpar)\rightarrow 1)\geq2/3\right\}.
		\end{align*}
	\fi

	By the hypothesis in~\cref{eqn: condition_removingoracle}, with probability at
	least \(1-\varepsilon(\lambda)\) over \(\Phi\leftarrow\sigma\), either
	\[\Pr[A^{S^\Phi}\rightarrow 1]\geq 2/3\quad\text{or}\quad\Pr[A^{S^\Phi}\rightarrow 0]\geq 2/3,\]
	thus
	\begin{equation}
		\label{eqn:gap-upper}
		\sigma(X\setminus(S_0\cup S_1)) \leq \varepsilon(\lambda).
	\end{equation}

	Choose arbitrary \(\Phi\in S_0\) and \(\Psi\in S_1\).
	The diamond distance between \(\cS^\Phi\) and \(\cS^\Psi\) can be bounded by
	\begin{align*}
		\|S^\Phi(\cdot)-S^{\Psi}(\cdot)\|_{\diamond} & \le 2\|S^\Phi - S^{\Psi}\|_{op}                                                                                   \\
		                                             & =  2\|\sum_x \ketbra{x} \otimes (S_{\ket{\phi_x}}-S_{\ket{\psi_x}})\|_{op}                                        \\
		                                             & = 2 \max_x \| (S_{\ket{\phi_x}}-S_{\ket{\psi_x}})\|_{op}                                                          \\
                                                     &= 4 \max_x \|\ketbra{\phi_x-} - \ketbra{\psi_x-}\|_{op} 
                                                     \\
                                                     &
                                                     \le 4\max_x \|{\phi_x} - {\psi_x}\|_{2}
                                                     = 4 d_{\infty}(\Phi,\Psi)
	\end{align*}
	where we use \ifnum\final=1
    \else\cref{eqn: diamond_bound_by_operator_norm,eqn:operator_norm_maximum} and
    \fi
    \begin{align*}
        \|\ketbra{\phi_x-} - \ketbra{\psi_x-}\|_{op} \le
        \|\ket{\phi_x - } - \ket{\psi_x - }\|_2 \le \|\phi_x -\psi_x \|_2.
    \end{align*}

	This implies that the diamond distance of the unitary oracles \(S^\Phi\) and
	\(S^\Psi\) is bounded by \(\|S^\Phi(\cdot)-S^{\Psi}(\cdot)\|_{\diamond} \leq 4d_{\infty}(\Phi,\Psi)\).
	On one hand, if the algorithm \(A\) makes \(T\) queries to the oracles, the
	subadditivity of the diamond norm under composition implies that
	\[\|A^{S^\Phi}-A^{S^\Psi}\|_\diamond\leq4T\cdot d_{\infty}(\Phi,\Psi).\]
	On the other hand, by the definition of the diamond distance, we can lower bound this quantity by
	\begin{align*}
		\|A^{S^\Phi}-A^{S^\Psi}\|_\diamond\geq\left|\Pr(A^{S^\Phi}\rightarrow1)-\Pr(A^{S^{\Psi}}\rightarrow1)\right|\geq\frac{1}{3},
	\end{align*}
	where the last inequality is obtained from \(\Phi\in S_0\) and
	\(\Psi\in S_1\).
    Given it holds for arbitrary \(\Phi\in S_0\) and
	\(\Psi\in S_1\),
	this implies that
	\begin{equation}\label{eqn:proof_delta}
		\Delta=d_{\infty}(S_0,S_1)  \ge \frac{1}{12 T}
	\end{equation}
	which is inverse-polynomially bounded.
	Let \(\Gamma:=\min(\sigma(S_0),\sigma(S_1))\).
	We find therefore ourselves in the hypothesis of~\cref{thm:state_conjecture},
	thus there exists an absolute constant \(\kappa>0\) such that
	\begin{equation}\label{eq:barrier}
		\sigma\bigl(X\setminus(S_0\cup S_1)\bigr) \geq \kappa\cdot \Gamma\cdot \Delta.
	\end{equation}
	We upper bound \(\Gamma\) by contradiction. Fix a sufficiently large absolute constant \(K>0\) and suppose that
	\begin{equation}\label{eq:Gamma-contrad}
		\Gamma\ >\ K\,T\,\varepsilon.
	\end{equation}
	Combining~\eqref{eq:Gamma-contrad} with~\eqref{eqn:proof_delta} and~\eqref{eq:barrier} gives
	\[
		\sigma\bigl(X\setminus(S_0\cup S_1)\bigr)
		\ >\ \kappa\cdot (K\,T\,\varepsilon)\cdot \frac{1}{12T}
		\ =\ \frac{\kappa K}{12}\,\varepsilon.
	\]
	Choosing \(K\) large enough so that \(\frac{\kappa K}{12}>1\) yields
	\(\sigma(X\setminus(S_0\cup S_1))>\varepsilon\), contradicting~\eqref{eqn:gap-upper}.
	Hence~\eqref{eq:Gamma-contrad} is false, i.e.,
	\begin{equation}
		\label{eq:Gamma-ub}
		\Gamma\ =\ \min(\sigma(S_0),\sigma(S_1))\ \le\ \bigO{T\varepsilon}.
	\end{equation}
	Define \(b(\secpar)\in\arg\max_{b\in\bin}\Pr_{\Phi\leftarrow\sigma}\!\left[b=b_\Phi\right]\).
	Since \(S_0\) and \(S_1\) are disjoint and \(\sigma(S_0\cup S_1)\ge 1-\varepsilon\) by~\eqref{eqn:gap-upper}, we have
	\begin{align*}
		\Pr_{\Phi\leftarrow\sigma}[\,b=b_\Phi\,]
		\ge\sigma(S_b)
		&=\sigma(S_0\cup S_1)-\sigma(S_{1-b}) \\
		&\ge (1-\varepsilon)-\min(\sigma(S_0),\sigma(S_1))\\
		&\ge 1-\varepsilon-\bigO{T\varepsilon},
	\end{align*}
	where the last step uses~\eqref{eq:Gamma-ub}.
	This concludes the proof.
\end{proof}

We also need the following lemma on \(t\)-designs, which will be used in the proof
of \cref{lemma:QPRGnotsecure} to simulate the CHFS oracle.

\begin{definition}[Approximate design \cite{AE07}]
	\label{def:state_t_design}
	A probability distribution \(S\) over \(\mathbb S(N)\) is an \emph{\(\eps\)-approximate \(t\)-design} state if:
	\[(1 - \eps)\Exp_{\ket{\psi} \sim \sigma_N} \left[(\ket{\psi}\bra{\psi})^{\otimes t}\right] \preceq \Exp_{\ket{\psi} \sim S} \left[(\ket{\psi}\bra{\psi})^{\otimes t}\right] \preceq (1 + \eps)\Exp_{\ket{\psi} \sim \sigma_N} \left[(\ket{\psi}\bra{\psi})^{\otimes t}\right].\]
\end{definition}

\begin{definition}[Phase-invariant design]
    An \(\epsilon\)-approximate \(t\)-design state \(S\) is called a phase-invariant design if: For all \(0\le i,j \le t\) and \(i\neq j\), with convention \(\ket{\psi}^0=\ket{0}\) and \(\bra{\psi}^0=\bra{0}\),
    \[
    \Exp_{\ket{\psi} \sim S} \left[\ket{\psi}^{\otimes i}\bra{\psi}^{\otimes j}\right]=0.
    \]
\end{definition}

The following lemma is shown by the corresponding results of \cite{Kre21} using the phase-invariant unitary designs instead of the standard designs.
\begin{lemma}[\cite{Kre21}, adapted]
	\label{lem:efficient_state_designs}
	For each \(n, t \in \mathbb N\) and \(\eps > 0\), there exists \(m \le \poly[n,t,\log \frac{1}{\eps}]\) and a \(\poly[n,t,\log \frac{1}{\eps}]\)-time classical algorithm \(\mathcal{T}\) that takes as input a random string \(x \sim \{0,1\}^m\) and outputs a description of a quantum state on \(n\) qubits such that the states sampled from \(\mathcal{T}\) form a phase-invariant \(\eps\)-approximate \(t\)-design over \(\mathbb S(2^n)\).
\end{lemma}
We additionally need the following CHFS oracle simulation lemma using phase-invariant designs. 
\ifnum\submission=1
The proof is deferred to
\ifnum\final=1
the full version~\cite{EPRINT:BHMV25}.
\else
\cref{app:proofs}.
\fi
\fi
\begin{lemma}[CHFS design simulation]
\label{lem:chfs-design-simulation}
Let \(\mathcal D\) be the distribution where the states
\(\{|\psi_x\rangle\}_{x\in\bit^*}\) are independent phase-invariant
\(\delta\)-approximate \(2q\)-design states, and let \(S^\Psi\) be the
corresponding CHFS oracle. 
Then every quantum algorithm \(A\) making at most
\(q\) coherent queries up to input length \(q\) satisfies
\[
\left|
\Pr_{\Phi\leftarrow\mathcal S}\!\left[A^{S^\Phi}\to1\right]
-
\Pr_{\Psi\leftarrow\mathcal D}\!\left[A^{S^\Psi}\to1\right]
\right|
\le
O\left(q\cdot 
C^{q^2} \cdot \delta\right)
\]
for some universal constant \(C>0\) (not depending on \(A\)),
where \(\mathcal S\) is the uniform CHFS oracle distribution.
The same conclusion holds if the design seeds are only \(2q\)-wise independent
across labels.
\end{lemma}
\begin{proof}[Proof of \cref{lem:chfs-design-simulation}]
Consider a \(D\)-dimensional unitary \(U\) and let
\[
    \ket{J(U)}=(I\otimes U)\ket{\Omega}
\]
be its normalized Choi state where \(\ket{\Omega } = \frac{1}{\sqrt D} \sum_i \ket{ii}\). 
By \cite[Lemmas 23, 24]{Kre21},
for every \(q\)-query oracle algorithm
\(A^U\), there exists some algorithm \(B\) such that
\[
    \Pr[A^U\to1] = D^{2q} \Pr[B(\ket{J(U)}^{\otimes q}) \to 1].
\]
This gives, by the monotonicity of the trace distance, we have
\begin{align*}
&\left|
\Pr_{\Phi\leftarrow\mathcal S}\!\left[A^{S^\Phi}\to1\right]
-
\Pr_{\Psi\leftarrow\mathcal D}\!\left[A^{S^\Psi}\to1\right]
\right| 
\\&
= O\left(
D^{2q} \cdot 
\left\|
\Exp_{\Phi\leftarrow\mathcal S}
\ketbra{J(S^\Phi)}^{\otimes q}
-
\Exp_{\Psi\leftarrow\mathcal D}
\ketbra{J(S^\Psi)}^{\otimes q}
\right\|_1 
\right),
\end{align*}
where \(S^{\Phi}\) and \(S^{\Psi}\) are concatenated up to the input length \(q\), which gives \(D=2^{O(q)}.\)
It remains to bound the Choi-moment difference.

Note that we have less than \(2^{q+1}=:M\) choices of \(x \in \bit^{\leq q}\).
For each \(x\),
\[
    S_{\phi_x}
    =
    I-
    \ketbra{0}
    -
    \ketbra{\phi_x}{\phi_x}
    +\ketbra{\phi_x}{0}
    +\ketbra{0}{\phi_x}.
\]
Write \(T_{x,j}\) to denote the \(j\)-th term in the above equation for \(j\in [5]\), i.e., \(T_{x,1}=I,T_{x,2}=\ketbra{0}\), etc.
Let \(\mathcal I\) be the set of all pairs of \((x,j)\), so that \(|\mathcal I|\le 5M\).
Expanding \(\ket{J(S^\Phi)}^{\otimes q}\) can be written as the linear summation of
\[
\bigotimes_{i=1}^q (I\otimes \ketbra{x_i} \otimes T_{x_i,j_i}) \ket{\Omega}^{\otimes q}
\]
of (less than) \((5M)^q\) terms.
A similar result holds when expanding \(\bra{J(S^\Phi)}^{\otimes q}\), resulting \((5M)^{2q}\) terms total.

Fix one expanded term. It involves at most \(2q\) label occurrences. For each
label \(x\), the contribution of it is a moment of the form
\[
    \mathbb E\left[
        |\phi_x\rangle^{\otimes a_x}
        \langle\phi_x|^{\otimes b_x}
    \right],
    \qquad
    a_x,b_x\le 2q .
\]
If \(a_x\ne b_x\), both the Haar moment and the design moment vanish by the phase-invariant property. If \(a_x=b_x=s\), then \(s\le 2q\), and the \(\delta\)-approximate
\(2q\)-design gives
\[
\left\|
\Exp_{\psi_x}
\ketbra{\psi_x}^{\otimes s}
-
\Exp_{\phi_x}
\ketbra{\phi_x}^{\otimes s}
\right\|_1
\le 2\delta .
\]
Since the fixed expanded term contains at most \(2q\) relevant labels, a simple hybrid shows that the distance is at most \(4q\delta\).

By linearity of expectation, the difference of the two Choi moments is the
sum, over all expanded monomials, of the corresponding differences of
monomial expectations. 
Let \(\mathsf J(A):=(I\otimes A)|\Omega\rangle\). For
\({\bf \alpha},{\bf \beta}\in\mathcal I^q\), define
\[
T_{{\bf \alpha},{\bf \beta}}(\Phi)
:=
\left(\bigotimes_{i=1}^q \mathsf J(B_{\alpha_i}(\Phi))\right)
\left(\bigotimes_{i=1}^q \mathsf J(B_{\beta_i}(\Phi))\right)^\dagger .
\]

For each fixed expanded monomial
\(T_{{\bf \alpha},{\bf \beta}}\), the previous argument shows
\[
\left\|
\Exp_{\Phi\gets \mathcal S}T_{{\bf \alpha},{\bf \beta}}(\Phi)
-
\Exp_{\Psi\gets \mathcal D}T_{{\bf \alpha},{\bf \beta}}(\Psi)
\right\|_1
\le 4q\delta .
\]
There are at most \((5M)^{2q}\) such monomials. Hence, by
the triangle inequality for the trace norm,
\begin{align*}    
&\left\|
\Exp_{\Phi\leftarrow\mathcal S}
\ketbra{J(S^\Phi)}^{\otimes q}
-
\Exp_{\Psi\leftarrow\mathcal D}
\ketbra{J(S^\Psi)}^{\otimes q}
\right\|_1 \\
&\le \sum_{{\bf \alpha},{\bf \beta}} 
\left\|
\Exp_{\Phi\gets \mathcal S}T_{{\bf \alpha},{\bf \beta}}(\Phi)
-
\Exp_{\Psi\gets \mathcal D}T_{{\bf \alpha},{\bf \beta}}(\Psi)
\right\|_1
\\&
\le (5M)^{2q} \cdot 4q\delta,
\end{align*}
thus the overall bound becomes
\[
O\left(
4q\delta \cdot (5DM)^{2q} 
\right)
=
O\left(
q \cdot C^{q^2}  \cdot \delta
\right)
\]
for some \(C>0\), because \(D=2^{O(q)}\) and \(M\le 2^{q+1}\).
This proves the main upper bound.
The final statement is clear because the proof only ever uses independence among the at most \(2q\) labels
appearing in a fixed expanded term.
\end{proof} 
\begin{proof}[Proof of~\cref{lemma:QPRGnotsecure}]
	Recall \cref{def:QPRG} for the definition of the pseudodeterministic QPRG.
	Consider a QPRG candidate \(G=(G_\lambda)_\lambda\) that takes an \(n\)-bit seed and outputs an \(m\)-bit string relative to the CHFS oracles and \(\pspace\) that runs in polynomial queries. Let \(p(\lambda)\) be the upper bound of the running time of QPT \(G_\lambda\).

	We say that an input \(x\in \{0,1\}^{n(\lambda_i)}\) is
	\(\epsilon_\star\)-\emph{good} if
	\begin{equation}\label{eqn:pseudodeterministic x}
		\Pr_{S}\Big[\exists\, y_{S,x} \text{ s.t. } \Pr\big[G^{S,\pspace}(x)\to y_{S,x}\big]\ge 1-\epsilon_\star\Big] \ge 1-\epsilon_\star.
	\end{equation}
	Furthermore, if the inner event holds for a particular \(S\) (i.e., there exists such a \(y_{S,x}\)),
	we say \((S,x)\) is \emph{good}.

	We write \(I_{S,x}\) to denote the indicator that \((S,x)\) is good.
	By the hypothesis that \(G\) satisfies \((1-\epsilon(\lambda_i))\)-pseudodeterminism for \((1-\epsilon(\lambda_i))\)-fraction of \(O\),
	\(\Pr_x[I_{S,x}=1]\ge 1-\epsilon(\lambda_i)\), and therefore
	\[
		\Exp_{S,x}[I_{S,x}] \ge (1-\epsilon(\lambda_i))^2 \ge 1-2\epsilon(\lambda_i).
	\]
	Fix any \(\epsilon_\star= (2\epsilon)^{0.5}\). By Markov's inequality (or the standard averaging argument),
	\[
		\Pr_x\Big[\Exp_S[I_{S,x}] \leq 1-\epsilon_\star\Big]
		= \Pr_x\Big[1-\Exp_S[I_{S,x}] \geq \epsilon_\star\Big]
		\le
		\frac{1-\Exp_{S,x}[I_{S,x}]}{\epsilon_\star}
		\le \frac{2\epsilon}{\epsilon_\star} = \epsilon_\star.
	\]
	Hence, for at least a \(1-\epsilon_\star\) fraction of \(x\) we have
	\(\Exp_S[I_{S,x}] \ge 1-\epsilon_\star\), i.e., \(x\) is \(\epsilon_\star\)-good.
	We call such \(x\) \emph{semi-deterministic}.

	We will construct \(F^{\pspace}\) that agrees with \(G^{O,\pspace}\) with high probability on all semi-deterministic \(x\). We divide \(O\) into the input length \(< a_i\) (which is already fixed by \(S^{(<a_i)}\)) and \(\ge a_i\).

	We
	first show that \(S^{(<a_i)}\) can be
	simulated up to negligible error
	in \(\poly[\lambda_i]\) time using tomography.
	Recall that the unitary CHFS oracle is indexed by an oracle-input length \(t<a_i\),
	and that on input \(x \in \bin^t\), it applies a reflection
	\(S_{t,x}=S_{\ket{\psi_x}\ket{1}}\) on \(\ell(t)+1\) qubits.
	That is, \(S_{t,x}\) acts on dimension \(d(t)\;=\;2^{\ell(t)+1}\le 2^{t+1}\le 2^{a_i} = \poly[\lambda_i]\) because \(a_i=O(\log \lambda_i).\)\footnote{We use \(\ell(\lambda)\le \lambda\) here. It can be relaxed to \(\ell=O(\lambda)\).\label{footnote:linear ell}}

	For each fixed pair \((t,x)\) with \(t<a_i\), applying \ifnum\final=1
    the process tomography from \cite{FOCS:HKOT23}
    \else
    \cref{thm: process_tomography}\fi to the black-box unitary \(S_{t,x}\) with
	\(\varepsilon_{\sf tom} \coloneqq \lambda_i^{-c'}\) (for a constant \(c'>0\) to be determined later) and
	\(\delta_{\sf tom} \coloneqq 2^{-2\lambda_i}\) yields an algorithm that outputs a classical
	description of a unitary \(\widetilde S_{t,x}\) such that, with probability at
	least \(1-\delta_{\sf tom}\),
	\[
		\abs{S_{t,x}-\widetilde S_{t,x}}_\diamond \le \varepsilon,
	\]
	using
	\(\bigO{\frac{d(t)^2}{\varepsilon_{\sf tom}}\log\frac{1}{\delta_{\sf tom}}} = \poly[\lambda_i]\) queries and time \(\poly[d(t),\frac 1{\varepsilon_{\sf tom}},\log(\frac1{\delta_{\sf tom}})]=\poly[\lambda_i]\).

	The number of such pairs \((t,x)\) is
	\(N_i = \sum_{t=1}^{a_i-1}2^t \le 2^{a_i} = \poly[\lambda_i]\).
	By a union bound over all \(N_i\) invocations, with probability at least
	\(1-N_i\delta_{\sf tom} \ge 1-\poly[\lambda_i]\cdot 2^{-2\lambda_i} = 1-\negl[\lambda_i]\),
	all tomography subroutines succeed simultaneously, producing descriptions
	\(\{\widetilde S_{t,x}\}_{t < a_i, x \in \bin^t}\) with diamond-error at most \(\varepsilon_{\sf tom}\) each.
	This shows that the trace distance between the output distribution can be bounded by
	\begin{align}\label{eqn: tomography}
		\|G^S(\cdot) - G^{\widetilde S}(\cdot)\|_1 \le p(\lambda_i) \cdot \varepsilon_{\sf tom} = p(\lambda_i) \cdot \lambda_i^{-c'}
	\end{align}
	for any input,
	where we omit the other oracle queries that are identical for the left and right cases. We choose \(c'\) so that \( \lambda_i^{-c'} \le \frac{1}{100p(\lambda_i)^2 \lambda_i^2}\).

	Now consider the oracle query length \(\ge a_i\).
	Let \(\delta=\frac{1}{100D \lambda_i^2 p(\lambda_i)^2 C^{p(\lambda_i)^2}}\) where \(C\) and \(D\) are constants from \cref{lem:chfs-design-simulation}.
	Let \(\mathcal T(\cdot)\) be the algorithm from~\Cref{lem:efficient_state_designs} that, on input a
	uniform seed \(r\in\bit^{s(q)}\), outputs a (description of a) phase-invariant \(q\)-qubit state whose output state forms
	a \(\delta\)-approximate \(2p(\lambda_i)\)-design.\footnote{The seed length \(s(q)\) depends also on
		\(p(\lambda_i)\) and \(\log(1/\delta)\); we suppress these dependencies for simplicity.}

	For each \(t\in[p(\lambda_i)]\), we sample a function \(h_t:\bit^t \to \bit^{s(\ell(t))}\) from a
	\(2p(\lambda_i)\)-wise independent family (independently for different \(t\)).
	By~\cite{C:Zhandry12},
	we can assume that \(h_t\) is sampled from a truly random function as \(F\) and \(G\) make at most
	\(p(\lambda_i)\) quantum queries.

	We now describe how \(F\) simulates \(G\)'s access to the unitary CHFS oracle \(S\), given the above data.
	On an oracle query with input \(x\in\bit^t\) (possibly in superposition), \(F\) proceeds as follows.
	\begin{enumerate}
		\item It uses \(\widetilde S\) instead of \(S\) for the input length \(t< a_i\). For the input length \(t\ge a_i\), it proceeds to the next step.
		\item Coherently compute \(r_x := h_t(x)\in\bit^{s(\ell(t))}\).
		\item
		      Coherently compute \(\ket{\psi_x}:=T(r_x)\).
		      Note that the parameters are such that on input \(x\in\{0,1\}^t\), \(h_t(x)\) is of size \(s(\ell(t))\), which is the size of the randomness needed to generate a \(\ell(t)\)-qubit \(\delta\)-error \(2p(\lambda_i)\)-design using \(\mathcal T(\cdot)\).
		\item Apply the reflection \(S_{\ket{\psi_x}\ket{1}}\) on the \(\ell(t)+1\)-qubit
		      register \({\bf Y}\) (cf.~\cref{def:unitary_chfs}).
	\end{enumerate}

	Equivalently, instead of querying the unitary CHFS oracle, \(F\) uses
	\[
		O':= \sum_{x\in \bit^{t}}\ketbra{x}_{\bf X} \otimes S_{\ket{\psi_x}\ket{1}},
	\]
	on \(\rho_{\bf XYZ}\) for \(|x|\ge a_i\), which is indistinguishable from the CHFS oracle up to some errors that we will now analyze.

	We write
	\(f_z(x)\) for the \(z\)-th bit of \(F^{\pspace}(x)\),
	\(g_z(x)\) for the \(z\)-th bit of \(G^{S,\pspace}(x)\).

	Fix a semi-deterministic \(x\).
	Applying~\cref{lemma:conj_learning_queries} to \(g_z(x)\) (which makes at most \(T\le p(\lambda)\) oracle queries)
	given the fact that \(G(x)\) is pseudodeterministic with probability at least \((1-\epsilon_\star)\) over the oracles,
	yields that \(g_z(x)\) must be fixed with probability at least \((1-O(p(\lambda_i) \epsilon_\star))\) over the CHFS oracles.

	Combining this with the \(\delta\)-approximate \(2p(\lambda_i)\)-design simulation error gives
	\begin{align*}
		\Pr\left[g_z(x)=f_z(x) \right]
		 & \ge\; 1-\bigO{p(\lambda_i)\epsilon_\star} - 
         O(p(\lambda_i)\cdot C^{p(\lambda_i)^2} \cdot \delta)
		-\frac{1}{100\lambda_i^2 p(\lambda_i)}
		\\&\ge\; 1-\bigO{p(\lambda_i)\epsilon_\star} - \frac{1}{100\lambda_i^2 p(\lambda_i)}-\frac{1}{100 \lambda_i^2p(\lambda_i)},
	\end{align*}
	where we use \cref{lem:chfs-design-simulation} and
    the term \(\frac{1}{100 \lambda_i^2p(\lambda_i)}\) is from the tomography error in \cref{eqn: tomography}.
	By a union bound over \(z\in[m]\) and using \(m\le p(\lambda)\), we obtain
	\begin{equation}\label{eq:closeFG}
		\Pr\left[G^{S,\pspace}(x)=F^{\pspace}(x)\right]
		\ \ge\ 1-O({p(\lambda)^2\epsilon_\star})-\frac{1}{50\lambda_i^2},
	\end{equation}
	for all \(\epsilon_\star\)-semi-deterministic \(x\).

	Recall the condition \(\epsilon_\star(\lambda) = (2\epsilon(\lambda))^{0.5} \le \frac{\sqrt{2c}}{p(\lambda)^4}\) for some small \(c>0\).
	We choose \(c\) so that the term \(O({p(\lambda)^2\epsilon_\star})\) is bounded above by \(\frac{1}{50\lambda_i^2}\).
	This shows that the right-hand side of \cref{eq:closeFG} is at least \(1-\frac{1}{4\lambda_i^2}\) for all sufficiently large \(\lambda\).

	Now, fix the initialization randomness \(r\) of \(F\) for the tomography and designs.
    Below, we assume that \(F\) uses the randomness \(r\), and consider the language
	\[
		L_{F,r} := \left\{y\in\bit^{m(\lambda)} : \exists x\in\bit^{n(\lambda)} \text{ s.t. } \Pr\left[F^{\pspace}(x)=y\right]>1-\frac{1}{4\lambda^2}\right\}.
	\]
	For fixed \((x,y)\), a \(\pspace\) machine can approximate \(\Pr\left[F^{\pspace}(x)=y\right]\) to additive error \(o(\lambda^{-2})\)
	(since \(\bqp^{\pspace}\subseteq\pspace\)), and hence decide whether it exceeds \(1-1/4\lambda^2\).
	It follows that \(L\in\pspace\): on input \(y\), enumerate all \(x\in\bit^n\) using polynomial space and accept iff any \(x\) satisfies \(\Pr\left[F^{\pspace}(x)=y\right]>1-1/4\lambda^2\).
	Therefore, there exists a \(\pspace\)-oracle algorithm \(\cB_0^{\pspace}\) that can decide membership in \(L\).

	We define our adversary \(\cB\) formally: it first computes the tomography for the input length \(<a_i\), \(2p(\lambda_i)\)-wise independent functions, and state designs described above as initialization. Then, given this, it runs \(\cB_0^{\pspace}\).

	We consider the advantage of \(\cB\) against the pseudorandomness of \(G\).
	By~\cref{eq:closeFG}, for every semi-deterministic \(x\) it holds that
	\begin{align*}
		\Pr_S\left[\cB^{\pspace}(G^{S,\pspace}(x))=1\right]\ge 1- \frac{1}{4\lambda_i^2}.
	\end{align*}
    
	As a \((1-O(1/\lambda_i^2))\)-fraction of \(x\), with a sufficiently small constant, satisfies the above, we have
	\[
		\Pr_{S,x}\left[\cB^{\pspace}(G^{S,\pspace}(x))=1\right]\ge 1- \frac{1}{3\lambda_i^2}.
	\]
    
	The standard averaging argument says that
	\[
		\Pr_{S}\left[\left[\cB^{\pspace}(G^{S,\pspace}(x))=1\right]\ge \frac 23\right] \ge 1- \frac{1}{\lambda_i^2}.
	\]

	On the other hand, \(|L_F|\le 2^n\) (each \(x\) can contribute at most one such \(y\)), so for uniform \(y\gets\bit^m\) we have
	\(\Pr\left[\cB^{\pspace}(y)=1\right]\le 2^{n-m}\le 1/2\)
	(using that \(m>n\) for a QPRG).
	This gives a constant distinguishing advantage of \(\cB\) against \(G\) for the parameter \(\lambda_i\).\qedhere
\end{proof}

\ifnum\submission=0
\begin{proof}[Proof of \cref{lem:chfs-design-simulation}]
Consider a \(D\)-dimensional unitary \(U\) and let
\[
    \ket{J(U)}=(I\otimes U)\ket{\Omega}
\]
be its normalized Choi state where \(\ket{\Omega } = \frac{1}{\sqrt D} \sum_i \ket{ii}\). 
By \cite[Lemmas 23, 24]{Kre21},
for every \(q\)-query oracle algorithm
\(A^U\), there exists some algorithm \(B\) such that
\[
    \Pr[A^U\to1] = D^{2q} \Pr[B(\ket{J(U)}^{\otimes q}) \to 1].
\]
This gives, by the monotonicity of the trace distance, we have
\begin{align*}
&\left|
\Pr_{\Phi\leftarrow\mathcal S}\!\left[A^{S^\Phi}\to1\right]
-
\Pr_{\Psi\leftarrow\mathcal D}\!\left[A^{S^\Psi}\to1\right]
\right| 
\\&
= O\left(
D^{2q} \cdot 
\left\|
\Exp_{\Phi\leftarrow\mathcal S}
\ketbra{J(S^\Phi)}^{\otimes q}
-
\Exp_{\Psi\leftarrow\mathcal D}
\ketbra{J(S^\Psi)}^{\otimes q}
\right\|_1 
\right),
\end{align*}
where \(S^{\Phi}\) and \(S^{\Psi}\) are concatenated up to the input length \(q\), which gives \(D=2^{O(q)}.\)
It remains to bound the Choi-moment difference.

Note that we have less than \(2^{q+1}=:M\) choices of \(x \in \bit^{\leq q}\).
For each \(x\),
\[
    S_{\phi_x}
    =
    I-
    \ketbra{0}
    -
    \ketbra{\phi_x}{\phi_x}
    +\ketbra{\phi_x}{0}
    +\ketbra{0}{\phi_x}.
\]
Write \(T_{x,j}\) to denote the \(j\)-th term in the above equation for \(j\in [5]\), i.e., \(T_{x,1}=I,T_{x,2}=\ketbra{0}\), etc.
Let \(\mathcal I\) be the set of all pairs of \((x,j)\), so that \(|\mathcal I|\le 5M\).
Expanding \(\ket{J(S^\Phi)}^{\otimes q}\) can be written as the linear summation of
\[
\bigotimes_{i=1}^q (I\otimes \ketbra{x_i} \otimes T_{x_i,j_i}) \ket{\Omega}^{\otimes q}
\]
of (less than) \((5M)^q\) terms.
A similar result holds when expanding \(\bra{J(S^\Phi)}^{\otimes q}\), resulting \((5M)^{2q}\) terms total.

Fix one expanded term. It involves at most \(2q\) label occurrences. For each
label \(x\), the contribution of it is a moment of the form
\[
    \mathbb E\left[
        |\phi_x\rangle^{\otimes a_x}
        \langle\phi_x|^{\otimes b_x}
    \right],
    \qquad
    a_x,b_x\le 2q .
\]
If \(a_x\ne b_x\), both the Haar moment and the design moment vanish by the phase-invariant property. If \(a_x=b_x=s\), then \(s\le 2q\), and the \(\delta\)-approximate
\(2q\)-design gives
\[
\left\|
\Exp_{\psi_x}
\ketbra{\psi_x}^{\otimes s}
-
\Exp_{\phi_x}
\ketbra{\phi_x}^{\otimes s}
\right\|_1
\le 2\delta .
\]
Since the fixed expanded term contains at most \(2q\) relevant labels, a simple hybrid shows that the distance is at most \(4q\delta\).

By linearity of expectation, the difference of the two Choi moments is the
sum, over all expanded monomials, of the corresponding differences of
monomial expectations. 
Let \(\mathsf J(A):=(I\otimes A)|\Omega\rangle\). For
\({\bf \alpha},{\bf \beta}\in\mathcal I^q\), define
\[
T_{{\bf \alpha},{\bf \beta}}(\Phi)
:=
\left(\bigotimes_{i=1}^q \mathsf J(B_{\alpha_i}(\Phi))\right)
\left(\bigotimes_{i=1}^q \mathsf J(B_{\beta_i}(\Phi))\right)^\dagger .
\]

For each fixed expanded monomial
\(T_{{\bf \alpha},{\bf \beta}}\), the previous argument shows
\[
\left\|
\Exp_{\Phi\gets \mathcal S}T_{{\bf \alpha},{\bf \beta}}(\Phi)
-
\Exp_{\Psi\gets \mathcal D}T_{{\bf \alpha},{\bf \beta}}(\Psi)
\right\|_1
\le 4q\delta .
\]
There are at most \((5M)^{2q}\) such monomials. Hence, by
the triangle inequality for the trace norm,
\begin{align*}    
&\left\|
\Exp_{\Phi\leftarrow\mathcal S}
\ketbra{J(S^\Phi)}^{\otimes q}
-
\Exp_{\Psi\leftarrow\mathcal D}
\ketbra{J(S^\Psi)}^{\otimes q}
\right\|_1 \\
&\le \sum_{{\bf \alpha},{\bf \beta}} 
\left\|
\Exp_{\Phi\gets \mathcal S}T_{{\bf \alpha},{\bf \beta}}(\Phi)
-
\Exp_{\Psi\gets \mathcal D}T_{{\bf \alpha},{\bf \beta}}(\Psi)
\right\|_1
\\&
\le (5M)^{2q} \cdot 4q\delta,
\end{align*}
thus the overall bound becomes
\[
O\left(
4q\delta \cdot (5DM)^{2q} 
\right)
=
O\left(
q \cdot C^{q^2}  \cdot \delta
\right)
\]
for some \(C>0\), because \(D=2^{O(q)}\) and \(M\le 2^{q+1}\).
This proves the main upper bound.
The final statement is clear because the proof only ever uses independence among the at most \(2q\) labels
appearing in a fixed expanded term.
\end{proof} \fi  
\ifnum\submission=0

\section{On Separating PRUs from PRFSGs}\label{sec:separation_pru}

In this section, we consider the length-\(\ell\) quantum-accessible unitarized CHFS
oracle
\(\mathcal S=\mathcal S_{\ell}\)
for \(\ell(|x|)=|x|\) and the \(\qpspace\) oracle.
We will prove the following theorem, which is the main result of this section.

\begin{theorem}
	\label{thm:pru-vs-prfs}
	There exist adaptively-secure quantum-accessible PRFSGs but there do not
	exist non-adaptive PRUs {whose implementation uses neither
ancillary registers nor intermediate measurements}, relative to
	\((\mathcal S,\qpspace)\).
\end{theorem}
Formally, the PRUs whose implementation uses neither
ancillary registers nor intermediate measurements can be specified by an alternating application of unitaries and oracle queries.
The existence of the adaptively-secure quantum-accessible PRFSGs relative to the
oracles is proven by~\Cref{thm:PRFS_in_CHFS}.
What remains is to prove that PRUs without ancilla do not exist in this model.

\begin{lemma}
	\label{lemma:no-pru}
	Non-adaptive PRUs
	{whose implementation uses neither
ancillary registers nor intermediate measurements}
	do not exist with probability 1 relative to the oracle \((\mathcal S,\qpspace)\).
\end{lemma}
\begin{proof}
	We prove the lemma by contradiction.
	Assume that
	\(\{G^{\mathcal S,\qpspace}_{\secpar}(\cdot)\}_{\secpar}\) is a secure \(n(\secpar)\)-PRU construction relative to \((\mathcal S,\qpspace)\) for \(n(\secpar)=\omega(\log \secpar).\)
	For simplicity,
	we drop the \(\qpspace\) oracle and \(\secpar\) in notations and write \(G_{k}^{\mathcal S}\) to denote \(G^{\mathcal S,\qpspace}_{|k|}(k)\).
	The adversary is given oracle access to the oracle \((V,\mathcal S,\qpspace)\) where \(V\) is either \(G_{k^{*}}^{\mathcal S}\) for some \(k^{*}\) or a Haar random unitary of the same size, and tries to determine which is the case with non-negligible probability.

	We write \(\mathcal S = (S_{d})_{d\in \mathbb N}\) where \(S_{d}=\sum_{x\in \bit^{d}} \ketbra{x} \otimes S_{\ket{\phi_{x}}}\) to denote the unitary CHFS oracle, where \(S_{\ket{\phi_{x}}}\) denotes the swap oracle defined in \cref{def:swap_oracle} for some \(d\)-qubit Haar random quantum state \(\ket{\phi_{x}}\) and \(S_{d}\) acts on a \((2d+1)\)-qubit space.\footnote{In the proof below, we consider the oracle queries to \(S_{d}\). The same proof can be extended to the oracle queries to \( S_{\ket{\phi_{x}}}\) for each \(x\), or more general cases. e.g., queries to \(\ketbra{0}\otimes S_{\ket{\phi_{x_{0}}}}+\ketbra{1}\otimes S_{\ket{\phi_{x_{1}}}}\) for any \(x_{1},x_{2}\) of the same length. We focus on the queries to \(S_{d}\) because it is the most complicated.}

	Let \(m=\poly\) be the maximum number of oracle queries to \(\mathcal S\) that \(G^{\mathcal S}\) makes.
	We show that distinguishing \(G_{k}^{\mathcal S}\) from a Haar random unitary can be done efficiently based on swap tests.
	More concretely, we prove that
	the following algorithm \(\adv^{V,\mathcal S,\qpspace}\) can guess with non-negligible probability whether \(V\) is \(G_{k}^{\mathcal S}\) for some random \(k\), (in which case it outputs \(1\)), or a truly Haar random unitary (in which case it outputs \(0\)).
	For simplicity, we omit the oracle notation and write \(\adv\) for \(\adv^{V,\mathcal S,\qpspace}\).

	\begin{algorithm}
		\label{alg:prusep}
		\(\adv\) chooses an \(n\)-qubit Haar random state \(\ket{\rho}\)
		and does the following on input oracle access to \(V\).\footnote{To be efficient, \(\adv\) can use \(s\)-design for large \(s\) instead of Haar random state.}
		\begin{enumerate}
			\item
			      \(\adv\)
			      executes the purity test \(16\secpar^{2}\) times on \(V(\ketbra \rho)\), which makes \(32\lambda^2\) non-adaptive queries to \(V\).
			      If the tests fails at least \(4\secpar\) times, \(\adv\) returns 1 and abort.
			      Otherwise, it proceeds to the next step.
			\item
			      Let \(\tau=2\log(16m) < n\).
			      For all \(i\le \tau\), \(\adv\) runs \(\Tom\) as defined in \cref{thm: process_tomography} on the oracles \(S_{i}\) with parameters \(\epsilon=\frac{2}{2^{\tau/2}},\delta=\frac{1}{2^{2\secpar}}\) and obtains \(S_{i}'\) that approximates \(S_{i}\).
			      Then it defines a new simulated oracle
			      \[
				      \tilde{S}_{d}\coloneqq
				      \begin{cases}
					      I        & \text{ if }d>\tau, \\
					      {S}'_{d} & \text{otherwise}.
				      \end{cases}
			      \]
			      We write \(\tilde{\mathcal S}=(\tilde{S}_{d})_{d\in \mathbb N}.\)
			      Let \(r= 1200\secpar\).
			      For each \(k\), we define the following sub-protocol \(P_{k}\) that takes as input a state \(\Psi\) over the register \({\bf A}_{1}{\bf A}_{1}'\dots{\bf A}_{r}{\bf A}_{r}'\):
			      \begin{description}
				      \item[\(P_{k}\):]
				            Apply \((G_{k}^{\tilde{\mathcal S}}\otimes I)^{\otimes r}(\Psi)\), where each \(G_{k}^{\tilde{\mathcal S}}\) acts on \({\bf A}_{i}\).
				            Then, apply the swap test on \({\bf A}_{i}{\bf A}_{i}'\) for each \(i\in[r]\).
				            Return 1 if at least \(2r/3=800\secpar\) tests passes, and return 0 otherwise.
			      \end{description}
			      Note that computing \(G_{k}^{\tilde{\mathcal S}}\) does not require any queries to the CHFS reflection oracle \(\mathcal S\), neither does \(P_{k}\).
			\item
			      \(\adv\) prepares the following state
			      \[
				      \Psi\coloneqq\bigotimes_{i\in [r]}\left(\ketbra{\rho}_{{\bf A}_{i}}\otimes V(\ketbra{\rho})_{{\bf A}_{i}'}\right).
			      \]
			      \(\adv\) applies \cref{lemma:quantum_or} on input state \(\Psi\) and the family of POVMs induced by \(\{P_{k}\}_{k\in \bit^{\secpar}}\), and outputs the same output as the OR tester.
		\end{enumerate}
	\end{algorithm}

	The following claims summarize the main analysis of the algorithm, which will be proven at the end of this section.
	\begin{claim}\label{claim: Sep1AlgBQP}
		\Cref{alg:prusep} is a \(\bqp^{V,\mathcal S,\qpspace}\) algorithm, which makes non-adaptive queries to \(V\).
	\end{claim}
	\begin{claim}\label{claim: Sep1Gk}
		If \(V=G_{k}^{\mathcal S}\) for some \(k\) and \(\Tr(G_{k}^{\mathcal S}(\rho)^{2})\ge 1-1/\secpar\), then
		\(\Pr[P_{k}(\Psi) \to 1] \ge 1-2^{-\secpar}\) holds with probability at least \(1-\frac{m+\tau}{2^{2\secpar}}\) over the randomness of the algorithm for sufficiently large \(\secpar\).
	\end{claim}
	\begin{claim}\label{claim: Sep1Haar}
		If \(V\gets \mu_{n}\), then \(\Pr[P_{k}(\Psi) \to 1]  \le 2^{-2\secpar}\) holds for all \(k\) with probability at least \(1-2^{-\secpar}\) over the randomness of the algorithm for sufficiently large \(\secpar\).
	\end{claim}

	The efficiency of the algorithm relative to \(\mathcal S,\qpspace\) is provided by \cref{claim: Sep1AlgBQP}.
	Also note that the algorithm breaks the non-adaptive security of PRUs, as the queries to \(V\) only occur in the first step and to prepare \(\Psi\) which are all non-adaptive queries.

	The correctness of the algorithm can be shown by the case analysis. If \(V=G_{k}^{\mathcal S}\) for some \(k\) and if \(\Tr(G_{k}^{\mathcal S}(\rho)^{2})\le 1-1/\secpar\), the first step of \(\adv\) outputs 1 with probability at least \(1-2^{-\secpar}\) as shown in \cref{lem: purity_test}.

	The other case, i.e., \(V=G_{k}^{\mathcal S}\) and \(\Tr(G_{k}^{\mathcal S}(\rho)^{2})\ge 1-1/\secpar\) or \(V\) is a true Haar random unitary is dealt with by the quantum OR lemma.
	In this case, by \cref{claim: Sep1Gk} and \cref{claim: Sep1Haar}, the POVMs \(\{P_{k}\}_{k\in \bit^{\secpar}}\) and \(\Psi\) satisfy the conditions of the quantum OR lemma (\cref{lemma:quantum_or}) unless with probability \( \frac{m+\tau}{2^{2\secpar}} +2^{-\secpar} \le 2/2^{\secpar}\) for large enough \(\secpar\),
	\(\epsilon=1/2^{\secpar}\) and \(\delta=1/2^{2\secpar}\).
	Therefore, \(\adv\) outputs 1 with probability at least \(1/8\) if \(V=G_{k}^{\mathcal S}\) for some \(k\), but it outputs 1 with probability at most \(4/2^{\secpar}\) if \(V\gets \mu_{n}\), that is, \(\adv\) breaks the PRU security of \(\{G_{k}^{\mathcal S}\}\).
	This concludes the proof.\end{proof}

We now prove the claims.
\begin{proof}[Proof of \cref{claim: Sep1AlgBQP}]
	The first step takes polynomial time and \(32\secpar^{2}\) non-adaptive queries to the oracle \(V\) for the \(16\secpar^2\) purity tests.
	The second step has time and query complexity (to \(\mathcal S\)) equal to \(\tau \times {\sf poly}\left(d,\frac{1}{\epsilon},\log \frac{1}{\delta}\right) =
	{\sf poly}\left(2^{\tau}, 2^{\tau}, \secpar\right) =\poly\).
	Note that it is clear that \(P_{k}\) can be executed by a quantum polynomial space machine. In the final step, the quantum OR tester can be executed by a \(\qpspace\)-aided BQP machine as noted in ``Moreover'' part of \cref{lemma:quantum_or} with inputs the descriptions of \(S_{i}'\) for \(i\le \tau\) (prepared by the first step) as \(P_{k}\) can be implemented by a quantum polynomial-space machine.
\end{proof}

\begin{proof}[Proof of \cref{claim: Sep1Gk}]
	We will show that \(G_{k}^{\mathcal S}(\ketbra\rho)\)
	and \(G_{k}^{\tilde{\mathcal S}}(\ketbra\rho)\) are
	close with high probability, for a Haar random input state \(\ket{\rho}\) of size
	\(n\)-qubit.
	We will write \(\rho = \ketbra{\rho}\) as a short-hand.

	We begin with the following two closeness properties for \(S_{d}\) and \(\tilde S_{d}\) from the later steps of the algorithm.
	First, for small dimensions \(d \le \tau < n\), \cref{thm: process_tomography} ensures that
	\begin{align}\label{eqn: good2_PRUPRFS}
		\|\tilde{S}_{d}\otimes I(\rho) - {S}_{d}\otimes I(\rho)\|_{tr}=
		\|S_{d}'\otimes I(\rho) - {S}_{d}\otimes I(\rho)\|_{tr} \le\frac{\epsilon}2= \frac{1}{2^{\tau/2}},
	\end{align}
	holds for any quantum state \(\rho\) with probability \(1-\delta = 1-\frac{1}{2^{2\secpar}}\).
	Thus all tomography outputs are \(1/2^{\tau/2}\)-close to the target unitaries with overwhelming probability \(1-p_{1}\), for \(p_{1}=\tau/2^{2\secpar}\).
	In the following, we assume it is the case.

    Now we discuss the large dimensions \(d>\tau\), where we want to show that \(S_d\) acts almost as the identity with high probability for a pure Haar quantum state \(\ket\rho\).
    Observe that for the reflection \(R=I-2\Pi\) such that \(\|\Pi\ket{\rho}\|^2=p\),
    \begin{align*}
        & \|\rho - R(\rho)\|^2_{tr} = \|\ketbra\rho - R \ketbra \rho R^{\dagger} \|^2_{tr}
        = {1-|\bra{\rho} R \ket{\rho}|^2} = 4{p(1-p)} \le 4p.
    \end{align*}
    Given this, for \(S_d = I - 2\sum_{x\in \bit^d} \ketbra{x}\otimes \ketbra{\phi_x -} \otimes I_{n-2d-1}  = I- 2\Pi_d\) for the rank \(2^{n-d-1}\) projector \(\Pi_d\),
    \begin{align*}
        \Exp_{\rho}
        \|\tilde{S}_{d} \otimes I(\rho)-S_{d}\otimes I(\rho)\|^2 _{tr}  
        =\Exp_{\rho}
        \|I(\rho)-S_{d}\otimes I(\rho)\|^2 _{tr}
        \le 4\Exp_{\rho} \| \Pi_d \ket{\rho}\|^2 \le \frac{1}{2^{d-1}} \le \frac{1}{2^{\tau}}
    \end{align*}
    where we use \cref{lem: Haarproject}.
    Note that \(\| \Pi_d \ket{\rho}\|^2 \) is the success probability of an 1-query algorithm to a Haar random unitary \(U\) by letting \(\rho = U\ket0\). 
    Applying \cref{cor: stateHaarconcentration} with \(t=1/2^{\tau+1}\), we have
    \[
    \Pr_{\rho} \left[
        \| \Pi_d \ket{\rho}\|^2 \ge \frac{1}{2^{\tau}}
    \right]\le \exp \left(
    -\frac{2^n-2}{24\cdot 2^{2\tau+2}}
    \right) \le \frac{1}{2^{2\lambda}}
    \]
    for sufficiently large \(n\). In other words, except for the probability \(1/2^{2\lambda}\), it holds that
    \begin{align}\label{eqn: close_large_d}
        \|\tilde{S}_{d} \otimes I(\rho)-S_{d}\otimes I(\rho)\| _{tr}  \le \frac{2}{2^{\tau/2}} =\epsilon
    \end{align}

	To bound the trace distance between \(G_{k}^{\mathcal S}(\ketbra{\rho})\) and \(G_{k}^{\tilde{\mathcal S}}(\ketbra{\rho})\), we use the hybrid argument using the above two observations.
	Let \({\Phi_{j}}\) for \(0\le j \le m\) be equal to
	the outcome of \(G_{k}\) on input \(\ketbra{\rho}\) with the first \(j\) oracle queries are answered using \(\tilde {\mathcal S}\) and the other \(m-j\) queries are answered using \(\mathcal S\).
	We have that \({\Phi_{0}}\) is the state \(G_{k}^{\mathcal S}(\rho)\) and \({\Phi_{m}}\) is the state \(G_{k}^{\tilde{\mathcal S}}({\rho})\).

	Let \(\ket{\phi_{j}}\) be the intermediate state right after \(j\)-th oracle query when computing \(G_{k}^{\tilde{\mathcal S}}(\ketbra{\rho})\). 
    Note that for large \(d\), we apply the identity instead of \(S_d\), thus the intermediate state is independent of \(S_d\).
    Therefore we can use \cref{eqn: close_large_d} for large \(d\) with high probability over \(\phi_{j}\).
    For small \(d\), \cref{eqn: good2_PRUPRFS} ensures the upper bound.
	Thus, by the subadditivity and monotonicity of the trace distance, we have
	\begin{align*}
		\left\|G_{k}^{\mathcal S} (\ketbra{\rho}) - G_{k}^{\tilde{\mathcal S}} (\ketbra{\rho}) \right\|_{tr} &
		\le \sum_{j=0}^{m-1}\|S_{d_{j}^{(k)}}\otimes I(\phi_{j})-\tilde{S}_{d_{j}^{(k)}}\otimes I(\phi_{j})\| _{tr}
		\\&
		\le \sum_{j=0}^{m-1}  \max(\frac{\epsilon}{2},\epsilon)
        = \frac{2m}{2^{\tau/2}} = \frac{1}{8},
	\end{align*}
	with probability \(1-p_{2}\) for \(p_{2}=\frac{m}{2^{2\secpar}}\); we again focus on this case.

	We finally analyze the success probability of a single swap test between \(G_{k}^{\mathcal S}(\ketbra{\rho})\) and \(G_{k}^{\tilde{\mathcal S}}(\ketbra{\rho})\) succeeds in subroutine \(P_{k}\).
	Since
	\ifnum\llncs=1
		\begin{align*}
			 & \|G_{k}^{{\mathcal S}}(\ketbra{\rho})\otimes G_{k}^{\tilde{\mathcal S}}(\ketbra{\rho})-G_{k}^{{\mathcal S}}(\ketbra{\rho})\otimes G_{k}^{{\mathcal S}}(\ketbra{\rho})\|_{tr} \\
			 & \quad\quad=\|G_{k}^{\tilde{\mathcal S}}(\ketbra{\rho}) - G_{k}^{{\mathcal S}}(\ketbra{\rho})\|_{tr}
			\le 1/8,
		\end{align*}
	\else
		\begin{align*}
			\|G_{k}^{{\mathcal S}}(\ketbra{\rho})\otimes G_{k}^{\tilde{\mathcal S}}(\ketbra{\rho})-G_{k}^{{\mathcal S}}(\ketbra{\rho})\otimes G_{k}^{{\mathcal S}}(\ketbra{\rho})\|_{tr} =\|G_{k}^{\tilde{\mathcal S}}(\ketbra{\rho}) - G_{k}^{{\mathcal S}}(\ketbra{\rho})\|_{tr}
			\le 1/8,
		\end{align*}
	\fi
	we have
	\begin{align*}
		\left|
		\frac{1+\Tr(G_{k}^{\tilde{\mathcal S}}(\ketbra{\rho})G_{k}^{{\mathcal S}}(\ketbra{\rho}))}{2}
		- \frac{1+\Tr(G_{k}^{{\mathcal S}}(\ketbra{\rho})^{2})}{2}
		\right|
		\le
		\frac{1}{8},
	\end{align*}
	and using the fact that \(\Tr(G_{k}^{{\mathcal S}}(\ketbra{\rho})^{2})\ge 1-1/\secpar\), we have
	\[
		\frac{1+\Tr(G_{k}^{\tilde{\mathcal S}}(\ketbra{\rho})G_{k}^{{\mathcal S}}(\ketbra{\rho}))}{2} \ge \frac{7}{8} - \frac{1}{2\secpar} \ge \frac34.
	\]
	Therefore, by Chernoff's inequality, the probability that at least \(\frac{2r}{3}\) tests succeed among \(r\) swap tests is bounded by
	\[
		1-\exp\left(-\frac{3r}{2\cdot 4\cdot 12^{2}}\right)=1-\exp\left(-\frac{r}{384}\right) \ge 1-2^{-\secpar}.
	\]
	Overall, if \(V=G_{k}^{{\mathcal S}}\) for some \(k\) and \(\Tr(G_{k}^{\mathcal S}(\rho)^{2})\ge 1-1/\secpar\), then it holds that
	\(\Pr[P_{k}(\Psi) \to 1] \ge 1-2^{-\secpar}\) with probability at least \(1-p_{1}-p_{2}=1-\frac{m+\tau}{2^{2\secpar}}\).
\end{proof}

\begin{proof}[Proof of \cref{claim: Sep1Haar}]
	In this case, we can regard \(V({\ketbra{\rho}})\) as an independent Haar random pure state \(\ket{\psi}\).
	By \cref{lem: swap}, the expected success probability of the swap test between \(G_{k}^{\tilde{\mathcal S}}({\ketbra{\rho}})\) and \(V({\ketbra{\rho}})\) is
	\ifnum\llncs=1
		\begin{align*}
			\underset{V\gets \mu_{n}}{\Exp}\left[\frac{1+\Tr[G_{k}^{\tilde{\mathcal S}}(\ketbra{\rho})V(\ketbra{\rho})]}{2}\right] & =
			\underset{\psi \gets \sigma_{n}}{\Exp}\left[\frac{1+\Tr[G_{k}^{\tilde{\mathcal S}}(\ketbra{\rho})\ketbra{\psi}]}{2}\right]                                   \\
			                                                                                                                       & = \frac{1}{2} + \frac{1}{2^{n+1}}{}
		\end{align*}
	\else
		\begin{align*}
			\underset{V\gets \mu_{n}}{\Exp}\left[\frac{1+\Tr[G_{k}^{\tilde{\mathcal S}}(\ketbra{\rho})V(\ketbra{\rho})]}{2}\right] =
			\underset{\psi \gets \sigma_{n}}{\Exp}\left[\frac{1+\Tr[G_{k}^{\tilde{\mathcal S}}(\ketbra{\rho})\ketbra{\psi}]}{2}\right] = \frac{1}{2} + \frac{1}{2^{n+1}}{}
		\end{align*}
	\fi
	where we use \cref{lem: Haarproject} in the last equality.
	Applying \cref{cor: stateHaarconcentration} for \(t=1/13\), we have
	\[
		\Pr\left[
			\frac{1+\Tr[G_{k}^{\tilde{\mathcal S}}(\ketbra{\rho})V(\ketbra{\rho})]}{2} \ge \frac{7}{12}
			\right] \le \exp\left(-\frac{2^{n}-2}{4056}\right) \le \frac{1}{2^{2\secpar}}
	\]
	for sufficiently large \(n\).
	In other words, with probability at least \(1-\frac{1}{2^{\secpar}}\),
	the swap test between \(G_{k}^{\tilde{\mathcal S}}(\ketbra{\rho})\) and \(V(\ketbra{\rho})\) succeeds with probability at most \(7/12\) for all \(k\) by the union bound. We only focus on such a case below.
	Chernoff inequality gives that
	\(\Pr[P_{k}(\Psi) \to 1]\) is at most
	\[
		\exp\left(
		-\frac{7r/12\cdot (1/12)^{2}}{(2+1/12)}
		\right)=
		\exp\left(
		-\frac{7r}{3600}
		\right) \le 2^{-2\secpar},
	\]
	for all \(k\). Therefore, if \(V\) is truly Haar random unitary, then it holds that \(\Tr\left[P_{k} \ket{\Psi}\right] \le 2^{-2\secpar}\) for all \(k\) with probability at least \(1-2^{-\secpar}.\)
\end{proof}

\section{On Length Extension of PRSGs}\label{sec:separation_prs}
In this section we show that the output size of a pseudorandom state may be
relevant, i.e., there exist short-PRSGs but PRSGs in a certain form do not
exist.

We will consider a particular type of quantum algorithms, which we call ``\algoname{} quantum algorithms'', where a quantum algorithm \(A\) acts on two registers: the output register \(\bf A\), and the ancilla register \(\bf B\) \footnote{Wlog, the input register can be part of \(\bf A\) and \(\bf B\).},  and the output of \(A\) is of the following form:
\begin{equation}
	A(\cdot) = \Tr_{\bf B}({\rho_{\bf A B}}), \text{ where } {\rho_{\bf A B}} = {\psi}_{\bf A} \otimes \ketbra{0}_{\bf B}.
\end{equation}

\begin{remark}
	We focus on algorithms that reset the ancilla to their initial values.
	More generally, we allow any algorithm that applies only reversible computation to the ancilla, i.e., maps it to a state independent of the oracle.
	In this case, one can assume without loss of generality that the ancilla are uncomputed back to \(\ket{0}\) at the end of the computation.
\end{remark}
We now show the following theorem, which is the main result of this section.
\begin{theorem}\label{thm:no_long_PRSG}
	There exists an isometry oracle \(\oracle\) relative to which
  (classical-accessible) short-PRFSGs exist, but long-PRSGs with \algoname{}
  generation algorithms \modified{making non-adaptive oracle queries and using only constant-size ancilla} do not.
\end{theorem}

The separating oracle \(\oracle\) consists of two oracles: the
classical-accessible isometry CHFS oracle \(O_\ell\) for \(\ell(\lambda)=\lfloor 2\log \lambda\rfloor\) and the
\(\qpspace\) oracle.
The existence of short-PRFSGs follows immediately from \cref{thm:PRFS_in_CHFS}.
It remains to break long PRSGs with \algoname{} generation algorithm
\modified{making non-adaptive oracle queries and using only constant-size ancilla}.

\begin{proof}[Proof of~\cref{thm:no_long_PRSG}]

	By contradiction, assume there exists a PRSG \(\gen(\cdot)\) with an \algoname{} generation algorithm relative to \(\oracle\).
	Let \(u_k\) be the length of the ancilla register.
	We can assume w.l.o.g. that \(u_k = u = \bigO{1}\) is independent of \(k\), by considering \(u = \max_k{u_k}\) and adding ancilla that will not be used for the \(k\) such that \(u_k<u\).
	Because the generation algorithm is \algoname{}, the ancilla registers are reset to \(\ket{0}\) after the computation.
	We write \(d(\secpar)\) and \(\kappa(\secpar)\) to denote the output length and the key length of the PRSG.
	Since the \(\qpspace\) oracle is unitary, we can embed them in the unitaries and write the output state of the algorithm (before tracing out the ancilla)  by
	\begin{equation}\label{eq:decomp}
        \widetilde\rho^{(k)} = U_k\left(\bigotimes_{i=1}^{t_k} O_{x_i^{(k)}}\right)\ket{0^{m(\lambda)}},
	\end{equation}
	where \(m(\secpar) = d(\secpar) + u\) is the number of qubits of the whole space where the computations are made, and \(x_1^{(k)}, \ldots, x_{t_k}^{(k)}\) are chosen before any oracle query.
	We omit the superscript \((k)\) when it is clear from the context.

	Let us denote by \(\widetilde\rho^{(k)}=\rho^{(k)}\otimes\ketbra{0}^{\otimes u}\) the final state before tracing out the ancilla.
	We consider the following adversary \(\adv\), given the polynomial copies of either \(\rho=\rho^{(k)}\) for some \(k\) (in which case it outputs 1) or Haar random state \(\rho\) (in which case it outputs 0).
	In the following, let \(r=C\kappa(\secpar)^2\) for a sufficient large universal constant \(C\) and \(T \ge 100r^2(2td+1)^3\) where \(t = \max_k t_k\). %

	\begin{algorithm}
		\label{alg:PRSattack}
		\(\adv\) does the following on input multiple copies of a state \(\rho\).
		\begin{enumerate}
			\item \(\adv\) executes the purity test \(16T{\kappa(\secpar)}\) times on \(\rho\). If the test fails at least \(4{\kappa(\secpar)}\) times, \(\adv\) returns \(1\) and aborts. Otherwise, it proceeds to the next step.
			\item 
			      We define the following sub-protocol \(P_k\) that takes as input a state \(\Psi=((\rho\otimes\ketbra{0}^{\otimes u})^{\otimes 2})^{\otimes r}\):
			      \begin{description}
				      \item[\(P_k\):]
				            For each \(i\in[r]\), compute \( U_k^\dagger\otimes  U_k^\dagger(\rho\otimes\ketbra{0}^{\otimes u}\otimes\rho\otimes\ketbra{0}^{\otimes u})\) and apply the product test for \(\ell(\abs{x_1}),\dots,\ell(\abs{x_{t_k}}),1,\dots,1\) qubits, where the number of \(1\) is \(s_k = d(\secpar) + u -\sum_{i\in[t_k]} \ell(\abs{x_i})\).
				            Let \(m_k = t_k + s_k\) be the total number of swap tests used in the product test.
				            Return 1 if all tests pass, and return 0 otherwise.
			      \end{description}
			      Then \(\adv\) runs the quantum OR tester with \(\{P_k\}_{k\in\bit^\kappa}\) on \(\Psi=((\rho\otimes\ketbra{0}^{\otimes u})^{\otimes 2})^{\otimes r}\), and returns the same output.
		\end{enumerate}
	\end{algorithm}

	We note that the sub-protocol \(P_k\) can be implemented in polynomial time.

	\begin{claim}\label{clm: rhorhok}
		If \(\rho=\rho^{(k)}\) and \(\Tr(\rho^2)\ge 1-1/T\),
		then
		\(\Pr[P_k(\Phi)]
			\ge 4/5.\)
	\end{claim}
	\begin{claim}\label{clm: rhoHaar}
		If \(\rho\) is a Haar random state, then \(\Pr[P_k(\Phi)\to 1] \le 1/2^{2\kappa(\secpar)}\) for all \(k\) with probability at least \(1-1/2^{\kappa(\secpar)}\).
	\end{claim}
	The same argument as in \cref{sec:separation_pru}
	concludes the proof.
	Indeed, if
	\(\rho=\rho^{(k)}\) for some \(k\) and \(\Tr(\rho^2) \le 1-1/T\), then \cref{lem: purity_test} asserts that the first step outputs \(1\) with probability
	\(1-2^{-{\kappa(\secpar)}}\).

	The other case, i.e., \(\rho=\rho^{(k)}\) and \(\Tr(\rho^2)\ge 1-1/T\) or \(\rho\)
	is a true Haar random state is dealt by the quantum OR lemma.
	In this case, by \cref{clm: rhorhok} and \cref{clm: rhoHaar}, the POVMs
	\(\{P_k\}_{k\in \bit^{\kappa(\secpar)}}\) and \(\Psi\) satisfy the conditions of the
	quantum OR lemma (\cref{lemma:quantum_or})
	unless with probability \(1/2^{{\kappa(\secpar)}} \).
	Therefore, \(\adv\) outputs 1 with probability at least \(1/20\) if
	\(\rho=\rho^{(k)}\) for some \(k\), but it outputs 1 with probability at most
	\(4/2^{{\kappa(\secpar)}} = \negl\) if \(\rho\gets \nu_n\), that is, \(\adv\) breaks the PRSG
	security of \(\gen(\cdot)\).\qedhere
\end{proof}
\begin{proof}[Proof of \cref{clm: rhorhok}]
	By the purity-test analysis, we can assume that \(\Tr(\rho^2)\ge 1-1/T\), otherwise \Cref{alg:PRSattack} would have terminated at step 1 with probability at least \(1-2^{-{\kappa(\secpar)}}\).
	Note that \(\Pr[P_k((\tilde{\rho}_t^{(k)})^{\otimes 2r})\to 1]=1\) by \cref{lemma:product_test_mixed}.
	This implies that \(P_k\) outputs 1 on input \(\Phi= (\rho^{(k)}\otimes\ketbra{0}^{\otimes u})^{\otimes 2r}\) with probability at least
	\[1-\frac{1}{T} \ge 4/5.\qedhere\]
\end{proof}

\begin{proof}[Proof of \cref{clm: rhoHaar}]
Since the CHFS output length is \(\ell(n) = \lfloor 2\log n\rfloor\), we therefore have \(\ell{\abs{x_i}} \leq 2\log \abs{x_i} \leq 2 c_G \log \secpar\), for some constant \(c_G\) for all sufficiently large \(\secpar\).
We thus have
	\begin{align*}
		m_k & = t + s_k
		\geq \frac{t \cdot 2c_G\log\lambda + s_k}{2c_G \log\secpar}
        \geq \frac{d(\secpar) + u}{2c_G\log \secpar}
		= \frac{\omega(\log\secpar)}{2c_G\log \secpar} = \omega(1),
	\end{align*}
	where we used the fact that the candidate PRSG has output dimension \(d(\secpar) = \omega(\log\secpar)\).
    Together with the hypothesis that \(u(\lambda)=\bigO{1}\), for all sufficiently large
  \(\lambda\) and all keys \(k\),
    this shows that there exists a constant \(m_0(u)\) such that \(4^u\cdot 2 \cdot (3/4)^{m_0(u)} \leq 0.05\).

	By \cref{lemma:product_test_haar} and~\cref{lemma:product-test-subspace}, we have that a single
  product test (for key \(k\)) succeeds with expected probability at most
  \(4^{u}\cdot 2 \cdot (3/4)^{m_{k}} \leq 0.05\).
  By the concentration inequality, we can show that with probability at least
  \(1-1/2^{{2\kappa(\secpar)}}\) over Haar random states, a single product test for \(k\) succeeds
  with probability at most \(0.1\).
	Using Chernoff's inequality, we conclude that for each \(k\),
  \(\Pr[P_k(\Phi)\to 1] \le 1/2^{2{\kappa(\secpar)}}\).
  Finally, taking a union bound over all \(2^{\kappa(\secpar)}\) keys \(k\), we obtain
  that the above bound holds for every \(k\in\{0,1\}^{\kappa(\secpar)}\) except with
  probability at most \( 2^{\kappa(\secpar)}\cdot 2^{-2{\kappa(\secpar)}}=2^{-{\kappa(\secpar)}} \).
  This proves the claim.
\end{proof}

 \fi

\bibliographystyle{alpha}

\begin{thebibliography}{GMMY26}

\bibitem[AA11]{STOC:AarArk11}
Scott Aaronson and Alex Arkhipov.
\newblock The computational complexity of linear optics.
\newblock In Lance Fortnow and Salil~P. Vadhan, editors, {\em 43rd ACM STOC},
  pages 333--342. {ACM} Press, June 2011.

\bibitem[Aar04]{aaronson2004limitations}
Scott Aaronson.
\newblock Limitations of quantum advice and one-way communication.
\newblock In {\em Proceedings. 19th IEEE Annual Conference on Computational
  Complexity, 2004.}, pages 320--332. IEEE, 2004.

\bibitem[Aar16]{aaronson2016complexity}
Scott Aaronson.
\newblock The complexity of quantum states and transformations: from quantum
  money to black holes.
\newblock {\em arXiv preprint arXiv:1607.05256}, 2016.

\bibitem[ABGL25]{EC:ABGL25}
Prabhanjan Ananth, John Bostanci, Aditya Gulati, and Yao-Ting Lin.
\newblock Pseudorandomness in the (inverseless) haar random oracle model.
\newblock In Serge Fehr and Pierre-Alain Fouque, editors, {\em EUROCRYPT~2025,
  Part~VII}, volume 15607 of {\em {LNCS}}, pages 138--166. Springer, Cham, May
  2025.

\bibitem[AE07]{AE07}
Andris Ambainis and Joseph Emerson.
\newblock Quantum t-designs: t-wise independence in the quantum world.
\newblock In {\em Twenty-Second Annual IEEE Conference on Computational
  Complexity (CCC'07)}, pages 129--140, 2007.

\bibitem[AGKL24]{EC:AGKL24}
Prabhanjan Ananth, Aditya Gulati, Fatih Kaleoglu, and Yao-Ting Lin.
\newblock Pseudorandom isometries.
\newblock In Marc Joye and Gregor Leander, editors, {\em EUROCRYPT~2024,
  Part~IV}, volume 14654 of {\em {LNCS}}, pages 226--254. Springer, Cham, May
  2024.

\bibitem[AGL24]{TCC:AnaGulLin24}
Prabhanjan Ananth, Aditya Gulati, and Yao-Ting Lin.
\newblock Cryptography in the common {Haar} state model: Feasibility results
  and separations.
\newblock In Elette Boyle and Mohammad Mahmoody, editors, {\em TCC~2024,
  Part~II}, volume 15365 of {\em {LNCS}}, pages 94--125. Springer, Cham,
  December 2024.

\bibitem[AGQY22]{TCC:AGQY22}
Prabhanjan Ananth, Aditya Gulati, Luowen Qian, and Henry Yuen.
\newblock Pseudorandom (function-like) quantum state generators: New
  definitions and applications.
\newblock In Eike Kiltz and Vinod Vaikuntanathan, editors, {\em TCC~2022,
  Part~I}, volume 13747 of {\em {LNCS}}, pages 237--265. Springer, Cham,
  November 2022.

\bibitem[AIK22]{CCC:AarIngKre22}
Scott Aaronson, DeVon Ingram, and William Kretschmer.
\newblock The acrobatics of {BQP}.
\newblock In {\em Proceedings of the 37th Computational Complexity Conference},
  pages 1--17, 2022.

\bibitem[ALY24]{ITCS:AnaLinYue24}
Prabhanjan Ananth, Yao-Ting Lin, and Henry Yuen.
\newblock Pseudorandom strings from pseudorandom quantum states.
\newblock In Venkatesan Guruswami, editor, {\em ITCS 2024}, volume 287, pages
  6:1--6:22. {LIPIcs}, January~/~February 2024.

\bibitem[AQY22]{C:AnaQiaYue22}
Prabhanjan Ananth, Luowen Qian, and Henry Yuen.
\newblock Cryptography from pseudorandom quantum states.
\newblock In Yevgeniy Dodis and Thomas Shrimpton, editors, {\em CRYPTO~2022,
  Part~I}, volume 13507 of {\em {LNCS}}, pages 208--236. Springer, Cham, August
  2022.

\bibitem[Aub98]{Aub98}
Thierry Aubin.
\newblock {\em Some nonlinear problems in Riemannian geometry}.
\newblock Springer Science \& Business Media, 1998.

\bibitem[Bar26]{Bar25}
Mohammed Barhoush.
\newblock Separating pseudorandom generators from logarithmic pseudorandom
  states.
\newblock In {\em Advances in Cryptology -- ASIACRYPT 2026}, 2026.
\newblock To appear.

\bibitem[BBBV97]{BBBV97}
Charles~H Bennett, Ethan Bernstein, Gilles Brassard, and Umesh Vazirani.
\newblock Strengths and weaknesses of quantum computing.
\newblock {\em SIAM journal on Computing}, 26(5):1510--1523, 1997.

\bibitem[BBO{\etalchar{+}}25]{AC:BBOSS25}
Mohammed Barhoush, Amit Behera, Lior Ozer, Louis Salvail, and Or~Sattath.
\newblock Signatures from pseudorandom states via \(\bot \)-{PRFs}.
\newblock In Goichiro Hanaoka and Bo-Yin Yang, editors, {\em ASIACRYPT~2025,
  Part~VIII}, volume 16252 of {\em {LNCS}}, pages 320--349. Springer,
  Singapore, December 2025.

\bibitem[BCN25]{EC:BosCheNeh25}
John Bostanci, Boyang Chen, and Barak Nehoran.
\newblock Oracle separation between quantum commitments and quantum
  one-wayness.
\newblock In Serge Fehr and Pierre-Alain Fouque, editors, {\em EUROCRYPT~2025,
  Part~VII}, volume 15607 of {\em {LNCS}}, pages 3--22. Springer, Cham, May
  2025.

\bibitem[BCQ23]{ITCS:BraCanQia23}
Zvika Brakerski, Ran Canetti, and Luowen Qian.
\newblock On the computational hardness needed for quantum cryptography.
\newblock In Yael~Tauman Kalai, editor, {\em ITCS 2023}, volume 251, pages
  24:1--24:21. {LIPIcs}, January 2023.

\bibitem[BFV20]{ITCS:BouFefVaz20}
Adam Bouland, Bill Fefferman, and Umesh~V. Vazirani.
\newblock Computational pseudorandomness, the wormhole growth paradox, and
  constraints on the {AdS}/{CFT} duality (abstract).
\newblock In Thomas Vidick, editor, {\em ITCS 2020}, volume 151, pages
  63:1--63:2. {LIPIcs}, January 2020.

\bibitem[BM24]{BM24}
Samuel Bouaziz{-}{-}Ermann and Garazi Muguruza.
\newblock Quantum pseudorandomness cannot be shrunk in a black-box way.
\newblock {\em arXiv preprint arXiv:2402.13324}, 2024.

\bibitem[BMM{\etalchar{+}}25]{EC:BMMMY25}
Amit Behera, Giulio Malavolta, Tomoyuki Morimae, Tamer Mour, and Takashi
  Yamakawa.
\newblock A new world in the depths of microcrypt: Separating {OWSGs} and
  quantum money from {QEFID}.
\newblock In Serge Fehr and Pierre-Alain Fouque, editors, {\em EUROCRYPT~2025,
  Part~VII}, volume 15607 of {\em {LNCS}}, pages 23--52. Springer, Cham, May
  2025.

\bibitem[BNY25]{AC:BarNisYam25}
Mohammed Barhoush, Ryo Nishimaki, and Takashi Yamakawa.
\newblock {MicroCrypt} assumptions with quantum input sampling and
  pseudodeterminism: Constructions and separations.
\newblock In Goichiro Hanaoka and Bo-Yin Yang, editors, {\em ASIACRYPT~2025,
  Part~VIII}, volume 16252 of {\em {LNCS}}, pages 516--548. Springer,
  Singapore, December 2025.

\bibitem[BS20]{C:BraShm20}
Zvika Brakerski and Omri Shmueli.
\newblock Scalable pseudorandom quantum states.
\newblock In Daniele Micciancio and Thomas Ristenpart, editors, {\em
  CRYPTO~2020, Part~II}, volume 12171 of {\em {LNCS}}, pages 417--440.
  Springer, Cham, August 2020.

\bibitem[BT06]{BT06}
Andrej Bogdanov and Luca Trevisan.
\newblock Average-case complexity.
\newblock {\em Foundations and Trends{\textregistered} in Theoretical Computer
  Science}, 2(1):1--106, 2006.

\bibitem[Bus82]{Bus82}
Peter Buser.
\newblock A note on the isoperimetric constant.
\newblock In {\em Annales scientifiques de l'{\'E}cole normale sup{\'e}rieure},
  volume~15, pages 213--230, 1982.

\bibitem[CCS25]{EC:CheColSat25}
Boyang Chen, Andrea Coladangelo, and Or~Sattath.
\newblock The power of a single haar random state: Constructing and separating
  quantum pseudorandomness.
\newblock In Serge Fehr and Pierre-Alain Fouque, editors, {\em EUROCRYPT~2025,
  Part~VII}, volume 15607 of {\em {LNCS}}, pages 108--137. Springer, Cham, May
  2025.

\bibitem[CGG24]{C:ChuGolGra24}
Kai-Min Chung, Eli Goldin, and Matthew Gray.
\newblock On central primitives for quantum cryptography with classical
  communication.
\newblock In Leonid Reyzin and Douglas Stebila, editors, {\em CRYPTO~2024,
  Part~VII}, volume 14926 of {\em {LNCS}}, pages 215--248. Springer, Cham,
  August 2024.

\bibitem[CGG{\etalchar{+}}25]{CGG+23}
Bruno Cavalar, Eli Goldin, Matthew Gray, Peter Hall, Yanyi Liu, and Angelos
  Pelecanos.
\newblock On the computational hardness of quantum one-wayness.
\newblock {\em Quantum}, 9:1679, 2025.

\bibitem[CGLQ20]{FOCS:CGLQ20}
Kai-Min Chung, Siyao Guo, Qipeng Liu, and Luowen Qian.
\newblock Tight quantum time-space tradeoffs for function inversion.
\newblock In {\em 61st FOCS}, pages 673--684. {IEEE} Computer Society Press,
  November 2020.

\bibitem[CM24]{quantum:CM24}
Lijie Chen and Ramis Movassagh.
\newblock Quantum merkle trees.
\newblock {\em Quantum}, 8:1380, June 2024.

\bibitem[DLS24]{DLS24}
Fr{\'e}d{\'e}ric Dupuis, Philippe Lamontagne, and Louis Salvail.
\newblock Fiat-shamir for proofs lacks a proof even in the presence of shared
  entanglement.
\newblock {\em Quantum}, 8:1568, 2024.

\bibitem[GGM86]{JACM:GolGolMic86}
Oded Goldreich, Shafi Goldwasser, and Silvio Micali.
\newblock How to construct random functions.
\newblock {\em Journal of the ACM (JACM)}, 33(4):792--807, 1986.

\bibitem[GLMY25]{cryptoeprint:2025/1864}
Aditya Gulati, Yao-Ting Lin, Tomoyuki Morimae, and Shogo Yamada.
\newblock Black-box separation between pseudorandom unitaries, pseudorandom
  isometries, and pseudorandom function-like states.
\newblock Cryptology {ePrint} Archive, Paper 2025/1864, 2025.

\bibitem[GMMY26]{GMMY24}
Eli Goldin, Tomoyuki Morimae, Saachi Mutreja, and Takashi Yamakawa.
\newblock {CountCrypt}: Quantum cryptography between {QCMA} and {PP}.
\newblock In {\em Theory of Cryptography -- TCC 2026}, 2026.
\newblock To appear.

\bibitem[GZ25]{C:GolZha25}
Eli Goldin and Mark Zhandry.
\newblock Translating between the common haar random state model and the
  unitary model.
\newblock In Yael~Tauman Kalai and Seny~F. Kamara, editors, {\em CRYPTO~2025,
  Part~II}, volume 16001 of {\em {LNCS}}, pages 269--300. Springer, Cham,
  August 2025.

\bibitem[HILL99]{HILL99}
Johan H{\aa}stad, Russell Impagliazzo, Leonid~A. Levin, and Michael Luby.
\newblock A pseudorandom generator from any one-way function.
\newblock {\em {SIAM} Journal on Computing}, 28(4):1364--1396, 1999.

\bibitem[HKOT23]{FOCS:HKOT23}
Jeongwan Haah, Robin Kothari, Ryan O'Donnell, and Ewin Tang.
\newblock Query-optimal estimation of unitary channels in diamond distance.
\newblock In {\em 64th FOCS}, pages 363--390. {IEEE} Computer Society Press,
  November 2023.

\bibitem[HLM17]{HLM17}
Aram~W Harrow, Cedric Yen-Yu Lin, and Ashley Montanaro.
\newblock Sequential measurements, disturbance and property testing.
\newblock In {\em Proceedings of the Twenty-Eighth Annual ACM-SIAM Symposium on
  Discrete Algorithms}, pages 1598--1611. SIAM, 2017.

\bibitem[HM10]{FOCS:HarMon10}
Aram~Wettroth Harrow and Ashley Montanaro.
\newblock An efficient test for product states with applications to quantum
  {Merlin}-{Arthur} games.
\newblock In {\em 51st FOCS}, pages 633--642. {IEEE} Computer Society Press,
  October 2010.

\bibitem[HXY19]{AC:HhaXagYam19}
Minki Hhan, Keita Xagawa, and Takashi Yamakawa.
\newblock Quantum random oracle model with auxiliary input.
\newblock In Steven~D. Galbraith and Shiho Moriai, editors, {\em
  ASIACRYPT~2019, Part~I}, volume 11921 of {\em {LNCS}}, pages 584--614.
  Springer, Cham, December 2019.

\bibitem[HY25]{TCC:HhaYam25}
Minki Hhan and Shogo Yamada.
\newblock Pseudorandom function-like states from common haar unitary.
\newblock In Benny Applebaum and Huijia~(Rachel) Lin, editors, {\em TCC~2025,
  Part~III}, volume 16270 of {\em {LNCS}}, pages 134--165. Springer, Cham,
  December 2025.

\bibitem[JLS18]{C:JiLiuSon18}
Zhengfeng Ji, Yi-Kai Liu, and Fang Song.
\newblock Pseudorandom quantum states.
\newblock In Hovav Shacham and Alexandra Boldyreva, editors, {\em CRYPTO~2018,
  Part~III}, volume 10993 of {\em {LNCS}}, pages 126--152. Springer, Cham,
  August 2018.

\bibitem[KN96]{kobayashi1996foundations}
Shoshichi Kobayashi and Katsumi Nomizu.
\newblock {\em Foundations of differential geometry, volume 2}, volume~2.
\newblock John Wiley \& Sons, 1996.

\bibitem[KQST23]{STOC:KQST23}
William Kretschmer, Luowen Qian, Makrand Sinha, and Avishay Tal.
\newblock Quantum cryptography in algorithmica.
\newblock In Barna Saha and Rocco~A. Servedio, editors, {\em 55th ACM STOC},
  pages 1589--1602. {ACM} Press, June 2023.

\bibitem[KQT25]{STOC:KreQiaTal25}
William Kretschmer, Luowen Qian, and Avishay Tal.
\newblock Quantum-computable one-way functions without one-way functions.
\newblock In Michal Kouck\'{y} and Nikhil Bansal, editors, {\em 57th ACM STOC},
  pages 189--200. {ACM} Press, June 2025.

\bibitem[Kre21]{Kre21}
William Kretschmer.
\newblock Quantum pseudorandomness and classical complexity.
\newblock In {\em 16th Conference on the Theory of Quantum Computation,
  Communication and Cryptography (TQC 2021)}. Schloss Dagstuhl-Leibniz-Zentrum
  f{\"u}r Informatik, 2021.

\bibitem[Lic58]{Lic58}
Andr{\'e} Lichnerowicz.
\newblock {\em G{\'e}om{\'e}trie des groupes de transformations}, volume III of
  {\em Travaux et Recherches Math{\'e}matiques}.
\newblock Dunod, Paris, 1958.

\bibitem[Liu23]{EC:Liu23}
Qipeng Liu.
\newblock Non-uniformity and quantum advice in the quantum random oracle model.
\newblock In Carmit Hazay and Martijn Stam, editors, {\em EUROCRYPT~2023,
  Part~I}, volume 14004 of {\em {LNCS}}, pages 117--143. Springer, Cham, April
  2023.

\bibitem[Lub78]{Lubkin}
Elihu Lubkin.
\newblock {Entropy of an n‐system from its correlation with a k‐reservoir}.
\newblock {\em Journal of Mathematical Physics}, 19(5):1028--1031, 05 1978.

\bibitem[LV24]{LV24}
Romi Levy and Thomas Vidick.
\newblock Prs length expansion, 2024.

\bibitem[Mec19]{Mec19}
Elizabeth~S Meckes.
\newblock {\em The random matrix theory of the classical compact groups},
  volume 218.
\newblock Cambridge University Press, 2019.

\bibitem[Mel24]{mele2024introduction}
Antonio~Anna Mele.
\newblock Introduction to haar measure tools in quantum information: A
  beginner's tutorial.
\newblock {\em Quantum}, 8:1340, 2024.

\bibitem[MH25]{STOC:MaHua25}
Fermi Ma and Hsin-Yuan Huang.
\newblock How to construct random unitaries.
\newblock In Michal Kouck\'{y} and Nikhil Bansal, editors, {\em 57th ACM STOC},
  pages 806--809. {ACM} Press, June 2025.

\bibitem[MY22]{C:MorYam22}
Tomoyuki Morimae and Takashi Yamakawa.
\newblock Quantum commitments and signatures without one-way functions.
\newblock In Yevgeniy Dodis and Thomas Shrimpton, editors, {\em CRYPTO~2022,
  Part~I}, volume 13507 of {\em {LNCS}}, pages 269--295. Springer, Cham, August
  2022.

\bibitem[MY24a]{MY22b}
Tomoyuki Morimae and Takashi Yamakawa.
\newblock One-wayness in quantum cryptography.
\newblock In {\em 19th Conference on the Theory of Quantum Computation,
  Communication and Cryptography (TQC 2024)}, pages 4--1. Schloss
  Dagstuhl--Leibniz-Zentrum f{\"u}r Informatik, 2024.

\bibitem[MY24b]{C:MorYam24}
Tomoyuki Morimae and Takashi Yamakawa.
\newblock Quantum advantage from one-way functions.
\newblock In Leonid Reyzin and Douglas Stebila, editors, {\em CRYPTO~2024,
  Part~V}, volume 14924 of {\em {LNCS}}, pages 359--392. Springer, Cham, August
  2024.

\bibitem[Nao91]{JC:Naor91}
Moni Naor.
\newblock Bit commitment using pseudorandomness.
\newblock {\em Journal of Cryptology}, 4(2):151--158, January 1991.

\bibitem[NC10]{nielsen2010quantum}
Michael~A Nielsen and Isaac~L Chuang.
\newblock {\em Quantum computation and quantum information}.
\newblock Cambridge university press, 2010.

\bibitem[Qia24]{C:Qian24}
Luowen Qian.
\newblock Unconditionally secure quantum commitments with preprocessing.
\newblock In Leonid Reyzin and Douglas Stebila, editors, {\em CRYPTO~2024,
  Part~VII}, volume 14926 of {\em {LNCS}}, pages 38--58. Springer, Cham, August
  2024.

\bibitem[Rit23]{Rit23}
Manuel Ritor{\'e}.
\newblock The isoperimetric profile of compact manifolds.
\newblock In {\em Isoperimetric Inequalities in Riemannian Manifolds}, pages
  127--155. Springer, 2023.

\bibitem[Rom90]{STOC:Rompel90}
John Rompel.
\newblock One-way functions are necessary and sufficient for secure signatures.
\newblock In {\em 22nd ACM STOC}, pages 387--394. {ACM} Press, May 1990.

\bibitem[Wee06]{TCC:Wee06}
Hoeteck Wee.
\newblock Finding pessiland.
\newblock In Shai Halevi and Tal Rabin, editors, {\em TCC~2006}, volume 3876 of
  {\em {LNCS}}, pages 429--442. Springer, Berlin, Heidelberg, March 2006.

\bibitem[Yao82]{FOCS:Yao82}
Andrew~Chi{-}Chih Yao.
\newblock Theory and applications of trapdoor functions (extended abstract).
\newblock In {\em 23rd Annual Symposium on Foundations of Computer Science,
  Chicago, Illinois, USA, 3-5 November 1982}, pages 80--91. {IEEE} Computer
  Society, 1982.

\bibitem[Zha12]{C:Zhandry12}
Mark Zhandry.
\newblock Secure identity-based encryption in the quantum random oracle model.
\newblock In Reihaneh Safavi-Naini and Ran Canetti, editors, {\em CRYPTO~2012},
  volume 7417 of {\em {LNCS}}, pages 758--775. Springer, Berlin, Heidelberg,
  August 2012.

\bibitem[Zha25]{C:Zhandry25}
Mark Zhandry.
\newblock How to model unitary oracles.
\newblock In Yael~Tauman Kalai and Seny~F. Kamara, editors, {\em CRYPTO~2025,
  Part~II}, volume 16001 of {\em {LNCS}}, pages 237--268. Springer, Cham,
  August 2025.

\end{thebibliography}
\newcommand{\etalchar}[1]{\(^{#1}\)}

\ifnum\final=1
\else
\appendix
\crefalias{section}{appendix}
\ifnum\submission=1
	\section{Preliminaries}
  \label{app:pre}
\subsection{Quantum states, operations, and trace}

A \(d\)-dimensional quantum state is a positive semi-definite Hermitian density matrix \(\rho = \sum_{x\in[d]} p_x\ketbra{\phi_x}\), where the pure states \(\ketbra{\phi_x}\) have trace one, and \(p_1,\dots,p_d\) is a probability distribution, i.e., \(p_1,\dots,p_d \ge 0\) and \(p_1+\dots+p_d=1\).
Pure states are the rank-1 quantum state that can be written as \(\ketbra{\phi}\). We sometimes write \(\ket{\phi}\) or just \(\phi\) to denote the pure state \(\ketbra{\phi}\) for simplicity.
We can consider any positive semi-definite Hermitian matrix (with arbitrary unit trace) as an unnormalized quantum state, e.g., \(\Pi \rho \Pi\) for some projection \(\Pi\) and quantum state \(\rho\), and call them unnormalized states.

A quantum operation \(\Phi\) is a completely positive and trace-nonincreasing operator, that can be represented by matrices \(B_1,\dots,B_k\) satisfying
\[
	I-\sum_{i=1}^k B_i^\dagger B_i \ge0.
\]

The matrices \(B_1,\dots,B_k\) are the Kraus operators of the operations, and with this notation, \(\Phi\) maps a quantum state \(\rho\) to
\(
\Phi(\rho) = \sum_{i=1}^k B_i \rho B_i^\dagger
\).
Quantum operations can represent unitary operations, projective measurements, or applying a projection \(\Pi\). We write the composition of two quantum operations \(\Phi,\Psi\) by \(\Phi\circ\Psi.\)
For a unitary \(U\), the corresponding operation is represented by \(U(\rho)=U \rho U^{\dagger}\) or sometimes the calligraphic font \(\mathcal U(\rho)\).

The trace norm of a Hermitian matrix \(A\) is defined by
\(\|A\|_1 := \sum_{i=1}^d |\lambda_i|\),
where \(\lambda_1,\dots,\lambda_d\) are the eigenvalues of \(A\). If \(A\) is positive semi-definite, we can write \(\|A\|_1 = {\Tr(A)}.\)
This induces the \emph{trace distance} \(\|\rho-\sigma\|_{tr} = \frac 12\|\rho - \sigma\|_1\) between two (possibly unnormalized) mixed states, which forms a distance over (unnormalized) mixed states.
A quantum operation \(\Phi\) does not increase the trace norm. That is, for any Hermitian matrix \(A\), it holds that \(\|\Phi(A)\|_1 \le \|A\|_1\).
In particular, we have \(\Tr(\Phi(A)) \le \Tr(A)\) for any positive semi-definite matrix \(A\).
For any two (possibly unnormalized) states \(\rho,\sigma\),
\begin{align}\label{eqn:channel_does_not_decrease_trace_distance}
	\|\Phi(\rho)-\Phi(\sigma)\|_{tr}=\frac{1}{2}\|\Phi(\rho-\sigma)\|_1 \le
	\frac12\|\rho-\sigma\|_1 = \|\rho-\sigma\|_{tr}.
\end{align}

For a positive semi-definite matrix \(A\), it holds that
\begin{align}\label{eqn: TrA2}
	\Tr(A^2) \le \Tr(A)^2.
\end{align}

We stress that most of the facts on the trace norm and distance also hold for unnormalized states, i.e., positive semi-definite Hermitian matrices.

\subsection{Cryptographic primitives}
We define cryptographic primitives relative to an oracle \(O\).
\begin{definition}[PRSGs]
	We say that an oracle QPT algorithm \(\gen^O\) is a secure pseudorandom
	state generator (PRSG) in the CHFS model if the following holds for some functions \(\kappa,n:\mathbb N\to \mathbb N\) such that \(\kappa=\omega(\log \secpar)\):
	\begin{itemize}
		\item \textbf{State Generation:} For any \(\secpar \in \NN\) and
		      \(k \in \bin^{\kappa(\secpar)}\), the algorithm \(\gen^O(k)\) outputs
		      an \(n(\secpar)\)-qubit state.
		\item \textbf{Pseudorandomness:} For any polynomial \(t(\cdot)\) and any
		      oracle QPT adversary \(\adv^O = \{\adv^O_{\secpar}\}_{\secpar \in \NN}\), there
		      exists a negligible function \(\varepsilon(\cdot)\) such that for all
		      \(\secpar \in \NN\):
		      \begin{equation*}
			      \abs{\prob[k \gets \bin^{\kappa(\secpar)}]{1 \gets \adv^O_{\secpar}(\gen^O(k)^{\otimes t(\secpar)})} -
			      \prob[\ket{\psi} \gets \sigma_{n(\secpar)}]{1 \gets \adv^O_{\secpar}(\ket{\psi}^{\otimes t(\secpar)})}} \leq \varepsilon(\secpar).
		      \end{equation*}
	\end{itemize}
	We say that \(\gen^O\) is a \(n(\secpar)\)-PRSG to indicate that
	its output length is \(n(\secpar)\).
	We further say that a PRSG is a \emph{short PRSG} when its output length is \(\bigTheta{\log \secpar}\), and a \emph{(long) PRSG} when its output length is \(\omega(\log \secpar)\).
\end{definition}
From now on we will use PRSGs to refer to long PRSGs and short PRSGs for logarithmic output.

For pseudorandom unitaries, we only consider the super-logarithmic output length and without inverse oracle access. Unlike~\cite{C:JiLiuSon18}, we allow PRUs to not be unitary.

\begin{definition}[PRUs]\label{def: PRU}
	We say that
	an oracle QPT algorithm \(G^O\) is a pseudorandom unitary in the CHFS model if
	the following holds for some \(n:\mathbb N \to \mathbb N\) such that \(n=\omega(\log \secpar)\):
	\begin{itemize}
		\item \textbf{Quantum operation:}
		      For any \(\secpar\in\NN\) and \(k\in \bit^\secpar\),
		      \(G^O_k\) takes as input an \(n(\secpar)\)-qubit (mixed) state \(\rho\) and outputs
		      an \(n(\secpar)\)-qubit state \(G^O_k(\rho)\).
		\item \textbf{Pseudorandomness}: For any oracle QPT adversary \(\adv^O = \{\adv^O_{\secpar}\}_{\secpar \in \NN}\), there exists a negligible
		      function \(\varepsilon\) such that for all \(\secpar \in \NN\),
		      \begin{equation*}
			      \abs{\prob[k \gets \bin^{\secpar}]{1 \gets \adv_{\secpar}^{O,G^O_k}} -
			      \prob[\mathcal{U} \gets \mu_{n(\secpar)}]{1 \gets \adv_{\secpar}^{O,\mathcal{U}}}} \leq \varepsilon(\secpar).
		      \end{equation*}
	\end{itemize}
	When the adversary makes non-adaptive queries to \(G^O\) and \(\mathcal U\), we say that \(G\) is non-adaptively secure.
\end{definition}

\subsection{QPSPACE oracle}
We recall the definition of the QPSPACE oracle that implements the arbitrary unitary operation described by polynomial size input \cite{EC:CheColSat25,EC:BMMMY25}.
\begin{definition}[\(\qpspace\) Oracle]\label{def:qpspace}
	The \emph{unitary} QPSPACE machine oracle, denoted by \(\qpspace\), is defined as follows: it takes a pair \((\rho,M,t,b)\) of an \(\ell\)-qubit quantum state \(\rho\), a classical Turing machine \(M\), and an integer \(t\in\mathbb N\), and \(b\in\bit\) denotes the abort bit.
	The oracle runs \(M\) for \(t\) steps to obtain the description of a unitary quantum circuit \(C\) that operates on \(\ell\) qubits; if \(M\) does not terminate after \(t\) steps or the output is not described as above, the oracle flips the abort register and halts. Otherwise, the oracle applies \(C\) on \(\rho\) and returns the output quantum state without measurement.
\end{definition}

The quantum access to the QPSPACE oracle is done by allowing coherent \((M,t)\).
For any unitary quantum circuit \(C\) that is output by a machine \(M\) after \(t\) steps, there is a QPT algorithm with \(\qpspace\) oracle that implements \(C^{-1}(\rho)\) on input \(\rho\) \cite[Proposition 3.5]{EC:BMMMY25}.

\subsection{State property tests}
\subsubsection{Swap test}
We review the basic results of the swap test, which can be used to test the purity of a state.
We provide some lemmas about the swap test on a state that is close to pure states, which are essential to obtain our results.

For two quantum states \(\sigma,\rho\) stored in two different registers \(\rA,\rB\), the swap test is executed on the registers \(\rA,\rB\) and a control register \(\rC\) initialized to \(\ketbra{1}\).
It applies Hadamard on \(\rC\), swaps \(\rA\) and \(\rB\) conditioned on \(\rC\), and measures \(\rC\) on the Hadamard basis.
\begin{lemma}[Swap test]
	\label{lem: swap}
	The swap test on input \((\sigma,\rho)\) outputs 1 with probability
	\[
		\frac{1+\Tr(\rho\sigma)}{2},
	\]
	in which case we say that it passes the swap test. For pure states \(\ket\sigma,\ket\rho\), it equals \(\frac{1+|\braket{\rho}{\sigma}|^{2}}{2}\).

\end{lemma}

When \(\sigma=\rho\), we sometimes call it a \text{purity test} on \(\rho\), which outputs \(1\) with certainty if and only if \(\rho\) is a pure state.
\begin{lemma}\label{lem: purity_test}
	Suppose that \(\Tr(\rho^{2}) \le 1-1/T\) for some state \(\rho\) and \(T\in \mathbb N\).
	Let \(\lambda\in \mathbb N\).
	If we run the purity test \(16T\lambda\) times on \(\rho\), then the probability that at least \(4\lambda\) tests fail among \(16T\lambda\) is at least \(1-2^{-\lambda}\).
\end{lemma}
\begin{proof}
	Note that each test succeeds with probability \((1+\Tr(\rho^{2}))/2\le 1-1/2T\), and is independent of each other.
	Applying Chernoff's inequality (\cref{lem:Cher}) for \(\delta=1/2\), we obtain the desired result.
\end{proof}

\subsubsection{Product test}
We first recall the product test to determine whether an \(n\)-partite state
\(\ket{\phi}\) is a product state or far from any product state from
\cite{FOCS:HarMon10}, then give a bound on the success of the product test on Haar-random
states.
\begin{lemma}[{\cite[Lemma 3]{FOCS:HarMon10}}, Product test for mixed states]\label{lemma:product_test_mixed}
	Let \(m\in\mathbb{N}\) and \(d_{1},\ldots,d_{m}\) be the local dimensions of a \(n\)-qubit system,
	i.e.\ \(\prod_{i\in[m]}d_{i}=2^{n}\).
	Let \(\rho\) be a mixed state of \(n\)-qubits and for every \(S\subseteq[m]\), denote
	by \(\rho_{S}\) the state after tracing out the subsystem \(\overline{S}:=[m]\setminus S\).
	Let \(\mathcal{A}_{\ptest}\) denote the algorithm that, given two copies of \(\rho\), performs the swap test on each
	of the \(m\) pairs of corresponding subsystems of the two copies of \(\rho\), and
	that outputs \(1\) if all the tests succeed, and \(0\) otherwise.
	Then, the probability that the algorithm \(\mathcal{A}_{\ptest}\) outputs \(1\) when
	applied to two copies of \(\rho\) is equal to
	\begin{equation*}
		\Pr(1\gets\mathcal{A}_{\ptest}(\ket{\phi}^{\otimes 2}))=\frac{1}{2^{m}}\sum_{S\subseteq[m]}\Tr[\rho_{S}^{2}].
	\end{equation*}
\end{lemma}

For Haar-random states, the above formula is explicitly calculated
for any partition \(S\cup\overline{S}\) of \([m]\) by~\cite{Lubkin}:
\begin{equation*}
	\underset{\ket{\psi}\gets\sigma}{\mathbb{E}}\Tr[\rho_{S}^{2}]=\frac{d_{S}+d_{\overline{S}}}{d_{S}\cdot d_{\overline{S}}+1}.
\end{equation*}
	As a consequence, we have the following bound for the success of the product
	test on Haar-random states.
\begin{lemma}[Product test for Haar-random states]\label{lemma:product_test_haar}
	Let \(m\in\mathbb{N}\) and \(\{d_{i}\}_{i\in[m]}\) be the local dimensions of a \(n\)-qubit system,
	i.e.\ \(\prod_{i\in[m]}d_{i}=2^{n}\).
	Then, the probability that the algorithm \(\mathcal{A}_{\ptest}\) outputs \(1\) when applied
	to two copies of a \(n\)-qubit Haar-random state \(\ket{\psi}\) satisfies:
	\begin{equation*}
		\underset{\ket{\psi}\gets\sigma}{\mathbb{E}}\Pr(1\gets\mathcal{A}_{\ptest}(\ket{\psi}^{\otimes 2}))\leq 2\left(\frac{3}{4}\right)^{m}.
	\end{equation*}
\end{lemma}
\ifnum\llncs=1
\begin{proof}[{Proof of~\cref{lemma:product_test_haar}}]%
\else
\begin{proof}
\fi
      For every partition \(S\cup\overline{S}\) of \([m]\), the local dimension of each partition is given by \(d_S=\prod_{i\in S}d_i\).
      \begin{align*}
        \underset{\ket{\psi}\gets\sigma}{\mathbb{E}}\Pr(1\gets\mathcal{A}_\ptest(\ket{\psi}^{\otimes 2}))&
        =\underset{\ket{\psi}\gets\sigma}{\mathbb{E}}\left[\frac{1}{2^m}\sum_{S\subseteq[m]}\Tr[\rho_S^2]\right]
        \\
        &=\frac{1}{2^m}\sum_{S\subseteq[m]}\frac{d_S+d_{\overline{S}}}{d_S\cdot d_{\overline{S}}+1}\le \frac{1}{2^m}\sum_{S\subseteq[m]}\frac{d_S+d_{\overline{S}}}{d_S\cdot d_{\overline{S}}}
        \\&=\frac{1}{2^m}\left(\sum_{S\subseteq[m]}\frac 1{d_S} + \frac1{d_{\overline{S}}}\right)
        =\frac{2}{2^m}\left(\sum_{S\subseteq[m]}\frac 1{d_S} \right)
        \\&=\frac{2}{2^m}\prod_{i\in[m]} \left(1+\frac1{d_i}\right)
        \leq \frac2{2^m}\prod_{i=1}^m\left(\frac{3}{2}\right)=2\left(\frac{3}{4}\right)^m,
  \end{align*}
  where we use the fact that each \(d_i\geq 2\) to obtain the last inequality.
\end{proof} %

We also need the following simple extension of the Haar product-test bound, in
which the Haar-random input is restricted to a subspace obtained by appending
zero ancillas.
\begin{lemma}
  \label{lemma:product-test-subspace}
  Let \(\mathcal H_0\) be a Hilbert space of dimension \(d=2^{n-t}\), and let
  \(\mathcal H=\mathcal H_0\otimes \mathbb C^{2^t}\), so that
  \(\dim(\mathcal H)=D=2^n\) for some \(0 \leq t < n\). 
  Let \(P\) be any quantum protocol that takes two copies of a pure state in
  \(\mathcal H\) and outputs a bit.
  Suppose that
  \begin{equation*}
    \prob[\ket{\phi} \leftarrow \sigma_{\mathcal H}]{P\left(\ket{\phi}^{\otimes 2}\right)=1} 
    \ge 1-\varepsilon.
  \end{equation*}
  Then
  \begin{equation*}
    \prob[\ket{\psi} \leftarrow \sigma_{\mathcal H_{0}}]{P\left(\left(\ket{\psi} \otimes \ket{0^{t}}\right)^{\otimes 2}\right)=1}
    \ge 1- 4^{t}\varepsilon .
  \end{equation*}
\end{lemma}

\begin{proof}
  Since \(P\) is a two-copy quantum protocol with binary output, there exists a
  POVM element \(0\le M\le I\) on \(\mathcal H^{\otimes 2}\) corresponding to
  the event that \(P\) outputs \(0\) such that, for every pure state
  \(\ket{\eta}\in\mathcal H\),
  \begin{equation*}
    \prob{P\left(\ket{\eta}^{\otimes 2}\right)=0}
    =
    \operatorname{Tr}\left(
      M \ketbra{\eta}^{\otimes 2}
    \right).
\end{equation*}
Let
\(\Omega_{\mathcal H} \coloneqq \mathbb E_{\ket{\phi}\leftarrow \sigma_{\mathcal H}} \left[ \ketbra{\phi}^{\otimes 2} \right]\).
By the Haar second-moment formula \cite{mele2024introduction},
\begin{equation*}
  \Omega_{\mathcal H}
  =
  \frac{\Pi_{\mathrm{sym}}^{\mathcal H}}{\binom{D+1}{2}}
  =
  \frac{2}{D(D+1)}\Pi_{\mathrm{sym}}^{\mathcal H},
\end{equation*}
where \(\Pi_{\mathrm{sym}}^{\mathcal H}\) is the projector onto the symmetric
subspace of \(\mathcal H^{\otimes 2}\).

Similarly, define the embedded subspace
\(W \coloneqq \mathcal H_0\otimes \ket{0^t}\subseteq \mathcal H\) and let
\(\Omega_{W} \coloneqq \mathbb E_{\ket{\psi}\leftarrow \sigma_{W}} \left[ \ketbra{\psi}^{\otimes 2} \right]\).
Again by the Haar second-moment formula, now applied inside the subspace \(W\),
\begin{equation*}
  \Omega_{W}
  =
  \frac{\Pi_{\mathrm{sym}}^{W}}{\binom{d+1}{2}}
  =
  \frac{2}{d(d+1)}\Pi_{\mathrm{sym}}^{W},
\end{equation*}
where \(\Pi_{\mathrm{sym}}^{W}\) is the projector onto the symmetric subspace of
\(W^{\otimes 2}\).

Since \(W\subseteq \mathcal H\), the symmetric subspace of \(W^{\otimes 2}\) is
a subspace of the symmetric subspace of \(\mathcal H^{\otimes 2}\). 
Hence
\(\Pi_{\mathrm{sym}}^{W}\le \Pi_{\mathrm{sym}}^{\mathcal H}\).
Therefore,
\begin{equation*}
  \Omega_{W}
  =
  \frac{2}{d(d+1)}\Pi_{\mathrm{sym}}^{W}
  \le
  \frac{2}{d(d+1)}\Pi_{\mathrm{sym}}^{\mathcal H}
  =
  \frac{D(D+1)}{d(d+1)}\Omega_{\mathcal H}.
\end{equation*}
By assumption,
\begin{equation*}
  \operatorname{Tr}(M\Omega_{\mathcal H})
  =
  \prob[\ket{\phi}\leftarrow \sigma_{\mathcal H}]
  {P\left(\ket{\phi}^{\otimes 2}\right)=0}
  \le \varepsilon .
\end{equation*}
Thus,
\begin{align*}
  \prob[\ket{\psi}\leftarrow \sigma_{\mathcal H_{0}}]
  {P\left(\left(\ket{\psi} \otimes \ket{0^{t}}\right)^{\otimes 2}\right) = 0}
  &=
  \operatorname{Tr}(M\Omega_{W}) \\
  &\le
  \frac{D(D+1)}{d(d+1)}
  \operatorname{Tr}(M\Omega_{\mathcal H}) \\
  &\le
    \frac{D(D+1)}{d(d+1)}\varepsilon \\
  &\le
    \left(\frac{D}{d}\right)^{2}\varepsilon = 4^{t}\varepsilon. \qedhere
\end{align*}
\end{proof}

\subsection{Quantum OR lemma}
\begin{lemma}[{\cite[Corollary 3.1]{HLM17}}, Quantum OR lemma]\label{lemma:quantum_or}
	Let \(\{\Pi_{i}\}_{i\in[N]}\) be binary-valued POVMs.
	Let \(0<\eps<1/2\) and \(\delta>0\).
	Let \(\Psi\) be a quantum state such that either
	\begin{enumerate}[label=\roman*)]
		\item there exists \(i\in[N]\) such that \(\Tr[\Pi_{i}\Psi]\geq1-\eps\), or
		\item for all \(i\in[N]\), \(\Tr[\Pi_{i}\Psi]\leq\delta\).
	\end{enumerate}
	Then, there is a quantum circuit \(C\), called ``OR tester'', such that
	measuring the first qubit in case \(i)\) yields
	\begin{equation*}
		\Pr(1\gets C(\Psi))\geq\frac{(1-\eps)^{2}}{7},
	\end{equation*}
	and in case \(ii)\),
	\begin{equation*}
		\Pr(1\gets C(\Psi))\leq 4N\delta.
	\end{equation*}

	Moreover, the circuit \(C\) can be implemented by a unitary quantum poly-space machine as long as each POVM \(\Pi_{i}\) can be implemented by a quantum poly-space machine and the set of measurements has a concise polynomial description.
	In other words, the quantum OR tester can be executed by a \(\qpspace\)-aided
	BQP algorithm, where the oracle \(\qpspace\) is defined in~\cref{def:qpspace}.
\end{lemma}
\begin{remark}\label{rem: POVMqpspace}
	``Moreover'' part of the above theorem for the projective measurements is shown in \cite[Appendix A]{EC:CheColSat25}, and the extension to the POVMs is observed in \cite[Lemma 5.2]{EC:BMMMY25}.
\end{remark}

\subsection{Useful lemmas}
\begin{lemma}[Almost as good as new lemma {\cite{aaronson2004limitations,aaronson2016complexity}}]\label{lem:almost-as-good-as-new}
	Let \(\cM=(\Pi_{0},\Pi_{1})\) be a binary measurement that acts as \(\cM(\rho)=\Pi_{0} \rho \Pi_{0} + \Pi_{1} \rho \Pi_{1}\). If \(\Tr[\Pi_{0} \rho]\ge 1-\epsilon\) for \(\epsilon>0\), then it holds that
	\(\|\rho - \cM(\rho)\|_{tr} \le \sqrt{\epsilon}.\)
\end{lemma}
\begin{corollary}\label{cor:gentle proj}
	In the same setting, \(\|\rho-\Pi_{0}\rho\Pi_{0}\|_{tr}\le \epsilon+\sqrt{\epsilon}\le 2\sqrt{\epsilon}.\)
\end{corollary}
\begin{proof}
	We have \(\|\cM(\rho)-\Pi_{0}\rho\Pi_{0}\|_{tr}=\|\Pi_{1}\rho \Pi_{1}\|_{tr}\le \epsilon,\) which gives the result.
\end{proof}

\subsubsection{Norms and Process tomography.}
For a matrix \(M\), the operator norm is defined by
\[
	\|M\|_{op}:= \sup_{\|\ket{\phi}\|_{2}=1} \|M\ket{\phi}\|_{2},
\]
which satisfies \(\|M+N\|_{op}= \max(\|M\|_{op},\|N\|_{op})\) if \(M\) and \(N\) act on the orthogonal space. In particular, for \(M=\sum_{x} \ketbra{x} \otimes M_{x}\), it holds that
\begin{align}\label{eqn:operator_norm_maximum}
	\|M\|_{op} = \max_{x} \|M_{x}\|_{op}.
\end{align}

The diamond norm of an operator \(A\), denoted by \(\|A\|_{\diamond}\), is defined by:
\[
	\|A(\cdot)\|_{\diamond}:= \sup_{\Tr(\rho)=1,\rho \ge 0} \|A\otimes I(\rho)\|_{1},
\]
where \(I\) denotes the identity acting with the same dimension as \(A\). We sometimes omit \((\cdot)\) if it is clear from the context.
We use the following fact about the diamond norm: for quantum channels
\(A,B\) and a density matrix \(\rho\), it holds that
\[
	\|A\otimes I(\rho) - B\otimes I(\rho)\|_{tr} \le\frac12 \|A(\cdot) -B(\cdot) \|_{\diamond}.
\]
We will also use the fact that for unitaries \(U,V\) and the corresponding channels \(\mathcal U,\mathcal V\), it holds that
\begin{equation}\label{eqn: diamond_bound_by_operator_norm}
	\|\mathcal U(\cdot) - \mathcal V(\cdot) \|_{\diamond} \le 2 \|U-V\|_{op}
\end{equation}
for the operator norm \(\|\cdot\|_{op}\)
because
\begin{align*}
	\|\mathcal U(\cdot) - \mathcal V(\cdot)\|_{\diamond} & =\sup_{\Tr(\rho)=1,\rho \ge 0} \|(U\otimes I)(\rho)(U^{\dagger} \otimes I)-(V\otimes I)(\rho)(V^{\dagger} \otimes I)\|_{1}                                                         \\
	                                                   & \le \sup_{\Tr(\rho)=1,\rho \ge 0} \|(U-V)\otimes I\|_{op} \|\rho\|_{1} \|U^{\dagger} \otimes I\|_{op} + \|V\otimes I\|_{op} \|\rho\|_{1} \|(U^{\dagger}- V^{\dagger}) \otimes I\|_{op}
	\\&\le 2\|U-V\|_{op}
\end{align*}
where we use \(\|A\rho B\|_{1} \le \|A\|_{op} \|\rho\|_{1} \|B\|_{op}\).

\begin{theorem}[{\cite{FOCS:HKOT23}}]\label{thm: process_tomography}
	There exists a quantum algorithm \(\Tom\) that, given black-box access to a unitary \(Z\) acting on the \(d\)-dimensional space, satisfies the following for any input \(\epsilon,\delta \in (0,1)\):
	\begin{description}
		\item[Accuracy:] It outputs a classical description of a unitary \(Z\) such that
		      \[
			      \Pr_{Z'\gets \Tom}\left[
			      \|\mathcal Z(\cdot ) - \mathcal Z' (\cdot) \|_{\diamond} \le \epsilon
			      \right] \ge 1-\delta.
		      \]
		\item[Efficiency:] It makes \(O\left(\frac{d^{2}}{\epsilon} \log \frac{1}{\delta}\right)\) queries to \(Z\), and takes \({\sf poly}(d,\frac{1}{\epsilon},\log\frac{1}{\delta})\) time.
	\end{description}
\end{theorem}

\subsubsection{Chernoff bounds.}

We use the following concentration inequalities.
\begin{lemma}[Multiplicative Chernoff bound]\label{lem:Cher}
	Let \(X_{1},\dots,X_{n}\) be some independent random variables over \(\bit\). Let \(X=\sum_{i=1}^{n} X_{i}\) and \(\mu=\Exp[X]\).
	It holds that
	\begin{itemize}
		\item \(\Pr[X \ge (1+\delta)\mu] \le \exp\left(-\frac{\mu\delta^{2} }{2+\delta}\right)\) for \(\delta\ge 0\), and
		\item \(\Pr[X \le (1-\delta) \mu] \le \exp\left(-\frac{\mu\delta^{2} }{2}\right)\) for \(0<\delta<1\).
	\end{itemize}
\end{lemma}
 \fi

\section{Formal Proof of the Barrier Theorem}
\label{sec:formal_proof_conjecture}
\ifnum\llncs=1
\begin{proof}

	Recall that \(\mathbb S(2^n):=\{\ket{\psi}\in \mathbb C^{2^n}:\|\psi\|_2=1\}\) denotes the space of \(n\)-qubit quantum states together with the global phase, i.e., the complex hypersphere in \(\mathbb C^{2^n}\). Our theorem is stated as follows.
	For two elements \(\Phi=(\ket{\phi_1},\ldots,\ket{\phi_k}),\Psi=(\ket{\psi_1},\ldots,\ket{\psi_k})\) in \(X=\mathbb S(2^{n_1})\times \cdots \times \mathbb S(2^{n_k})\), the max-\(\ell_2\) distance is defined by \(d_{\infty}(\Phi,\Psi)\eqdef\max_{i\in [k]} \|\phi_i - \psi_i\|_{2}\).
	\paragraph{Distances.}
	We consider various distances of quantum states.
	For two states in \(\mathbb S(2^{n_i})\), the distance is
	\[
        \|\phi -\psi\|_
		{2} = \sqrt{2-2\Re\braket{\phi}{\psi}}.
	\]
	On the complex unit sphere \(\mathbb S(2^{n_i})\), the round geodesic distance is defined by
	\[
		d_{geo}(\phi,\psi) = \arccos(\Re\braket{\phi}{\psi}).
	\]
	We have \(\|\phi -\psi\|_{2}= 2\sin (\frac{d_{geo}(\phi,\psi)}2) \le d_{geo}(\phi,\psi)\) in \(\mathbb S(2^{n_i})\).

	In the product space \(X\), we consider two distances. The first distance is the max-\(\ell_2\) distance \(d_{\infty}\) defined above. The other distance is the \(\ell_2\)-geodesic distance
	\begin{align}\label{eqn:defell2distance}
		d^{geo}_{2}(\Phi,\Psi) = \sqrt{\sum_{i=1}^k d_{geo,i} (\phi_i,\psi_i)^2}
	\end{align}
	where \(d_{geo,i}\) is the angular geodesic distance in \(\mathbb S(2^{n_i})\).
	We stress that the distance \(d^{geo}_2\) is defined for the geodesic distance, while \(d_\infty\) is defined for the \(\ell_2\) distance. We will later use the following inequality:
	\[d_{\infty}(\Phi,\Psi)  = \max_i \|\phi_i-\psi_i\|_2 \le \max_i d_{geo,i}(\phi_i,\psi_i) \le d^{geo}_2(\Phi,\Psi).\]

	\paragraph{Riemannian manifolds and metrics, and Ricci curvature.}
	A Riemannian manifold is a pair of \((M,g)\), where \(M\) is a smooth manifold and \(g\) is a Riemannian metric\footnote{We stress that two notions, \emph{distance} and \emph{metric}, are used differently; the metric is used only for the Riemannian metric.}, which assigns a positive-definite symmetric bilinear form \(g_p: T_pM\times T_pM \to \mathbb R\) for each point \(p\in M\), where \(T_pM\) denotes the tangent space of \(M\) at \(p\). The Riemannian manifold gives rise to the geodesic distance \(d_g\). In the complex sphere \(\mathbb{S}(2^n) \subseteq \mathbb C^{2^n} \subseteq \mathbb R^{2^{n+1}}\), we use the round
    Riemannian metric \(g_{\mathbb S}\) induced by the real Euclidean inner product
    \[
        \langle u,v\rangle_{\mathbb R}:=\Re\braket{u}{v}.
    \]
    The corresponding geodesic distance is, as discussed above,
    \[
        d_{\mathbb S}(\phi,\psi)
        =
        \arccos(\Re\braket{\phi}{\psi}).
    \]
    There is a natural notion of the products of Riemannian manifolds; the geodesic distance of the products of the complex spheres with the round Riemannian metric gives the \(\ell_2\) distance defined in \cref{eqn:defell2distance}. Looking ahead, we will prove the statement for \(d^{geo}_2\) (instead of \(d_\infty\)) using the results from differential geometry below.

	On the Riemannian manifold, the Ricci curvature tensor \(\Ric\) is uniquely determined and gives a symmetric bilinear form \(\Ric_p: T_pM\times T_pM \to \mathbb R\) for each point \(p\in M\). In the product of Riemannian manifolds \(S=\prod_{1\le i \le r}(M_i,g_i)\) and the point \({\bf p} \in S\), 
    the Ricci curvature on a tangent vector \({\bf v}=(v_1,...,v_r) \in T_{\bf p}S\) satisfies
	\begin{align}\label{eqn:Ricci_product}
		\Ric_S({\bf v},{\bf v}) = \sum_{i=1}^r {\Ric}_{M_i} (v_i,v_i).
	\end{align}
	The intrinsic Riemannian metric \({\bf g}=(g_1,...,g_r)\)\footnote{Occasionally written \(\oplus_i g_i\).} satisfies a similar equality \(g({\bf v},{\bf v}) = \sum_{i=1}^{r} g_i(v_i,v_i)\).

	For a constant \(c\), we occasionally use the notation
	\begin{align}\label{eqn:Lic_condition}
		\Ric \ge c \cdot g \Longleftrightarrow \Ric_p - c\cdot  g_p \geqslant 0~~ \forall p \in M
	\end{align}
	where \(\geqslant\) denotes the positive semi-definite inequality.
	Equivalently, it means
	\[
		\Ric_p(v,v) \ge c\cdot g_p(v,v)
	\]
	holds (as an inequality over real numbers) for all \(p \in M\) and \(v \in T_pM\).
	We note that \(\Ric \ge c \cdot g\) is usually denoted by \(\Ric \ge c\) in the literature.

	We only use Riemannian manifolds and Ricci curvatures for our underlying space \(X=\mathbb S(2^{n_1})\times \cdots \times \mathbb S(2^{n_k})\) regarding the above relations, where we recall \(\mathbb S(2^{n})\) can be understood as the real unit sphere of dimension \(m_n:=2^{n+1}-1\). 
    The round sphere of real dimension \(m\) satisfies \(\Ric = (m-1)\cdot g\) (i.e., an Einstein manifold)~\cite{kobayashi1996foundations}.
    Therefore \(\Ric_{\mathbb S(2^{n_i})} = (2^{n_i+1}-2)\cdot g_{i}\).

	For our main interest \(X\) together with the product round Riemannian metric, for \({\bf v} = (v_1,...,v_k) \in T_{\bf p} X\) for some \({\bf p} \in X\),
	it holds that
    \begin{align*}
        \Ric_X({\bf v},{\bf v}) &= \sum_{i=1}^k \Ric_{\mathbb S(2^{n_i})} (v_i,v_i)\\
        &=\sum_{i=1}^k (2^{n_i+1}-2)\cdot g_{i}(v_i,v_i) \\
        &\ge 2 \sum_i g_{i} (v_i,v_i) = 2{\bf g}({\bf v},{\bf v})
    \end{align*}
    where we use \(2^{n_i+1}-2 \ge 2\) for \(n_i \ge 1\). This means \(X\) satisfies \cref{eqn:Lic_condition} for \(c=2.\)

	Lichnerowicz's theorem~\cite{Lic58} states that (See e.g., \cite[Theorem 4.19]{Aub98}) for an \(n\)-dimensional compact Riemannian manifold \((M_n,g)\), if \(\Ric \ge c\cdot g\) for some \(c>0\), then the first nonzero eigenvalue of the Laplacian\footnote{Here, the Laplacian refers the Laplace-Beltrami operator of Riemannian manifold is defined as the negative of the divergence of the gradient, i.e., \(-\Delta f =-{\sf div}({\sf grad} f) .\) Negative sign is the analysts' convention which ensures the non-negativeness of the eigenvalues. The first nonzero eigenvalue is the smallest positive eigenvalue of \(-\Delta\), i.e., the smallest \(c>0\) such that \(-\Delta f = c f\) for some function \(f\).} satisfies
	\begin{align}
		\lambda_1 \ge \frac{nc}{n-1} \ge c.
	\end{align}
	In particular, for \(X\) together with the product round Riemannian metric, it holds that
	\begin{align}\label{eqn:minimum_nzeign}
		\lambda_1 \ge 2.
	\end{align}

	\paragraph{Cheeger isoperimetric constant.}
	We use the modern exposition for isoperimetric inequalities from \cite{Rit23}.
	Consider a Riemannian manifold \((M,g)\) with the geodesic distance \(d\) and probability measure \(\sigma\) (i.e., \(\sigma(M)=1\)).
	We define the outer Minkowski boundary measure (or the \emph{area}) of a measurable set \(E\) by
	\[
		\sigma^+(E)=\liminf_{r \downarrow 0} \frac{\sigma(E[r]) - \sigma(E)}{r}
	\]
	where \(E[r]:= \{p \in M: d(p,E)\le r\}\) is a closed neighborhood of \(E\).
	The Cheeger isoperimetric constant is
	\begin{equation}\label{def:Cheeger}
		h_M := \inf_{E:\text{measurable}, 0<\sigma(E)<1} \frac{\sigma^+(E)}{\min(\sigma(E),1-\sigma(E))}.
	\end{equation}

	Let \(\lambda_1\) be the smallest nonzero eigenvalue of the Laplacian of \(M\).
	Buser's inequality~\cite{Bus82} says that if \(\Ric\ge -(n-1)a^2\) for some \(a\ge 0\), then it holds that \(\lambda_1\le 2a(n-1)h_M + 10h_M^2\).
	If \(\Ric \ge 0 \), we have \(\lambda_1 \le 10h_M^2\).
	This implies that, for our \(X\),
	\begin{align}\label{eqn: hXminimum}
		h_X \ge \sqrt{1/5}
	\end{align}
	holds because \(\Ric \ge 2 {\bf g} \ge 0\) in \(X\) and \cref{eqn:minimum_nzeign}.

	\paragraph{Proof of~\cref{thm:state_conjecture}.}
	We first work on the product space \(X\) together with the product round Riemannian metric and the corresponding \(\ell_2\) distance \(d_2\).
	Let \(v(r):= \sigma(S_0[r])\) for \(0\le r\le \Delta\)
    where \(S_0[r]=\{x: d(x,S_0)\le r\}.\)
	Here, we assume that \(v\) is differentiable so that \(v'(r)=\sigma^+(S_0[r])\). This significantly simplifies the proof of \cref{eqn: after_integral}, and we provide the formal proof without this assumption in \cref{subsec: differentiability}.

	Note that in the definition of \(S_0[r]\) in \(X\), we use the natural distance \(d_2\) induced by the Riemannian manifold \(X\).
	Given the condition \( d_\infty(S_0,S_1) \ge \Delta\), we also have \(d^{geo}_2(S_0,S_1) \ge d_\infty(S_0,S_1) \ge \Delta\). For any \(0\le r <\Delta\), 
    \(v(r)\ge v(0) \ge \Gamma\). Also, 
    \(S_0[r] \cap S_1 = \emptyset\) holds so that \(v(r)= \sigma(S_0[r]) \le 1-\sigma(S_1)\) and \(1-v(r) \ge \sigma(S_1) \ge \Gamma\). This implies \(\min(v(r),1-v(r)) \ge \Gamma.\)

	The Cheeger constant ensures
	\begin{equation}\label{eqn:v'lowerbound}
		v'(r) \ge h_X \min (v(r),1-v(r)) \ge h_X \Gamma.
	\end{equation}
	Integrating \(v'\) from \(0\) to \(r<\Delta\) gives:
	\begin{align}\label{eqn: after_integral}
		1-\sigma(S_1) -\sigma(S_0) \ge \sigma(S_0[r]) - \sigma(S_0) = v(r)-v(0) \ge \int_0^{r} v'(x) dx \ge h_X \Gamma r.
	\end{align}
    By taking \(r\to \Delta\) from below, 
	this gives:
	\[
		\sigma(X\setminus(S_0 \cup S_1))=1-\sigma(S_0) - \sigma(S_1) \ge h_X \Gamma \Delta
	\]
	where \(h_X \ge  \sqrt{1/5}\) by \cref{eqn: hXminimum}. This proves the desired result.
\end{proof} \fi
\subsection{Removing differentiability assumption}\label{subsec: differentiability}
In this section, we show how to prove \cref{eqn: after_integral} without assuming the differentiability of \(v(r)=\sigma(S_0[r])\).
Recall that the Cheeger constant in \cref{def:Cheeger} and the condition \(\sigma(S_0,S_1)\ge \Gamma\) implies that
\[
	\sigma^+(S_0[r]) \ge h_X \Gamma
\]
for \(r\in [0,\Delta)\)
as in \cref{eqn:v'lowerbound}.
Let \(\epsilon>0\).
The outer Minkowski measure says that for each \(r\), there exists \(\delta_r\) such that for any \(h\in [0,\delta_r)\) and \(r+h\le \Delta\),
\begin{equation}\label{eqn:Minkow_eps}
	\frac{\sigma(S_0[r+h]) - \sigma(S_0[r])}h \ge (h_X \Gamma-\epsilon)
\end{equation}
where we use \(S_0[r][h]=S_0[r+h]\) in our ambient space \(X\).
Define the set
\[
	G_\epsilon := \left\{
	t\in[0,\Delta]: \sigma(S_0[t]) - \sigma(S_0) \ge t\cdot (h_X \Gamma-\epsilon)
	\right\}.
\]

We will prove that \(r \in G_\epsilon\) for all \(r<\Delta\). Assuming this, we have
\begin{equation}\label{eqn:G_eps_final}
	\sigma(S_0[r]) - \sigma(S_0) \ge r \cdot (h_X \Gamma-\epsilon)
\end{equation}
thus choosing \(\epsilon\to 0\) proves \cref{eqn: after_integral}.

Let \(a:=\sup G_\epsilon \cap [0,r]\), which is well-defined because \(0\) is included in \(G_\epsilon\). Let us assume that \(a\in G_\epsilon\), which will be proven later. If \(a<r\), for a sufficiently small \(h\) such that \(a+h\le r\), we have
\begin{align*}
	\sigma(S_0[a+h])-\sigma(S_0[0]) & =(\sigma(S_0[a+h])-\sigma(S_0[a]))+(\sigma(S_0[a])-\sigma(S_0[0]))
	\\& \ge h \cdot (h_X \Gamma-\epsilon) + a \cdot (h_X \Gamma-\epsilon)
	\\&= (h+a)  \cdot (h_X \Gamma-\epsilon)
\end{align*}
where, to prove the inequality, we use \cref{eqn:Minkow_eps} in the first term and the fact that \(a\in G_\epsilon\) in the second term.
This says that \(a+h\in G_\epsilon \cap [0,r]\), which contradicts to \(a= \sup G_\epsilon \cap[0,r]\). Therefore \(a=r \in G_\epsilon\cap[0,r]\), proving \cref{eqn:G_eps_final}.

It remains to prove that \(a\in G_\epsilon\cap[0,r]\).
By the definition of \(a\), there exists an increasing sequence \((a_n)_{n\in \mathbb N}\) that converges to \(a\). Since \(\sigma(S_0[r])\) is an increasing function, we have
\[
	\sigma(S_0[a]) - \sigma(S_0)\ge \sigma(S_0[a_n])  - \sigma(S_0)\ge a_n\cdot (h_X \Gamma-\epsilon)
\]
where the last inequality holds because \(a_n \in G_\epsilon\).
Taking \(n\to \infty\) gives
\[
	\sigma(S_0[a]) - \sigma(S_0)\ge a\cdot (h_X \Gamma-\epsilon),
\]
proving \(a\in G_\epsilon\). This concludes the proof.

\section{Average-case Hard Problem}\label{app:average-hard}
This section presents an average-case hard samplable promise problem in \(\qcma \cap \coqcma\) relative to the log-length unitary CHFS oracle with a sketch of the analysis.
The non-existence of QOWF is clear due to the main theorem. All results in this section is in the relativized world.

We first formally define the samplable problem distributions and related quantum complexity, adapting \cite{BT06}.
\begin{definition}
    We say that a probability ensemble \(\{D_n\}_{n\in \mathbb N}\) is (QPT) samplable if there is a quantum polynomial time sampler \(S\) such that \(\Pr[S(1^n)\to x]=D_n(x)\) for all \(n\) and \(x\in \{0,1\}^*\). We denote by \(\sampqcma \cap \sampcoqcma\) the class of distributional promise problems consisting of decisional promise problems in \(\qcma \cap \coqcma\) coupled with samplable probability ensembles whose outputs satisfy the promise with an overwhelming probability.
\end{definition}
We say that a samplable decisional problem \((\mathcal L,\{D_n\})\) is quantumly easy-on-average if there exists a BQP algorithm \(A\), such that
\[
\Pr_{x \gets D_n}\left[
    \Pr_A\left[A(x)\to \mathcal L(x)\right]\le 2/3
\right]=\negl[n].
\]
Otherwise, we say that \((\mathcal L,\{D_n\})\) is hard-on-average. 
Our goal is to exhibit a samplable problem \((\mathcal L,\{D_n\})\) in \(\sampqcma \cap \sampcoqcma\) that is hard-on-average. Note that the existence of quantum-computable one-way functions implies the existence of a quantumly hard-on-average problem in \(\sampqcma\) by the Goldreich-Levin theorem.\footnote{Similarly, the one-way permutation implies a hard-on-average samplable problem in \(\sampqcma \cap \sampcoqcma\).}
The converse of this implication does not hold in our world, where quantum-computable one-way functions do not exist, yet there exists a quantumly hard-on-average problem in \(\sampqcma\),
suggesting that our world is a quantum-computable version of Pessiland.

We consider the unitary CHFS oracle model with \(\ell(\lambda)= \lceil c\log \lambda\rceil\) for sufficiently large \(c\gg 1\). In other words, our CHFS oracle maps \(\ket{x,0,0}\) to \(\ket{x,1,\phi_x}\) for \(\ell (|x|)\)-qubit state \(\phi_x\). 
We have the following properties for the CHFS oracles.
\begin{itemize}
    \item The output states are almost orthogonal. 
    More precisely, with probability 1 over the choice of the CHFS oracle,
    \(|\braket{\phi_x}{\phi_{x'}}|^2\le 0.01\) holds 
    for all \((x,x') \in \{0,1\}^\lambda \times\{0,1\}^\lambda\) for all sufficiently large \(\lambda\). 
    This is because of \cref{lem: Haarproject} and \cref{cor: stateHaarconcentration}.\footnote{This requires \(c\gg 1\) when using the union bound.}
    \item The output states admit polynomial-length classical descriptions. 
    More precisely, there are two efficient procedures \(\sf Description\) that maps a (polynomial copies of) \(\ell(n)\)-qubit quantum state to a \(\poly[n]\)-length classical string and \(\sf Construct\) that maps a \(\poly[n]\)-length classical string to an \(\ell(n)\)-qubit quantum state such that:
    \[
    \|\rho -{\sf Construct}({\sf Description}(\rho))\|_1\le 2^{-|\rho|}.
    \]
    Furthermore, \({\sf Construct}\) is deterministic.\footnote{The description can be an approximation of all amplitudes.}
\end{itemize}

Given these observations, we define the distribution ensembles that sample from \(D^{yes}\) with probability \(1/2\) and from \(D^{no}\) with probability \(1/2\). Here we use \(\le\) to denote the lexicographical order.
\begin{description}
    \item[(\(D^{yes}\))] We pick a random pair \(x \le z\), and outputs \(({\sf Description}(\phi_x),z)\).
    \item[(\(D^{no}\))] We pick a random pair \(x > z\), and outputs \(({\sf Description}(\phi_x),z)\).
\end{description}

Define \[
\Pi_n :=
\left\{
(y,z):
\exists!x\in\{0,1\}^n
\text{ such that }
\|{\sf Construct}(y)-\phi_x\|_1\le 0.01
\right\}.
\]
For promised inputs \((y,z)\in\Pi\), let \(x(y)\) denote this unique string.
We define the language \(\mathcal L\) so that
\(\mathcal L(y,z)=1\) only if \(x(y) \le z\).
Equivalently,
\[
\Pi_{yes}=\{(y,z)\in\Pi:x(y)\le z\},\qquad
\Pi_{no}=\{(y,z)\in\Pi:x(y)>z\}.
\]
By definition, this language is in \(\sampqcma\) using such an \(x\) as the witness. It is also clear that \({\bf Supp}(D^{yes}) \subset \Pi_{yes}\) and \({\bf Supp}(D^{no}) \subset \Pi_{no}\) except with negligible probability due to the tomography errors.

To analyze, we recall that \(\ket{\phi_x}\) is almost orthogonal to \(\ket{\phi_x}\) for \(x'\neq x\), which implies that for each \(y\), there is at most a single \(x\) such that \(\|{\sf Construct}(y)-\phi_x\|_1 \le 0.01\). This shows that \(x\) can be a witness for \(\coqcma\) as well, showing that the language is in \(\sampcoqcma.\)

The proof of the quantum average-case hardness can be directly proven using the argument analogous to the proof of the PRFSG security in the main body and the BBBV lower bound for unstructured search~\cite{BBBV97}.

\ifnum\submission=1
\section{Missing proofs}
\label{app:proofs}
\begin{proof}[Proof of \cref{lem:chfs-design-simulation}]
Consider a \(D\)-dimensional unitary \(U\) and let
\[
    \ket{J(U)}=(I\otimes U)\ket{\Omega}
\]
be its normalized Choi state where \(\ket{\Omega } = \frac{1}{\sqrt D} \sum_i \ket{ii}\). 
By \cite[Lemmas 23, 24]{Kre21},
for every \(q\)-query oracle algorithm
\(A^U\), there exists some algorithm \(B\) such that
\[
    \Pr[A^U\to1] = D^{2q} \Pr[B(\ket{J(U)}^{\otimes q}) \to 1].
\]
This gives, by the monotonicity of the trace distance, we have
\begin{align*}
&\left|
\Pr_{\Phi\leftarrow\mathcal S}\!\left[A^{S^\Phi}\to1\right]
-
\Pr_{\Psi\leftarrow\mathcal D}\!\left[A^{S^\Psi}\to1\right]
\right| 
\\&
= O\left(
D^{2q} \cdot 
\left\|
\Exp_{\Phi\leftarrow\mathcal S}
\ketbra{J(S^\Phi)}^{\otimes q}
-
\Exp_{\Psi\leftarrow\mathcal D}
\ketbra{J(S^\Psi)}^{\otimes q}
\right\|_1 
\right),
\end{align*}
where \(S^{\Phi}\) and \(S^{\Psi}\) are concatenated up to the input length \(q\), which gives \(D=2^{O(q)}.\)
It remains to bound the Choi-moment difference.

Note that we have less than \(2^{q+1}=:M\) choices of \(x \in \bit^{\leq q}\).
For each \(x\),
\[
    S_{\phi_x}
    =
    I-
    \ketbra{0}
    -
    \ketbra{\phi_x}{\phi_x}
    +\ketbra{\phi_x}{0}
    +\ketbra{0}{\phi_x}.
\]
Write \(T_{x,j}\) to denote the \(j\)-th term in the above equation for \(j\in [5]\), i.e., \(T_{x,1}=I,T_{x,2}=\ketbra{0}\), etc.
Let \(\mathcal I\) be the set of all pairs of \((x,j)\), so that \(|\mathcal I|\le 5M\).
Expanding \(\ket{J(S^\Phi)}^{\otimes q}\) can be written as the linear summation of
\[
\bigotimes_{i=1}^q (I\otimes \ketbra{x_i} \otimes T_{x_i,j_i}) \ket{\Omega}^{\otimes q}
\]
of (less than) \((5M)^q\) terms.
A similar result holds when expanding \(\bra{J(S^\Phi)}^{\otimes q}\), resulting \((5M)^{2q}\) terms total.

Fix one expanded term. It involves at most \(2q\) label occurrences. For each
label \(x\), the contribution of it is a moment of the form
\[
    \mathbb E\left[
        |\phi_x\rangle^{\otimes a_x}
        \langle\phi_x|^{\otimes b_x}
    \right],
    \qquad
    a_x,b_x\le 2q .
\]
If \(a_x\ne b_x\), both the Haar moment and the design moment vanish by the phase-invariant property. If \(a_x=b_x=s\), then \(s\le 2q\), and the \(\delta\)-approximate
\(2q\)-design gives
\[
\left\|
\Exp_{\psi_x}
\ketbra{\psi_x}^{\otimes s}
-
\Exp_{\phi_x}
\ketbra{\phi_x}^{\otimes s}
\right\|_1
\le 2\delta .
\]
Since the fixed expanded term contains at most \(2q\) relevant labels, a simple hybrid shows that the distance is at most \(4q\delta\).

By linearity of expectation, the difference of the two Choi moments is the
sum, over all expanded monomials, of the corresponding differences of
monomial expectations. 
Let \(\mathsf J(A):=(I\otimes A)|\Omega\rangle\). For
\({\bf \alpha},{\bf \beta}\in\mathcal I^q\), define
\[
T_{{\bf \alpha},{\bf \beta}}(\Phi)
:=
\left(\bigotimes_{i=1}^q \mathsf J(B_{\alpha_i}(\Phi))\right)
\left(\bigotimes_{i=1}^q \mathsf J(B_{\beta_i}(\Phi))\right)^\dagger .
\]

For each fixed expanded monomial
\(T_{{\bf \alpha},{\bf \beta}}\), the previous argument shows
\[
\left\|
\Exp_{\Phi\gets \mathcal S}T_{{\bf \alpha},{\bf \beta}}(\Phi)
-
\Exp_{\Psi\gets \mathcal D}T_{{\bf \alpha},{\bf \beta}}(\Psi)
\right\|_1
\le 4q\delta .
\]
There are at most \((5M)^{2q}\) such monomials. Hence, by
the triangle inequality for the trace norm,
\begin{align*}    
&\left\|
\Exp_{\Phi\leftarrow\mathcal S}
\ketbra{J(S^\Phi)}^{\otimes q}
-
\Exp_{\Psi\leftarrow\mathcal D}
\ketbra{J(S^\Psi)}^{\otimes q}
\right\|_1 \\
&\le \sum_{{\bf \alpha},{\bf \beta}} 
\left\|
\Exp_{\Phi\gets \mathcal S}T_{{\bf \alpha},{\bf \beta}}(\Phi)
-
\Exp_{\Psi\gets \mathcal D}T_{{\bf \alpha},{\bf \beta}}(\Psi)
\right\|_1
\\&
\le (5M)^{2q} \cdot 4q\delta,
\end{align*}
thus the overall bound becomes
\[
O\left(
4q\delta \cdot (5DM)^{2q} 
\right)
=
O\left(
q \cdot C^{q^2}  \cdot \delta
\right)
\]
for some \(C>0\), because \(D=2^{O(q)}\) and \(M\le 2^{q+1}\).
This proves the main upper bound.
The final statement is clear because the proof only ever uses independence among the at most \(2q\) labels
appearing in a fixed expanded term.
\end{proof} \fi

\ifnum\submission=1

\section{On Separating PRUs from PRFSGs}\label{sec:separation_pru}

In this section, we consider the length-\(\ell\) quantum-accessible unitarized CHFS
oracle
\(\mathcal S=\mathcal S_{\ell}\)
for \(\ell(|x|)=|x|\) and the \(\qpspace\) oracle.
We will prove the following theorem, which is the main result of this section.

\begin{theorem}
	\label{thm:pru-vs-prfs}
	There exist adaptively-secure quantum-accessible PRFSGs but there do not
	exist non-adaptive PRUs {whose implementation uses neither
ancillary registers nor intermediate measurements}, relative to
	\((\mathcal S,\qpspace)\).
\end{theorem}
Formally, the PRUs whose implementation uses neither
ancillary registers nor intermediate measurements can be specified by an alternating application of unitaries and oracle queries.
The existence of the adaptively-secure quantum-accessible PRFSGs relative to the
oracles is proven by~\Cref{thm:PRFS_in_CHFS}.
What remains is to prove that PRUs without ancilla do not exist in this model.

\begin{lemma}
	\label{lemma:no-pru}
	Non-adaptive PRUs
	{whose implementation uses neither
ancillary registers nor intermediate measurements}
	do not exist with probability 1 relative to the oracle \((\mathcal S,\qpspace)\).
\end{lemma}
\begin{proof}
	We prove the lemma by contradiction.
	Assume that
	\(\{G^{\mathcal S,\qpspace}_{\secpar}(\cdot)\}_{\secpar}\) is a secure \(n(\secpar)\)-PRU construction relative to \((\mathcal S,\qpspace)\) for \(n(\secpar)=\omega(\log \secpar).\)
	For simplicity,
	we drop the \(\qpspace\) oracle and \(\secpar\) in notations and write \(G_{k}^{\mathcal S}\) to denote \(G^{\mathcal S,\qpspace}_{|k|}(k)\).
	The adversary is given oracle access to the oracle \((V,\mathcal S,\qpspace)\) where \(V\) is either \(G_{k^{*}}^{\mathcal S}\) for some \(k^{*}\) or a Haar random unitary of the same size, and tries to determine which is the case with non-negligible probability.

	We write \(\mathcal S = (S_{d})_{d\in \mathbb N}\) where \(S_{d}=\sum_{x\in \bit^{d}} \ketbra{x} \otimes S_{\ket{\phi_{x}}}\) to denote the unitary CHFS oracle, where \(S_{\ket{\phi_{x}}}\) denotes the swap oracle defined in \cref{def:swap_oracle} for some \(d\)-qubit Haar random quantum state \(\ket{\phi_{x}}\) and \(S_{d}\) acts on a \((2d+1)\)-qubit space.\footnote{In the proof below, we consider the oracle queries to \(S_{d}\). The same proof can be extended to the oracle queries to \( S_{\ket{\phi_{x}}}\) for each \(x\), or more general cases. e.g., queries to \(\ketbra{0}\otimes S_{\ket{\phi_{x_{0}}}}+\ketbra{1}\otimes S_{\ket{\phi_{x_{1}}}}\) for any \(x_{1},x_{2}\) of the same length. We focus on the queries to \(S_{d}\) because it is the most complicated.}

	Let \(m=\poly\) be the maximum number of oracle queries to \(\mathcal S\) that \(G^{\mathcal S}\) makes.
	We show that distinguishing \(G_{k}^{\mathcal S}\) from a Haar random unitary can be done efficiently based on swap tests.
	More concretely, we prove that
	the following algorithm \(\adv^{V,\mathcal S,\qpspace}\) can guess with non-negligible probability whether \(V\) is \(G_{k}^{\mathcal S}\) for some random \(k\), (in which case it outputs \(1\)), or a truly Haar random unitary (in which case it outputs \(0\)).
	For simplicity, we omit the oracle notation and write \(\adv\) for \(\adv^{V,\mathcal S,\qpspace}\).

	\begin{algorithm}
		\label{alg:prusep}
		\(\adv\) chooses an \(n\)-qubit Haar random state \(\ket{\rho}\)
		and does the following on input oracle access to \(V\).\footnote{To be efficient, \(\adv\) can use \(s\)-design for large \(s\) instead of Haar random state.}
		\begin{enumerate}
			\item
			      \(\adv\)
			      executes the purity test \(16\secpar^{2}\) times on \(V(\ketbra \rho)\), which makes \(32\lambda^2\) non-adaptive queries to \(V\).
			      If the tests fails at least \(4\secpar\) times, \(\adv\) returns 1 and abort.
			      Otherwise, it proceeds to the next step.
			\item
			      Let \(\tau=2\log(16m) < n\).
			      For all \(i\le \tau\), \(\adv\) runs \(\Tom\) as defined in \cref{thm: process_tomography} on the oracles \(S_{i}\) with parameters \(\epsilon=\frac{2}{2^{\tau/2}},\delta=\frac{1}{2^{2\secpar}}\) and obtains \(S_{i}'\) that approximates \(S_{i}\).
			      Then it defines a new simulated oracle
			      \[
				      \tilde{S}_{d}\coloneqq
				      \begin{cases}
					      I        & \text{ if }d>\tau, \\
					      {S}'_{d} & \text{otherwise}.
				      \end{cases}
			      \]
			      We write \(\tilde{\mathcal S}=(\tilde{S}_{d})_{d\in \mathbb N}.\)
			      Let \(r= 1200\secpar\).
			      For each \(k\), we define the following sub-protocol \(P_{k}\) that takes as input a state \(\Psi\) over the register \({\bf A}_{1}{\bf A}_{1}'\dots{\bf A}_{r}{\bf A}_{r}'\):
			      \begin{description}
				      \item[\(P_{k}\):]
				            Apply \((G_{k}^{\tilde{\mathcal S}}\otimes I)^{\otimes r}(\Psi)\), where each \(G_{k}^{\tilde{\mathcal S}}\) acts on \({\bf A}_{i}\).
				            Then, apply the swap test on \({\bf A}_{i}{\bf A}_{i}'\) for each \(i\in[r]\).
				            Return 1 if at least \(2r/3=800\secpar\) tests passes, and return 0 otherwise.
			      \end{description}
			      Note that computing \(G_{k}^{\tilde{\mathcal S}}\) does not require any queries to the CHFS reflection oracle \(\mathcal S\), neither does \(P_{k}\).
			\item
			      \(\adv\) prepares the following state
			      \[
				      \Psi\coloneqq\bigotimes_{i\in [r]}\left(\ketbra{\rho}_{{\bf A}_{i}}\otimes V(\ketbra{\rho})_{{\bf A}_{i}'}\right).
			      \]
			      \(\adv\) applies \cref{lemma:quantum_or} on input state \(\Psi\) and the family of POVMs induced by \(\{P_{k}\}_{k\in \bit^{\secpar}}\), and outputs the same output as the OR tester.
		\end{enumerate}
	\end{algorithm}

	The following claims summarize the main analysis of the algorithm, which will be proven at the end of this section.
	\begin{claim}\label{claim: Sep1AlgBQP}
		\Cref{alg:prusep} is a \(\bqp^{V,\mathcal S,\qpspace}\) algorithm, which makes non-adaptive queries to \(V\).
	\end{claim}
	\begin{claim}\label{claim: Sep1Gk}
		If \(V=G_{k}^{\mathcal S}\) for some \(k\) and \(\Tr(G_{k}^{\mathcal S}(\rho)^{2})\ge 1-1/\secpar\), then
		\(\Pr[P_{k}(\Psi) \to 1] \ge 1-2^{-\secpar}\) holds with probability at least \(1-\frac{m+\tau}{2^{2\secpar}}\) over the randomness of the algorithm for sufficiently large \(\secpar\).
	\end{claim}
	\begin{claim}\label{claim: Sep1Haar}
		If \(V\gets \mu_{n}\), then \(\Pr[P_{k}(\Psi) \to 1]  \le 2^{-2\secpar}\) holds for all \(k\) with probability at least \(1-2^{-\secpar}\) over the randomness of the algorithm for sufficiently large \(\secpar\).
	\end{claim}

	The efficiency of the algorithm relative to \(\mathcal S,\qpspace\) is provided by \cref{claim: Sep1AlgBQP}.
	Also note that the algorithm breaks the non-adaptive security of PRUs, as the queries to \(V\) only occur in the first step and to prepare \(\Psi\) which are all non-adaptive queries.

	The correctness of the algorithm can be shown by the case analysis. If \(V=G_{k}^{\mathcal S}\) for some \(k\) and if \(\Tr(G_{k}^{\mathcal S}(\rho)^{2})\le 1-1/\secpar\), the first step of \(\adv\) outputs 1 with probability at least \(1-2^{-\secpar}\) as shown in \cref{lem: purity_test}.

	The other case, i.e., \(V=G_{k}^{\mathcal S}\) and \(\Tr(G_{k}^{\mathcal S}(\rho)^{2})\ge 1-1/\secpar\) or \(V\) is a true Haar random unitary is dealt with by the quantum OR lemma.
	In this case, by \cref{claim: Sep1Gk} and \cref{claim: Sep1Haar}, the POVMs \(\{P_{k}\}_{k\in \bit^{\secpar}}\) and \(\Psi\) satisfy the conditions of the quantum OR lemma (\cref{lemma:quantum_or}) unless with probability \( \frac{m+\tau}{2^{2\secpar}} +2^{-\secpar} \le 2/2^{\secpar}\) for large enough \(\secpar\),
	\(\epsilon=1/2^{\secpar}\) and \(\delta=1/2^{2\secpar}\).
	Therefore, \(\adv\) outputs 1 with probability at least \(1/8\) if \(V=G_{k}^{\mathcal S}\) for some \(k\), but it outputs 1 with probability at most \(4/2^{\secpar}\) if \(V\gets \mu_{n}\), that is, \(\adv\) breaks the PRU security of \(\{G_{k}^{\mathcal S}\}\).
	This concludes the proof.\end{proof}

We now prove the claims.
\begin{proof}[Proof of \cref{claim: Sep1AlgBQP}]
	The first step takes polynomial time and \(32\secpar^{2}\) non-adaptive queries to the oracle \(V\) for the \(16\secpar^2\) purity tests.
	The second step has time and query complexity (to \(\mathcal S\)) equal to \(\tau \times {\sf poly}\left(d,\frac{1}{\epsilon},\log \frac{1}{\delta}\right) =
	{\sf poly}\left(2^{\tau}, 2^{\tau}, \secpar\right) =\poly\).
	Note that it is clear that \(P_{k}\) can be executed by a quantum polynomial space machine. In the final step, the quantum OR tester can be executed by a \(\qpspace\)-aided BQP machine as noted in ``Moreover'' part of \cref{lemma:quantum_or} with inputs the descriptions of \(S_{i}'\) for \(i\le \tau\) (prepared by the first step) as \(P_{k}\) can be implemented by a quantum polynomial-space machine.
\end{proof}

\begin{proof}[Proof of \cref{claim: Sep1Gk}]
	We will show that \(G_{k}^{\mathcal S}(\ketbra\rho)\)
	and \(G_{k}^{\tilde{\mathcal S}}(\ketbra\rho)\) are
	close with high probability, for a Haar random input state \(\ket{\rho}\) of size
	\(n\)-qubit.
	We will write \(\rho = \ketbra{\rho}\) as a short-hand.

	We begin with the following two closeness properties for \(S_{d}\) and \(\tilde S_{d}\) from the later steps of the algorithm.
	First, for small dimensions \(d \le \tau < n\), \cref{thm: process_tomography} ensures that
	\begin{align}\label{eqn: good2_PRUPRFS}
		\|\tilde{S}_{d}\otimes I(\rho) - {S}_{d}\otimes I(\rho)\|_{tr}=
		\|S_{d}'\otimes I(\rho) - {S}_{d}\otimes I(\rho)\|_{tr} \le\frac{\epsilon}2= \frac{1}{2^{\tau/2}},
	\end{align}
	holds for any quantum state \(\rho\) with probability \(1-\delta = 1-\frac{1}{2^{2\secpar}}\).
	Thus all tomography outputs are \(1/2^{\tau/2}\)-close to the target unitaries with overwhelming probability \(1-p_{1}\), for \(p_{1}=\tau/2^{2\secpar}\).
	In the following, we assume it is the case.

    Now we discuss the large dimensions \(d>\tau\), where we want to show that \(S_d\) acts almost as the identity with high probability for a pure Haar quantum state \(\ket\rho\).
    Observe that for the reflection \(R=I-2\Pi\) such that \(\|\Pi\ket{\rho}\|^2=p\),
    \begin{align*}
        & \|\rho - R(\rho)\|^2_{tr} = \|\ketbra\rho - R \ketbra \rho R^{\dagger} \|^2_{tr}
        = {1-|\bra{\rho} R \ket{\rho}|^2} = 4{p(1-p)} \le 4p.
    \end{align*}
    Given this, for \(S_d = I - 2\sum_{x\in \bit^d} \ketbra{x}\otimes \ketbra{\phi_x -} \otimes I_{n-2d-1}  = I- 2\Pi_d\) for the rank \(2^{n-d-1}\) projector \(\Pi_d\),
    \begin{align*}
        \Exp_{\rho}
        \|\tilde{S}_{d} \otimes I(\rho)-S_{d}\otimes I(\rho)\|^2 _{tr}  
        =\Exp_{\rho}
        \|I(\rho)-S_{d}\otimes I(\rho)\|^2 _{tr}
        \le 4\Exp_{\rho} \| \Pi_d \ket{\rho}\|^2 \le \frac{1}{2^{d-1}} \le \frac{1}{2^{\tau}}
    \end{align*}
    where we use \cref{lem: Haarproject}.
    Note that \(\| \Pi_d \ket{\rho}\|^2 \) is the success probability of an 1-query algorithm to a Haar random unitary \(U\) by letting \(\rho = U\ket0\). 
    Applying \cref{cor: stateHaarconcentration} with \(t=1/2^{\tau+1}\), we have
    \[
    \Pr_{\rho} \left[
        \| \Pi_d \ket{\rho}\|^2 \ge \frac{1}{2^{\tau}}
    \right]\le \exp \left(
    -\frac{2^n-2}{24\cdot 2^{2\tau+2}}
    \right) \le \frac{1}{2^{2\lambda}}
    \]
    for sufficiently large \(n\). In other words, except for the probability \(1/2^{2\lambda}\), it holds that
    \begin{align}\label{eqn: close_large_d}
        \|\tilde{S}_{d} \otimes I(\rho)-S_{d}\otimes I(\rho)\| _{tr}  \le \frac{2}{2^{\tau/2}} =\epsilon
    \end{align}

	To bound the trace distance between \(G_{k}^{\mathcal S}(\ketbra{\rho})\) and \(G_{k}^{\tilde{\mathcal S}}(\ketbra{\rho})\), we use the hybrid argument using the above two observations.
	Let \({\Phi_{j}}\) for \(0\le j \le m\) be equal to
	the outcome of \(G_{k}\) on input \(\ketbra{\rho}\) with the first \(j\) oracle queries are answered using \(\tilde {\mathcal S}\) and the other \(m-j\) queries are answered using \(\mathcal S\).
	We have that \({\Phi_{0}}\) is the state \(G_{k}^{\mathcal S}(\rho)\) and \({\Phi_{m}}\) is the state \(G_{k}^{\tilde{\mathcal S}}({\rho})\).

	Let \(\ket{\phi_{j}}\) be the intermediate state right after \(j\)-th oracle query when computing \(G_{k}^{\tilde{\mathcal S}}(\ketbra{\rho})\). 
    Note that for large \(d\), we apply the identity instead of \(S_d\), thus the intermediate state is independent of \(S_d\).
    Therefore we can use \cref{eqn: close_large_d} for large \(d\) with high probability over \(\phi_{j}\).
    For small \(d\), \cref{eqn: good2_PRUPRFS} ensures the upper bound.
	Thus, by the subadditivity and monotonicity of the trace distance, we have
	\begin{align*}
		\left\|G_{k}^{\mathcal S} (\ketbra{\rho}) - G_{k}^{\tilde{\mathcal S}} (\ketbra{\rho}) \right\|_{tr} &
		\le \sum_{j=0}^{m-1}\|S_{d_{j}^{(k)}}\otimes I(\phi_{j})-\tilde{S}_{d_{j}^{(k)}}\otimes I(\phi_{j})\| _{tr}
		\\&
		\le \sum_{j=0}^{m-1}  \max(\frac{\epsilon}{2},\epsilon)
        = \frac{2m}{2^{\tau/2}} = \frac{1}{8},
	\end{align*}
	with probability \(1-p_{2}\) for \(p_{2}=\frac{m}{2^{2\secpar}}\); we again focus on this case.

	We finally analyze the success probability of a single swap test between \(G_{k}^{\mathcal S}(\ketbra{\rho})\) and \(G_{k}^{\tilde{\mathcal S}}(\ketbra{\rho})\) succeeds in subroutine \(P_{k}\).
	Since
	\ifnum\llncs=1
		\begin{align*}
			 & \|G_{k}^{{\mathcal S}}(\ketbra{\rho})\otimes G_{k}^{\tilde{\mathcal S}}(\ketbra{\rho})-G_{k}^{{\mathcal S}}(\ketbra{\rho})\otimes G_{k}^{{\mathcal S}}(\ketbra{\rho})\|_{tr} \\
			 & \quad\quad=\|G_{k}^{\tilde{\mathcal S}}(\ketbra{\rho}) - G_{k}^{{\mathcal S}}(\ketbra{\rho})\|_{tr}
			\le 1/8,
		\end{align*}
	\else
		\begin{align*}
			\|G_{k}^{{\mathcal S}}(\ketbra{\rho})\otimes G_{k}^{\tilde{\mathcal S}}(\ketbra{\rho})-G_{k}^{{\mathcal S}}(\ketbra{\rho})\otimes G_{k}^{{\mathcal S}}(\ketbra{\rho})\|_{tr} =\|G_{k}^{\tilde{\mathcal S}}(\ketbra{\rho}) - G_{k}^{{\mathcal S}}(\ketbra{\rho})\|_{tr}
			\le 1/8,
		\end{align*}
	\fi
	we have
	\begin{align*}
		\left|
		\frac{1+\Tr(G_{k}^{\tilde{\mathcal S}}(\ketbra{\rho})G_{k}^{{\mathcal S}}(\ketbra{\rho}))}{2}
		- \frac{1+\Tr(G_{k}^{{\mathcal S}}(\ketbra{\rho})^{2})}{2}
		\right|
		\le
		\frac{1}{8},
	\end{align*}
	and using the fact that \(\Tr(G_{k}^{{\mathcal S}}(\ketbra{\rho})^{2})\ge 1-1/\secpar\), we have
	\[
		\frac{1+\Tr(G_{k}^{\tilde{\mathcal S}}(\ketbra{\rho})G_{k}^{{\mathcal S}}(\ketbra{\rho}))}{2} \ge \frac{7}{8} - \frac{1}{2\secpar} \ge \frac34.
	\]
	Therefore, by Chernoff's inequality, the probability that at least \(\frac{2r}{3}\) tests succeed among \(r\) swap tests is bounded by
	\[
		1-\exp\left(-\frac{3r}{2\cdot 4\cdot 12^{2}}\right)=1-\exp\left(-\frac{r}{384}\right) \ge 1-2^{-\secpar}.
	\]
	Overall, if \(V=G_{k}^{{\mathcal S}}\) for some \(k\) and \(\Tr(G_{k}^{\mathcal S}(\rho)^{2})\ge 1-1/\secpar\), then it holds that
	\(\Pr[P_{k}(\Psi) \to 1] \ge 1-2^{-\secpar}\) with probability at least \(1-p_{1}-p_{2}=1-\frac{m+\tau}{2^{2\secpar}}\).
\end{proof}

\begin{proof}[Proof of \cref{claim: Sep1Haar}]
	In this case, we can regard \(V({\ketbra{\rho}})\) as an independent Haar random pure state \(\ket{\psi}\).
	By \cref{lem: swap}, the expected success probability of the swap test between \(G_{k}^{\tilde{\mathcal S}}({\ketbra{\rho}})\) and \(V({\ketbra{\rho}})\) is
	\ifnum\llncs=1
		\begin{align*}
			\underset{V\gets \mu_{n}}{\Exp}\left[\frac{1+\Tr[G_{k}^{\tilde{\mathcal S}}(\ketbra{\rho})V(\ketbra{\rho})]}{2}\right] & =
			\underset{\psi \gets \sigma_{n}}{\Exp}\left[\frac{1+\Tr[G_{k}^{\tilde{\mathcal S}}(\ketbra{\rho})\ketbra{\psi}]}{2}\right]                                   \\
			                                                                                                                       & = \frac{1}{2} + \frac{1}{2^{n+1}}{}
		\end{align*}
	\else
		\begin{align*}
			\underset{V\gets \mu_{n}}{\Exp}\left[\frac{1+\Tr[G_{k}^{\tilde{\mathcal S}}(\ketbra{\rho})V(\ketbra{\rho})]}{2}\right] =
			\underset{\psi \gets \sigma_{n}}{\Exp}\left[\frac{1+\Tr[G_{k}^{\tilde{\mathcal S}}(\ketbra{\rho})\ketbra{\psi}]}{2}\right] = \frac{1}{2} + \frac{1}{2^{n+1}}{}
		\end{align*}
	\fi
	where we use \cref{lem: Haarproject} in the last equality.
	Applying \cref{cor: stateHaarconcentration} for \(t=1/13\), we have
	\[
		\Pr\left[
			\frac{1+\Tr[G_{k}^{\tilde{\mathcal S}}(\ketbra{\rho})V(\ketbra{\rho})]}{2} \ge \frac{7}{12}
			\right] \le \exp\left(-\frac{2^{n}-2}{4056}\right) \le \frac{1}{2^{2\secpar}}
	\]
	for sufficiently large \(n\).
	In other words, with probability at least \(1-\frac{1}{2^{\secpar}}\),
	the swap test between \(G_{k}^{\tilde{\mathcal S}}(\ketbra{\rho})\) and \(V(\ketbra{\rho})\) succeeds with probability at most \(7/12\) for all \(k\) by the union bound. We only focus on such a case below.
	Chernoff inequality gives that
	\(\Pr[P_{k}(\Psi) \to 1]\) is at most
	\[
		\exp\left(
		-\frac{7r/12\cdot (1/12)^{2}}{(2+1/12)}
		\right)=
		\exp\left(
		-\frac{7r}{3600}
		\right) \le 2^{-2\secpar},
	\]
	for all \(k\). Therefore, if \(V\) is truly Haar random unitary, then it holds that \(\Tr\left[P_{k} \ket{\Psi}\right] \le 2^{-2\secpar}\) for all \(k\) with probability at least \(1-2^{-\secpar}.\)
\end{proof}

\section{On Length Extension of PRSGs}\label{sec:separation_prs}
In this section we show that the output size of a pseudorandom state may be
relevant, i.e., there exist short-PRSGs but PRSGs in a certain form do not
exist.

We will consider a particular type of quantum algorithms, which we call ``\algoname{} quantum algorithms'', where a quantum algorithm \(A\) acts on two registers: the output register \(\bf A\), and the ancilla register \(\bf B\) \footnote{Wlog, the input register can be part of \(\bf A\) and \(\bf B\).},  and the output of \(A\) is of the following form:
\begin{equation}
	A(\cdot) = \Tr_{\bf B}({\rho_{\bf A B}}), \text{ where } {\rho_{\bf A B}} = {\psi}_{\bf A} \otimes \ketbra{0}_{\bf B}.
\end{equation}

\begin{remark}
	We focus on algorithms that reset the ancilla to their initial values.
	More generally, we allow any algorithm that applies only reversible computation to the ancilla, i.e., maps it to a state independent of the oracle.
	In this case, one can assume without loss of generality that the ancilla are uncomputed back to \(\ket{0}\) at the end of the computation.
\end{remark}
We now show the following theorem, which is the main result of this section.
\begin{theorem}\label{thm:no_long_PRSG}
	There exists an isometry oracle \(\oracle\) relative to which
  (classical-accessible) short-PRFSGs exist, but long-PRSGs with \algoname{}
  generation algorithms \modified{making non-adaptive oracle queries and using only constant-size ancilla} do not.
\end{theorem}

The separating oracle \(\oracle\) consists of two oracles: the
classical-accessible isometry CHFS oracle \(O_\ell\) for \(\ell(\lambda)=\lfloor 2\log \lambda\rfloor\) and the
\(\qpspace\) oracle.
The existence of short-PRFSGs follows immediately from \cref{thm:PRFS_in_CHFS}.
It remains to break long PRSGs with \algoname{} generation algorithm
\modified{making non-adaptive oracle queries and using only constant-size ancilla}.

\begin{proof}[Proof of~\cref{thm:no_long_PRSG}]

	By contradiction, assume there exists a PRSG \(\gen(\cdot)\) with an \algoname{} generation algorithm relative to \(\oracle\).
	Let \(u_k\) be the length of the ancilla register.
	We can assume w.l.o.g. that \(u_k = u = \bigO{1}\) is independent of \(k\), by considering \(u = \max_k{u_k}\) and adding ancilla that will not be used for the \(k\) such that \(u_k<u\).
	Because the generation algorithm is \algoname{}, the ancilla registers are reset to \(\ket{0}\) after the computation.
	We write \(d(\secpar)\) and \(\kappa(\secpar)\) to denote the output length and the key length of the PRSG.
	Since the \(\qpspace\) oracle is unitary, we can embed them in the unitaries and write the output state of the algorithm (before tracing out the ancilla)  by
	\begin{equation}\label{eq:decomp}
        \widetilde\rho^{(k)} = U_k\left(\bigotimes_{i=1}^{t_k} O_{x_i^{(k)}}\right)\ket{0^{m(\lambda)}},
	\end{equation}
	where \(m(\secpar) = d(\secpar) + u\) is the number of qubits of the whole space where the computations are made, and \(x_1^{(k)}, \ldots, x_{t_k}^{(k)}\) are chosen before any oracle query.
	We omit the superscript \((k)\) when it is clear from the context.

	Let us denote by \(\widetilde\rho^{(k)}=\rho^{(k)}\otimes\ketbra{0}^{\otimes u}\) the final state before tracing out the ancilla.
	We consider the following adversary \(\adv\), given the polynomial copies of either \(\rho=\rho^{(k)}\) for some \(k\) (in which case it outputs 1) or Haar random state \(\rho\) (in which case it outputs 0).
	In the following, let \(r=C\kappa(\secpar)^2\) for a sufficient large universal constant \(C\) and \(T \ge 100r^2(2td+1)^3\) where \(t = \max_k t_k\). %

	\begin{algorithm}
		\label{alg:PRSattack}
		\(\adv\) does the following on input multiple copies of a state \(\rho\).
		\begin{enumerate}
			\item \(\adv\) executes the purity test \(16T{\kappa(\secpar)}\) times on \(\rho\). If the test fails at least \(4{\kappa(\secpar)}\) times, \(\adv\) returns \(1\) and aborts. Otherwise, it proceeds to the next step.
			\item 
			      We define the following sub-protocol \(P_k\) that takes as input a state \(\Psi=((\rho\otimes\ketbra{0}^{\otimes u})^{\otimes 2})^{\otimes r}\):
			      \begin{description}
				      \item[\(P_k\):]
				            For each \(i\in[r]\), compute \( U_k^\dagger\otimes  U_k^\dagger(\rho\otimes\ketbra{0}^{\otimes u}\otimes\rho\otimes\ketbra{0}^{\otimes u})\) and apply the product test for \(\ell(\abs{x_1}),\dots,\ell(\abs{x_{t_k}}),1,\dots,1\) qubits, where the number of \(1\) is \(s_k = d(\secpar) + u -\sum_{i\in[t_k]} \ell(\abs{x_i})\).
				            Let \(m_k = t_k + s_k\) be the total number of swap tests used in the product test.
				            Return 1 if all tests pass, and return 0 otherwise.
			      \end{description}
			      Then \(\adv\) runs the quantum OR tester with \(\{P_k\}_{k\in\bit^\kappa}\) on \(\Psi=((\rho\otimes\ketbra{0}^{\otimes u})^{\otimes 2})^{\otimes r}\), and returns the same output.
		\end{enumerate}
	\end{algorithm}

	We note that the sub-protocol \(P_k\) can be implemented in polynomial time.

	\begin{claim}\label{clm: rhorhok}
		If \(\rho=\rho^{(k)}\) and \(\Tr(\rho^2)\ge 1-1/T\),
		then
		\(\Pr[P_k(\Phi)]
			\ge 4/5.\)
	\end{claim}
	\begin{claim}\label{clm: rhoHaar}
		If \(\rho\) is a Haar random state, then \(\Pr[P_k(\Phi)\to 1] \le 1/2^{2\kappa(\secpar)}\) for all \(k\) with probability at least \(1-1/2^{\kappa(\secpar)}\).
	\end{claim}
	The same argument as in \cref{sec:separation_pru}
	concludes the proof.
	Indeed, if
	\(\rho=\rho^{(k)}\) for some \(k\) and \(\Tr(\rho^2) \le 1-1/T\), then \cref{lem: purity_test} asserts that the first step outputs \(1\) with probability
	\(1-2^{-{\kappa(\secpar)}}\).

	The other case, i.e., \(\rho=\rho^{(k)}\) and \(\Tr(\rho^2)\ge 1-1/T\) or \(\rho\)
	is a true Haar random state is dealt by the quantum OR lemma.
	In this case, by \cref{clm: rhorhok} and \cref{clm: rhoHaar}, the POVMs
	\(\{P_k\}_{k\in \bit^{\kappa(\secpar)}}\) and \(\Psi\) satisfy the conditions of the
	quantum OR lemma (\cref{lemma:quantum_or})
	unless with probability \(1/2^{{\kappa(\secpar)}} \).
	Therefore, \(\adv\) outputs 1 with probability at least \(1/20\) if
	\(\rho=\rho^{(k)}\) for some \(k\), but it outputs 1 with probability at most
	\(4/2^{{\kappa(\secpar)}} = \negl\) if \(\rho\gets \nu_n\), that is, \(\adv\) breaks the PRSG
	security of \(\gen(\cdot)\).\qedhere
\end{proof}
\begin{proof}[Proof of \cref{clm: rhorhok}]
	By the purity-test analysis, we can assume that \(\Tr(\rho^2)\ge 1-1/T\), otherwise \Cref{alg:PRSattack} would have terminated at step 1 with probability at least \(1-2^{-{\kappa(\secpar)}}\).
	Note that \(\Pr[P_k((\tilde{\rho}_t^{(k)})^{\otimes 2r})\to 1]=1\) by \cref{lemma:product_test_mixed}.
	This implies that \(P_k\) outputs 1 on input \(\Phi= (\rho^{(k)}\otimes\ketbra{0}^{\otimes u})^{\otimes 2r}\) with probability at least
	\[1-\frac{1}{T} \ge 4/5.\qedhere\]
\end{proof}

\begin{proof}[Proof of \cref{clm: rhoHaar}]
Since the CHFS output length is \(\ell(n) = \lfloor 2\log n\rfloor\), we therefore have \(\ell{\abs{x_i}} \leq 2\log \abs{x_i} \leq 2 c_G \log \secpar\), for some constant \(c_G\) for all sufficiently large \(\secpar\).
We thus have
	\begin{align*}
		m_k & = t + s_k
		\geq \frac{t \cdot 2c_G\log\lambda + s_k}{2c_G \log\secpar}
        \geq \frac{d(\secpar) + u}{2c_G\log \secpar}
		= \frac{\omega(\log\secpar)}{2c_G\log \secpar} = \omega(1),
	\end{align*}
	where we used the fact that the candidate PRSG has output dimension \(d(\secpar) = \omega(\log\secpar)\).
    Together with the hypothesis that \(u(\lambda)=\bigO{1}\), for all sufficiently large
  \(\lambda\) and all keys \(k\),
    this shows that there exists a constant \(m_0(u)\) such that \(4^u\cdot 2 \cdot (3/4)^{m_0(u)} \leq 0.05\).

	By \cref{lemma:product_test_haar} and~\cref{lemma:product-test-subspace}, we have that a single
  product test (for key \(k\)) succeeds with expected probability at most
  \(4^{u}\cdot 2 \cdot (3/4)^{m_{k}} \leq 0.05\).
  By the concentration inequality, we can show that with probability at least
  \(1-1/2^{{2\kappa(\secpar)}}\) over Haar random states, a single product test for \(k\) succeeds
  with probability at most \(0.1\).
	Using Chernoff's inequality, we conclude that for each \(k\),
  \(\Pr[P_k(\Phi)\to 1] \le 1/2^{2{\kappa(\secpar)}}\).
  Finally, taking a union bound over all \(2^{\kappa(\secpar)}\) keys \(k\), we obtain
  that the above bound holds for every \(k\in\{0,1\}^{\kappa(\secpar)}\) except with
  probability at most \( 2^{\kappa(\secpar)}\cdot 2^{-2{\kappa(\secpar)}}=2^{-{\kappa(\secpar)}} \).
  This proves the claim.
\end{proof}

 \fi

\ifnum\submission=1
\fi
\ifnum\submission=1
\fi
\fi

\end{document}